\documentclass[preprintnumbers, superscriptaddress, showpacs, nofootinbib, amsfonts, amsmath, amssymb, aps, prd, notitlepage, floatfix]{revtex4-2}

\usepackage[utf8]{inputenc}
\usepackage{amsmath,amssymb,amsfonts,mathrsfs,bbold}
\usepackage{yfonts}
\usepackage{graphicx}
\usepackage[colorlinks=true,citecolor=blue,linkcolor=blue,urlcolor=blue]{hyperref}
\usepackage[dvipsnames]{xcolor}
\usepackage{lipsum}
\usepackage{slashed}
\usepackage[normalem]{ulem}
\usepackage{url}
\usepackage{physics}

\usepackage{amssymb, bm, amsmath, graphicx}
\usepackage{enumerate}
\usepackage{amsfonts}
\usepackage{mathtools}
\usepackage{latexsym}
\usepackage{epstopdf}
\usepackage{subfigure}
\usepackage{tikz}
\usetikzlibrary{decorations.pathmorphing, decorations.markings, shapes.geometric, shapes.misc, calc, math}
\tikzstyle{gluon}=[decorate, decoration={coil,aspect=0.8, amplitude=1.5pt,  segment length=3pt}]
\usepackage{stackrel}
\usepackage[most]{tcolorbox}

\def\sec#1{{Sec.~\ref{#1}}}
\def\eq#1{{Eq.~(\ref{#1})}}
\def\fig#1{{Fig.~\ref{#1}}}
\newcommand{\ben}{\begin{eqnarray*}}
\newcommand{\een}{\end{eqnarray*}}

\newcommand{\as}{\alpha_s}

\allowdisplaybreaks

\begin{document}

\preprint{JLAB-THY-23-3896}

\title{Global analysis of polarized DIS \hspace*{-1.7mm} \& \hspace*{-1.7mm} SIDIS data with improved small-$x$ helicity evolution}

    \makeatletter  
\def\@fnsymbol#1{\ensuremath{\ifcase#1\or *\or \dagger\or \ddagger\or
   \mathsection\or \mathparagraph\or \|\or **\or \dagger\dagger
   \or \ddagger\ddagger \or \mathsection\mathsection \else\@ctrerr\fi}}
    \makeatother
\author{Daniel~Adamiak }
\affiliation{Department of Physics, The Ohio State University, Columbus, Ohio 43210, USA}
\affiliation{Jefferson Lab, Newport News, Virginia 23606, USA}
\author{Nicholas~Baldonado}
\affiliation{Department of Physics, New Mexico State University, Las Cruces, New Mexico 88003, USA}
\author{Yuri~V.~Kovchegov}
\affiliation{Department of Physics, The Ohio State University, Columbus, Ohio 43210, USA}
\author{W.~Melnitchouk}
\affiliation{Jefferson Lab, Newport News, Virginia 23606, USA}
\author{Daniel~Pitonyak }
\affiliation{Department of Physics, Lebanon Valley College, Annville, Pennsylvania 17003, USA}
\author{Nobuo~Sato}
\affiliation{Jefferson Lab, Newport News, Virginia 23606, USA}
\author{Matthew~D.~Sievert}
\affiliation{Department of Physics, New Mexico State University, Las Cruces, New Mexico 88003, USA}

\author{Andrey~Tarasov}
\affiliation{Department of Physics, North Carolina State University, Raleigh, North Carolina 27695, USA}
\affiliation{
Joint BNL-SBU 
Center for Frontiers in Nuclear Science (CFNS) at Stony Brook University,\protect\\
Stony Brook, New York 11794, USA}

\author{Yossathorn~Tawabutr}
         \affiliation{\mbox{Department of Physics, University of Jyv\"askyl\"a,  P.O. Box 35, 40014 University of Jyv\"askyl\"a, Finland}}
         \affiliation{Helsinki Institute of Physics, P.O. Box 64, 00014 University of Helsinki, Finland \\
      \vspace*{0.2cm}
      {\bf Jefferson Lab Angular Momentum (JAM) Collaboration
        \vspace*{0.2cm}}
        }


\begin{abstract}

We analyze the world polarized deep-inelastic scattering (DIS) and semi-inclusive DIS (SIDIS) data at low values of $x < 0.1$, using small-$x$ evolution equations for the flavor singlet and nonsinglet helicity parton distribution functions (hPDFs), which resum all powers of  both $\as \, \ln^2 (1/x)$ and $\as \, \ln (1/x) \, \ln (Q^2/Q_0^2)$ with $\as$ the strong coupling constant. The hPDFs for quarks, antiquarks, and gluons are extracted and evolved to lower values of $x$ to make predictions for the future Electron-Ion Collider (EIC). We improve on our earlier work by employing the more realistic large-$N_c\, \& N_f$ limit of the revised small-$x$ helicity evolution, and incorporating running coupling corrections along with
SIDIS data into the fit. We find an anti-correlation between the signs of the gluon and $C$-even quark hPDFs as well as the $g_1$ structure function. While the existing low-$x$ polarized DIS and SIDIS data are insufficient to  constrain the initial conditions for the polarized dipole amplitudes in the helicity evolution equations, future EIC data will allow more precise predictions for hPDFs and the $g_1$ structure function for $x$ values beyond those probed at the EIC. Using the obtained hPDFs, we discuss the contributions to the proton spin from quark and gluon spins at small $x$. 

\end{abstract}

\date{\today}
\maketitle
\tableofcontents

%
\section{Introduction}
%

\subsection{General motivation}
%

The proton spin puzzle has been one of the most intriguing and profound mysteries in our understanding of the proton structure for over three decades (for reviews see Refs.~\cite{Aidala:2012mv, Accardi:2012qut, Leader:2013jra, Aschenauer:2013woa, Aschenauer:2015eha, Boer:2011fh, Proceedings:2020eah, Ji:2020ena, AbdulKhalek:2021gbh}).
The main challenge is to determine, both qualitatively and quantitatively, how the proton spin is distributed among the spins and orbital angular momenta (OAM) of its quark and gluon constituents.
The question is usually formulated in terms of spin sum rules, such as the Jaffe-Manohar sum rule~\cite{Jaffe:1989jz} (see also the Ji sum rule~\cite{Ji:1996ek}), that decompose the proton spin of 1/2 (in units of $\hbar$) into the sum of the quark ($S_q$) and gluon ($S_G$) spins and the OAM carried by the quarks ($L_q$) and gluons ($L_G$), 
\begin{equation}
S_q+L_q+S_G+L_G=\frac{1}{2}\,. 
\label{eqn:JM}
\end{equation}
Each of the contributions in Eq.~\eqref{eqn:JM} can, in turn, be written as the integral of a partonic function over the longitudinal momentum fraction $x$ carried by the parton. For example,
\begin{align}\label{eqn:SqSG}
    S_q(Q^2) = \frac{1}{2} \int\limits_0^1 \dd{x} \, \Delta\Sigma(x,Q^2)\,, \qquad\qquad
    S_G(Q^2) = \int\limits_0^1 \dd{x} \, \Delta G(x,Q^2)\,,
\end{align}
with similar expressions for the OAM contributions \cite{Bashinsky:1998if, Hagler:1998kg, Harindranath:1998ve, Hatta:2012cs, Ji:2012ba}, where $\Delta\Sigma(x,Q^2)$ is the flavor singlet combination of the quark helicity parton distribution functions  (hPDFs) $\Delta q(x,Q^2)$ (quark flavor $q$) and $\Delta G(x,Q^2)$ is the gluon hPDF~\cite{Jaffe:1989jz}. The goal of current research in the field of proton spin physics is to determine $\Delta\Sigma(x,Q^2)$, $\Delta G(x,Q^2)$, $L_q (x, Q^2)$, and $L_G (x, Q^2)$ across a broad range of $x$ and $Q^2$ in order to quantify how much of the proton spin is carried by the partons in different kinematic regions.  

The standard way to address the proton spin puzzle is by extracting the hPDFs $\Delta q(x,Q^2)$ and $\Delta G(x,Q^2)$ from experimental data using collinear factorization along with the spin-dependent Dokshitzer-Gribov-Lipatov-Altarelli-Parisi (DGLAP) evolution equations \cite{Gribov:1972ri, Altarelli:1977zs, Dokshitzer:1977sg} to relate observables at different $Q^2$. 
There has been a number of very successful extractions of hPDFs over the years within this approach~\cite{Gluck:2000dy, Leader:2005ci, deFlorian:2009vb, Leader:2010rb, Jimenez-Delgado:2013boa, Ball:2013lla, Nocera:2014gqa, deFlorian:2014yva, Leader:2014uua, Sato:2016tuz, Ethier:2017zbq, DeFlorian:2019xxt, Borsa:2020lsz, Zhou:2022wzm, Cocuzza:2022jye}. Nevertheless, the DGLAP-based methodology has a drawback:~since the DGLAP equations evolve PDFs in $Q^2$, they cannot truly predict the $x$ dependence of PDFs. The $x$ dependence is greatly affected by the functional form of the PDF parametrization at the initial momentum scale $Q_0^2$, which gives the initial conditions for the DGLAP evolution. The parameters are then determined by optimizing agreement between the theoretical calculations to the experimental measurements. In this way, the experimental data, in the $x$ range where it is available, make up for the inability of DGLAP evolution to predict the $x$ dependence of PDFs. Conversely, in the $x$ region which has not yet been probed experimentally, DGLAP-based predictions typically acquire a broad uncertainty band due to extrapolation errors. This is particularly true in the small-$x$ region. Since no experiment, present or future, can perform measurements down to $x=0$, further theoretical input is needed to constrain the hPDFs at low $x$. The benefit of small-$x$ helicity evolution is it makes a genuine prediction for the hPDFs at small $x$ given some initial conditions at a higher $x_0$.  Due to the integrals in Eq.~\eqref{eqn:SqSG}, precise control over the behavior of hPDFs at small $x$ is mandatory to resolving the proton spin puzzle. 

\subsection{Proton spin at small $x$}
%

The first resummation of hPDFs at small $x$ was performed in the pioneering work by Bartels, Ermolaev and Ryskin (BER) \cite{Bartels:1995iu,Bartels:1996wc}, who employed the infrared evolution equations (IREE) formalism from Refs.~\cite{Gorshkov:1966ht,Kirschner:1983di,Kirschner:1994rq,Kirschner:1994vc,Griffiths:1999dj}. The BER IREE resummed double logarithms of $x$, {\it i.e.}, powers of the parameter $\as \, \ln^2 (1/x)$ (with $\as$ the strong coupling constant), which is referred to as the double-logarithmic approximation (DLA). The leading small-$x$ asymptotics for the flavor singlet combination of quark hPDFs and the gluon hPDF can be written as 
\begin{align}\label{DSigmaDG_asympt}
    \Delta\Sigma(x,Q^2) \sim \Delta G(x,Q^2) \sim \left( \frac{1}{x} \right)^{\!\alpha_h}\,,
\end{align}
with $\alpha_h$ the helicity intercept.  BER found $\alpha_h = 3.66 \sqrt{\frac{\as N_c}{2 \pi}}$ in the pure-gluon case and $\alpha_h = 3.45 \sqrt{\frac{\as N_c}{2 \pi}}$ for $N_f =4$ (the numbers $3.66$ and $3.45$ were calculated numerically, the latter for $N_c =3$, with $N_c$/$N_f$ being the number of quark colors/flavors). These intercepts are numerically large, with $\alpha_h >1$ for realistic coupling $\as = 0.2 - 0.3$, making the integrals~\eqref{eqn:SqSG} divergent as $x \to 0$. One may hope that the higher-order corrections in $\as$, when calculated, would lower the intercept $\alpha_h$ below $1$, making the integrals~\eqref{eqn:SqSG}  convergent. In addition, at very small $x$, parton saturation corrections  (see Refs.~\cite{Gribov:1984tu, Iancu:2003xm, Weigert:2005us, JalilianMarian:2005jf, Gelis:2010nm, Albacete:2014fwa, Kovchegov:2012mbw, Morreale:2021pnn} for reviews) are likely to significantly modify the asymptotics \eqref{DSigmaDG_asympt} by slowing down (or completely stopping) the growth of hPDFs with decreasing $x$ (see, {\it e.g.}, \cite{Itakura:2003jp} for the impact of saturation effects on the unpolarized flavor nonsinglet evolution). Phenomenological applications of the BER IREE approach were developed in Refs.~\cite{Blumlein:1995jp,Blumlein:1996hb,Ermolaev:1999jx,Ermolaev:2000sg,Ermolaev:2003zx,Ermolaev:2009cq}. Recently, the BER approach has been applied to the OAM distributions as well \cite{Boussarie:2019icw}. 

Over the past decade a new approach to helicity evolution at small $x$ has been developed~\cite{Kovchegov:2015pbl, Hatta:2016aoc, Kovchegov:2016zex, Kovchegov:2016weo, Kovchegov:2017jxc, Kovchegov:2017lsr, Kovchegov:2018znm, Kovchegov:2019rrz, Cougoulic:2019aja, Kovchegov:2020hgb, Cougoulic:2020tbc, Chirilli:2021lif, Kovchegov:2021lvz, Cougoulic:2022gbk} employing the shock wave/$s$-channel evolution formalism originally constructed in Refs.~\cite{Mueller:1994rr,Mueller:1994jq,Mueller:1995gb,Balitsky:1995ub,Balitsky:1998ya,Kovchegov:1999yj,Kovchegov:1999ua,Jalilian-Marian:1997dw,Jalilian-Marian:1997gr,Weigert:2000gi,Iancu:2001ad,Iancu:2000hn,Ferreiro:2001qy} for unpolarized eikonal scattering. The main idea behind the works \cite{Kovchegov:2015pbl, Hatta:2016aoc, Kovchegov:2016zex, Kovchegov:2016weo, Kovchegov:2017jxc, Kovchegov:2017lsr, Kovchegov:2018znm, Kovchegov:2019rrz, Cougoulic:2019aja, Kovchegov:2020hgb, Cougoulic:2020tbc, Chirilli:2021lif, Kovchegov:2021lvz, Cougoulic:2022gbk}  is that the sub-eikonal, sub-sub-eikonal, etc., quantities obey small-$x$ evolution equations similar to the eikonal ones \cite{Balitsky:1995ub,Balitsky:1998ya,Kovchegov:1999yj,Kovchegov:1999ua,Jalilian-Marian:1997dw,Jalilian-Marian:1997gr,Weigert:2000gi,Iancu:2001ad,Iancu:2000hn,Ferreiro:2001qy}, resulting from an $s$-channel gluon (or quark) cascade (see Refs.~\cite{Altinoluk:2014oxa,Balitsky:2015qba,Balitsky:2016dgz, Kovchegov:2017lsr, Kovchegov:2018znm, Chirilli:2018kkw, Jalilian-Marian:2018iui, Jalilian-Marian:2019kaf, Altinoluk:2020oyd, Kovchegov:2021iyc, Altinoluk:2021lvu, Kovchegov:2022kyy, Altinoluk:2022jkk, Altinoluk:2023qfr,Altinoluk:2023dww, Li:2023tlw} for the formalism of sub-eikonal and sub-sub-eikonal evolution in high-energy scattering). The sub-eikonal quantities are suppressed by one power of $x$ compared to the eikonal ones, sub-sub-eikonal quantities are suppressed by two powers of $x$, etc.

The equations developed in Refs.~\cite{Kovchegov:2015pbl, Kovchegov:2016zex, Kovchegov:2017lsr, Kovchegov:2018znm, Cougoulic:2019aja, Cougoulic:2022gbk} were also derived in the DLA. Similar to the unpolarized evolution equations~\cite{Balitsky:1995ub,Balitsky:1998ya,Kovchegov:1999yj,Kovchegov:1999ua,Jalilian-Marian:1997dw,Jalilian-Marian:1997gr,Weigert:2000gi,Iancu:2001ad,Iancu:2000hn,Ferreiro:2001qy}, the helicity evolution equations~\cite{Kovchegov:2015pbl, Kovchegov:2016zex, Kovchegov:2017lsr, Kovchegov:2018znm, Cougoulic:2022gbk} only take on a closed form in the large-$N_c$  \cite{tHooft:1973alw} and large-$N_c \& N_f$ \cite{Veneziano:1976wm} limits.  In that case  they become the evolution equations for the so-called ``polarized dipole amplitudes," which are dipole scattering amplitudes with an insertion of one gluon or two quark operators at the sub-eikonal level into the light-cone Wilson lines~\cite{Kovchegov:2017lsr, Kovchegov:2018znm, Kovchegov:2021iyc, Cougoulic:2022gbk}. The earlier version of this evolution, constructed in Refs.~\cite{Kovchegov:2015pbl, Kovchegov:2016zex, Kovchegov:2017lsr} (which we will refer to as KPS) led to an intercept of $\alpha_h = \frac{4}{\sqrt{3}} \sqrt{\frac{\as N_c}{2 \pi}} \approx 2.31 \sqrt{\frac{\as N_c}{2 \pi}}$ in the large-$N_c$ limit \cite{Kovchegov:2016weo, Kovchegov:2017jxc}, significantly smaller than the intercept of $\alpha_h = 3.66 \sqrt{\frac{\as N_c}{2 \pi}}$ found by BER in the same limit. The KPS evolution has recently been augmented~\cite{Cougoulic:2022gbk} by inclusion of the operators which couple what can be interpreted as the OAM of the gluon probe (in $A^- =0$ light-cone gauge of the projectile) to the spin of the proton.\footnote{We thank Florian Cougoulic, Alex Kovner, and Feng Yuan for suggesting this interpretation of those operators.} The revised evolution equations, which we will refer to as the KPS-CTT equations~\cite{Kovchegov:2015pbl, Kovchegov:2018znm,Cougoulic:2022gbk}, have been solved at large $N_c$ both numerically~\cite{Cougoulic:2022gbk} and analytically~\cite{Borden:2023ugd}. While the former reference found the numerical value of the intercept to be $\alpha_h = 3.66 \sqrt{\frac{\as N_c}{2 \pi}}$, appearing to agree with BER, the analytic solution \cite{Borden:2023ugd} found that the BER and KPS-CTT intercepts at large~$N_c$ disagree in the third decimal point. Very recently, a numerical solution of the large-$N_c \& N_f$ version of the KPS-CTT evolution~\cite{Adamiak:2023okq} established a disagreement with BER (in the same limit) at the 2--3$\%$ level, with the discrepancy increasing with $N_f$. While the observed differences between the two sets of results appear to demand further theoretical investigation, they are sufficiently small to allow one to proceed with rigorous phenomenological applications of the KPS-CTT evolution equations~\cite{Kovchegov:2015pbl, Kovchegov:2016zex, Kovchegov:2017lsr, Kovchegov:2018znm,  Cougoulic:2022gbk}.

The first phenomenological application of the polarized dipole amplitude formalism, more precisely its KPS version, was performed by a subset of the present authors in  Ref.~\cite{Adamiak:2021ppq}. In that work a successful ``proof-of-principle" fit of the world polarized DIS data for $x < 0.1$ and $Q^2 > m_c^2$ (with $m_c$ the charm quark mass) based solely on small-$x$ helicity evolution was performed. Since the analysis of Ref.~\cite{Adamiak:2021ppq} was limited to DIS data, only the $g_1$ structure functions of the proton and neutron were extracted instead of the individual flavor hPDFs.
The impact of DIS data from the EIC on our ability to predict the $g_1$ structure function at small-$x$ was also estimated. In addition, in order to demonstrate that it is possible to extract the combinations $\Delta q^+ (x, Q^2) \equiv \Delta q (x, Q^2) + \Delta \bar{q} (x, Q^2)$ for $q=u, d, s$ using small-$x$ helicity evolution, parity-violating DIS EIC pseudodata was utilized. We refer to $\Delta q^+(x,Q^2)$ as the $C$-even hPDFs, whereas the flavor nonsinglet $C$-odd hPDFs are similarly defined as $\Delta q^-(x,Q^2) \equiv \Delta q(x,Q^2) - \Delta \bar{q}(x,Q^2)$.

\subsection{Subject of this work}
%

In the present paper we  perform, for the first time, a phenomenological analysis based on the KPS-CTT  version of small-$x$ helicity evolution with several other significant new features beyond the work of Ref.~\cite{Adamiak:2021ppq}. Instead of the large-$N_c$ limit of evolution employed in Ref.~\cite{Adamiak:2021ppq}, we base our analysis on the large-$N_c \& N_f$ limit. In addition to the polarized DIS data, we  also include in our analysis polarized SIDIS data. Since the SIDIS data is sensitive to the individual quark and anti-quark helicity PDFs, $\Delta q(x, Q^2)$ and $\Delta \bar{q}(x, Q^2)$, it is not sufficient to just use the flavor singlet helicity evolution from Ref.~\cite{Cougoulic:2022gbk}, which only yields the $\Delta q^+ (x, Q^2)$ combination (in addition to the gluon hPDF $\Delta G (x, Q^2)$). One also needs the flavor nonsinglet quark hPDFs $\Delta q^- (x, Q^2)$. Those are constructed using the large-$N_c$ small-$x$ helicity evolution equation for the flavor nonsinglet case from Ref.~\cite{Kovchegov:2016zex}. Finally, to make the calculation more realistic and avoid the integrals \eqref{eqn:SqSG} diverging at $x \to 0$, we include running coupling corrections into the kernel of the evolution equations (both flavor singlet and nonsinglet).  We make the coupling run with the daughter dipole size, which ends up effectively reducing the intercept $\alpha_h$ for $\Delta q^+$ and $\Delta G$ below 1. (The intercept of the flavor nonsinglet hPDFs is smaller than 1 even at fixed coupling in the realistic $\as = 0.2 - 0.3$ range; still, for consistency, we apply running coupling corrections to the flavor nonsinglet helicity evolution as well.) The analysis of SIDIS data also requires input for fragmentation functions, which are not specific to the small-$x$ evolution at hand; therefore, we employ the existing JAM fragmentation functions for pions, kaons, and unidentified hadrons from Ref.~\cite{Cocuzza:2022jye}.

The paper is structured as follows. We begin in Sec.~\ref{sec:meth} by outlining the polarized dipole amplitude formalism developed in Refs.~\cite{Kovchegov:2015pbl, Kovchegov:2016zex, Kovchegov:2017lsr, Kovchegov:2018znm,  Cougoulic:2022gbk} and explicitly writing out the flavor-singlet KPS-CTT large $N_c \& N_f$ DLA small-$x$ helicity evolution equations with running coupling corrections, along with the flavor nonsinglet helicity evolution equation derived in Ref.~\cite{Kovchegov:2016zex}.  We also present the details of our numerical methodology in solving these evolution equations. We describe the calculation of observables (double-longitudinal spin asymmetries) in DIS and SIDIS, particularly detailing the calculation of the polarized SIDIS cross section at small $x$. We explain our analysis of the world polarized DIS and SIDIS low-$x$ data and describe the implementation of the KPS-CTT evolution within the JAM Bayesian Monte Carlo framework. The results of our analysis are presented in Sec.~\ref{sec:results}, which include plots of data versus theory, the hPDFs, and the $g_1$ structure function as well as an estimate of how much of the proton spin is carried by the net spin of partons at small $x$. We also conduct an EIC impact study on the aforementioned quantities.  Conclusions and an outlook are given in Sec.~\ref{sec:conclusions}.

%
\section{Methodology}
%

\label{sec:meth}

%
\subsection{Flavor singlet evolution at small $x$}
%

The small-$x$ helicity formalism in the light-cone operator treatment (LCOT) framework along with the large-$N_c \& N_f$ small-$x$ evolution equations for helicity were revised in Ref.~\cite{Cougoulic:2022gbk}. In the new formalism, the (DIS) $g_1$ structure function is given by
\begin{align}\label{g1_Dq}
    g_1 (x, Q^2)  = \frac{1}{2} \sum_q \, e^2_q \, \Delta q^+ (x, Q^2)\,,
\end{align}
where $e_q$ is the quark electric charge as a fraction of the magnitude of the electron's charge. The $C$-even quark hPDFs in the DLA take the form~\cite{Kovchegov:2018znm,Cougoulic:2022gbk}
\begin{align}\label{q+}
    \Delta q^+ (x, Q^2) \equiv \Delta q (x, Q^2) + \Delta \bar{q} (x, Q^2) = - \frac{N_c}{2 \pi^3} \:  \int\limits_{\Lambda^2/s}^1 \frac{\dd{z}}{z} \,  \int\limits_{1/zs}^{\min \left[ {1}/{z Q^2},  {1}/{\Lambda^2} \right]} \frac{\dd x^{2}_{10}}{x_{10}^2}  \, \left[  Q_q (x^2_{10} , zs) + 2 \, G_2 (x^2_{10} , zs) \right].
\end{align}
The gluon hPDF in the DLA is \cite{Kovchegov:2017lsr}
\begin{align}\label{JM_DeltaG}
    \Delta G (x, Q^2) = \frac{2 N_c}{\as \pi^2} \, G_2\!\!\left(\!x_{10}^2 = \frac{1}{Q^2},  zs = \frac{Q^2}{x}\right)\, .
\end{align}
Note that the quark and gluon hPDFs $\Delta q^+$ and $\Delta G$ are expressed in terms of the impact-parameter-integrated polarized dipole amplitudes $Q_q$ and $G_2$, whose operator definitions can be found in Refs.~\cite{Kovchegov:2015pbl,Kovchegov:2018znm,Cougoulic:2022gbk} and Ref.~\cite{Kovchegov:2017lsr}, respectively. The dipole amplitudes depend on the transverse size of the dipole $x_{10} = |{\bf x}_1 - {\bf x}_0|$, where the ``polarized" (sub-eikonally interacting) line is located at ${\bf x}_1$ and the unpolarized (standard) Wilson line is at ${\bf x}_0$ in the transverse plane. The amplitudes also depend on the center-of-mass energy squared $s$ of the projectile--proton scattering. The dimensionless longitudinal momentum fraction $z$ can be thought of as the momentum fraction of the softest of the two lines in the dipole.  (However, this definition is somewhat imprecise, and it is more accurate to think of $z s$ as the effective energy of the dipole--proton scattering \cite{Kovchegov:2015pbl, Kovchegov:2016zex, Kovchegov:2021lvz}.) The momentum scale $\Lambda$ denotes our infrared (IR) cutoff and is the scale characterizing the proton. No dipole can be larger than $1/\Lambda$, that is, the transverse size $x_{10} < 1/\Lambda$.

At small $x$, \eq{g1_Dq} was derived in Refs.~\cite{Kovchegov:2015pbl,Kovchegov:2016zex,Kovchegov:2016weo}. However, the contribution of $G_2$ to $\Delta q^+$ in \eq{q+} was recognized only recently~\cite{Cougoulic:2022gbk}. Given that $G_2$ is closely related to the gluon hPDF $\Delta G$, as follows from \eq{JM_DeltaG}, Eqs.~\eqref{g1_Dq} and \eqref{q+} show that in our LCOT approach the contribution of $\Delta G$ to $g_1$ comes in through $\Delta q^+$ \cite{Cougoulic:2022gbk,Adamiak:2023okq} (see more on this below). We have also expanded the definition of the amplitude $Q_q$ to include dependence on the quark flavor $q=u,d,s$, such that we have three different amplitudes $Q_u$, $Q_d$ and $Q_s$ for the light flavors, which is necessary since the quark spinor field operators are flavor dependent. The operator definition for the three flavors is the same, but the flavor dependence can enter through the initial condition of the dipole amplitude evolution.

While \eq{g1_Dq} appears to correspond to the leading-order (LO) expression in the collinear factorization approach to polarized DIS (see, {\it e.g.}, Eq.~(4.5) in Ref.~\cite{Lampe:1998eu}), in the LCOT framework it contains more information than that. In collinear factorization at the next-to-leading order (NLO) and beyond, the expression for the $g_1$ structure function also involves the contribution of $\Delta G$. More precisely, one can write \cite{Altarelli:1977zs,Dokshitzer:1977sg,Zijlstra:1993sh,Mertig:1995ny,Moch:1999eb,vanNeerven:2000uj,Vermaseren:2005qc,Moch:2014sna,Blumlein:2021ryt,Blumlein:2021lmf,Davies:2022ofz,Blumlein:2022gpp} 
\begin{align}\label{g1_coef_ftns}
    g_1 (x, Q^2)  = \frac{1}{2} \sum_q \, e^2_q \, \left\{ \Delta q^+ (x, Q^2) + \int\limits_x^1 \frac{\dd{z}}{z} \, \left[ \Delta c_q (z) \,  \Delta q^+\! \left( \frac{x}{z} , Q^2\right) +  \Delta c_G (z) \,  \Delta G\! \left( \frac{x}{z} , Q^2\right) \right] \right\},
\end{align}
with the coefficient functions $\Delta c_q (z)$ and $\Delta c_G (z)$ calculated order-by-order in perturbation theory. In the $\overline{\text{MS}}$ scheme, the small-$x$ large-$N_c \& N_f$ coefficient functions are \cite{Zijlstra:1993sh} (see also \cite{Blumlein:2022gpp} for the three-loop contribution, which we do not show explicitly here)
\begin{subequations}\label{coef_functions}
\begin{align}
    & \Delta c_q (z) = \frac{\as N_c}{4 \pi} \, \ln \frac{1}{z} + \frac{5}{12} \, \left( \frac{\as N_c}{4 \pi} \right)^2 \left[ 1 - 4 \, \frac{N_f}{N_c} \right] \, \ln^3 \frac{1}{z} + {\cal O} (\as^3)\,, \label{Cq_full}  \\
    & \Delta c_G (z) = - \frac{\as}{2 \pi} \, \ln \frac{1}{z} - \frac{11}{2} \, \left( \frac{\as }{4 \pi} \right)^2 \, N_c \, \ln^3 \frac{1}{z} + {\cal O} (\as^3\,). \label{dCg}
\end{align}
\end{subequations}
Note that after the $z$-integration in \eq{g1_coef_ftns}, the contribution from the order-$\as$ terms in Eqs.~\eqref{coef_functions} becomes of the order $\as \, \ln^2 (1/x)$, the contribution from the order-$\as^2$ terms in Eqs.~\eqref{coef_functions} becomes of the order $[\as \, \ln^2 (1/x)]^2$, etc. Consequently, in  the collinear factorization power counting, the contributions from $\Delta c_q (z)$ and $\Delta c_G (z)$ in \eq{g1_coef_ftns} are NLO and beyond, allowing one to truncate the expansion at a given order in $\as$ determined by the accuracy of the calculation. In our DLA small-$x$ power counting, the leading small-$x$ parts of $\Delta c_q (z)$ and $\Delta c_G (z)$ are already included to all orders in the powers of $\as \, \ln^2 (1/x)$. This is precisely what \eq{g1_Dq} accomplishes \cite{Adamiak:2023okq}. While it appears to be just the LO part of \eq{g1_coef_ftns}, the fact that $\Delta q^+$ in it is evolved with the DLA small-$x$ helicity evolution~\cite{Kovchegov:2015pbl, Kovchegov:2016zex, Kovchegov:2017lsr, Kovchegov:2018znm,  Cougoulic:2022gbk}, resuming powers of both $\as \, \ln^2 (1/x)$ and $\as \, \ln (1/x) \, \ln (Q^2/Q_0^2)$, implies that \eq{g1_Dq} contains both the DLA DGLAP evolution of $\Delta q^+$, which mixes it with $\Delta G$ (by resumming the powers of $\as \, \ln (1/x) \, \ln (Q^2/Q_0^2)$), and the leading small-$x$ parts of the coefficient functions $\Delta c_q (z)$ and $\Delta c_G (z)$, resummed to all orders in $\as \, \ln^2 (1/x)$, bringing in the $\Delta G$ and additional $\Delta q^+$ contributions into $g_1$, as expected from \eq{g1_coef_ftns} (see \cite{Adamiak:2023okq} for a more detailed discussion). The fact that all these contributions are contained in \eq{g1_Dq}, which looks much simpler than \eq{g1_coef_ftns}, appears to suggest that we are working in the ``polarized DIS scheme" \cite{Adamiak:2023okq} for our hPDFs (cf. \cite{Altarelli:1979ub} for the standard DIS scheme), where $\Delta G$ does not contribute to $g_1$ directly, unlike the more widely used $\overline{\mbox{MS}}$ scheme from \eq{g1_coef_ftns}. Other small-$x$ calculations, such as the NLO BFKL evolution \cite{Fadin:1998py,Ciafaloni:1998gs} (in the small-$x$ power counting), result in the spin-independent GG anomalous dimension in the DIS scheme \cite{vanNeerven:2000uj}. This appears to be similar to our calculation giving a polarized DIS scheme result, with the difference between the anomalous dimensions in different schemes being proportional to $N_f$~\cite{vanNeerven:2000uj,Adamiak:2023okq}.  

The polarized dipole amplitudes $Q_q$ and $G_2$, which enter Eqs.~\eqref{g1_Dq}, \eqref{q+} and \eqref{JM_DeltaG}, are found by solving the small-$x$ evolution equations. The DLA large-$N_c \& N_f$ revised evolution equations at fixed coupling are given by Eqs.~(155) in Ref.~\cite{Cougoulic:2022gbk} (see also Refs.~\cite{Kovchegov:2015pbl,Kovchegov:2018znm}). Its existing numerical solution \cite{Adamiak:2023okq} (with fixed coupling) leads to a large intercept $\alpha_h$ for the flavor singlet hPDFs and for $\Delta q^+$ (see \eq{DSigmaDG_asympt} with the intercept values in the text following that equation), making the integrals in \eq{eqn:SqSG} divergent as $x \to 0$. As we discussed above, this divergence may be regulated by higher-order corrections and/or by the onset of saturation, which is likely to slow down the growth of hPDFs as $x \to 0$. As the unpolarized small-$x$ evolution \cite{Mueller:1994rr,Mueller:1994jq,Mueller:1995gb,Balitsky:1995ub,Balitsky:1998ya,Kovchegov:1999yj,Kovchegov:1999ua,Jalilian-Marian:1997dw,Jalilian-Marian:1997gr,Weigert:2000gi,Iancu:2001ad,Iancu:2000hn,Ferreiro:2001qy} is single-logarithmic, resumming powers of $\as \, \ln (1/x)$, a consistent inclusion of saturation effects is beyond the double-logarithmic approximation employed here. While, strictly-speaking, phenomenology based on small-$x$ evolution in the DLA should work with the high intercepts found in Ref.~\cite{Adamiak:2023okq}, it appears unphysical to perform an analysis of experimental  data with a formalism that would yield an infinite amount of spin at small $x$. While we cannot include the single-logarithmic (resumming powers of $\as \, \ln (1/x)$) corrections to the revised DLA evolution equations (155) from Ref.~\cite{Cougoulic:2022gbk}, since they have not been fully calculated yet (see Ref.~\cite{Kovchegov:2021lvz} for the single-logarithmic corrections to the earlier KPS evolution), we can include running-coupling corrections into the DLA evolution. A similar approximation was employed in the BER framework~\cite{Ermolaev:1999jx,Ermolaev:2003zx} and for the spin-independent eikonal small-$x$ evolution \cite{Albacete:2009fh,Albacete:2010sy}, resulting in successful phenomenology. 

In the DLA equations (155) from Ref.~\cite{Cougoulic:2022gbk}, the scale of the coupling could be given by either the ``parent" ($x_{10}$) or the ``daughter" ($x_{21}$ or $x_{32}$) dipole. The running coupling corrections to the (un-revised) KPS evolution, calculated in Ref.~\cite{Kovchegov:2021lvz} (along with other single-logarithmic corrections), indicate that at DLA the coupling runs with the daughter dipole size. For the neighbor dipole amplitudes $\overline{\Gamma}, \widetilde{\Gamma}$, and $\Gamma_2$, introduced in Refs.~\cite{Kovchegov:2015pbl,Kovchegov:2016zex,Kovchegov:2017lsr,Kovchegov:2018znm,Cougoulic:2019aja,Cougoulic:2022gbk} and also entering helicity evolution equations, the coupling runs with the dipole size $x_{32}$, which determines the next emission's lifetime and is integrated over in the kernel  \cite{Kovchegov:2021lvz}. Therefore, we proceeded by running the coupling with the daughter dipole size (or, more precisely, with the dipole size that we integrate over in the kernel) in all the terms of the KPS-CTT evolution. (See Refs.~\cite{Balitsky:2006wa,Gardi:2006rp,Kovchegov:2006vj, Kovchegov:2006wf, Albacete:2007yr} for calculations and analyses of the running coupling corrections in the unpolarized small-$x$ evolution case.) The resulting running-coupling version of the large-$N_c \& N_f$ helicity evolution equations~(155) from \cite{Cougoulic:2022gbk} reads
\begin{subequations}\label{eq_LargeNcNf}
\begin{align}
    & Q_q (x^2_{10},zs) = Q_q^{(0)}(x^2_{10},zs) + \frac{N_c}{2\pi} \int^{z}_{1/x^2_{10}s} \frac{\dd z'}{z'}   \int_{1/z's}^{x^2_{10}}  \frac{\dd x^2_{21}}{x_{21}^2}  \ \as \!\!\left( \frac{1}{x_{21}^2} \right) \, \left[ 2 \, {\widetilde G}(x^2_{21},z's) + 2 \, {\widetilde \Gamma}(x^2_{10},x^2_{21},z's) \right. \\
    &\hspace*{5cm}\left.+ \; Q_q (x^2_{21},z's) -  \overline{\Gamma}_q (x^2_{10},x^2_{21},z's) + 2 \, \Gamma_2(x^2_{10},x^2_{21},z's) + 2 \, G_2(x^2_{21},z's)   \right] \notag \\
    &\hspace*{3cm}+ \frac{N_c}{4\pi} \int_{\Lambda^2/s}^{z} \frac{\dd z'}{z'}   \int_{1/z's}^{\min \left[ x^2_{10}z/z', 1/\Lambda^2 \right]}  \frac{\dd x^2_{21}}{x_{21}^2} \ \as\!\! \left( \frac{1}{x_{21}^2} \right) \, \left[Q_q (x^2_{21},z's) + 2 \, G_2(x^2_{21},z's) \right] ,  \notag  \\[0.5cm]
    &\overline{\Gamma}_q (x^2_{10},x^2_{21},z's) = Q^{(0)}_q (x^2_{10},z's) + \frac{N_c}{2\pi} \int^{z'}_{1/x^2_{10}s} \frac{\dd z''}{z''}   \int_{1/z''s}^{\min[x^2_{10}, x^2_{21}z'/z'']}  \frac{\dd x^2_{32}}{x_{32}^2}   \ \as\!\! \left( \frac{1}{x_{32}^2} \right) \, \left[ 2\, {\widetilde G} (x^2_{32},z''s)  \right. \\
    &\hspace*{2.5cm}\left.+ \; 2\, {\widetilde \Gamma} (x^2_{10},x^2_{32},z''s) +  Q_q (x^2_{32},z''s) -  \overline{\Gamma}_q (x^2_{10},x^2_{32},z''s) + 2 \, \Gamma_2(x^2_{10},x^2_{32},z''s) + 2 \, G_2(x^2_{32},z''s) \right] \notag \\
    &\hspace*{3cm}+ \frac{N_c}{4\pi} \int_{\Lambda^2/s}^{z'} \frac{\dd z''}{z''}   \int_{1/z''s}^{\min \left[ x^2_{21}z'/z'', 1/\Lambda^2 \right] }  \frac{\dd x^2_{32}}{x_{32}^2} \ \as\!\! \left( \frac{1}{x_{32}^2} \right) \, \left[Q_q (x^2_{32},z''s) + 2 \, G_2(x^2_{32},z''s) \right] , \notag \\[0.5cm]
    & {\widetilde G}(x^2_{10},zs) = {\widetilde G}^{(0)}(x^2_{10},zs) + \frac{N_c}{2\pi}\int^{z}_{1/x^2_{10}s}\frac{\dd z'}{z'}\int_{1/z's}^{x^2_{10}} \frac{\dd x^2_{21}}{x^2_{21}} \ \as\!\! \left( \frac{1}{x_{21}^2} \right) \, \left[3 \, {\widetilde G}(x^2_{21},z's) + {\widetilde \Gamma}(x^2_{10},x^2_{21},z's) \right. \\
    &\hspace*{4cm}\left.  + \; 2\,G_2(x^2_{21},z's)  +  \left(2 - \frac{N_f}{2N_c}\right) \Gamma_2(x^2_{10},x^2_{21},z's) - \frac{1}{4N_c}\, \sum_q \overline{\Gamma}_q (x^2_{10},x^2_{21},z's)\right] \notag \\
    &\hspace*{3cm}- \frac{1}{8\pi}  \int_{\Lambda^2/s}^z \frac{\dd z'}{z'}\int_{\max[x^2_{10},\,1/z's]}^{\min \left[ x^2_{10}z/z' , 1/\Lambda^2 \right]} \frac{\dd x^2_{21}}{x^2_{21}} \ \as\!\! \left( \frac{1}{x_{21}^2} \right) \, \left[ \sum_q  Q_q (x^2_{21},z's) +     2 \, N_f \, G_2(x^2_{21},z's)  \right] , \notag \\[0.5cm]
    & {\widetilde \Gamma} (x^2_{10},x^2_{21},z's) = {\widetilde G}^{(0)}(x^2_{10},z's) + \frac{N_c}{2\pi}\int^{z'}_{1/x^2_{10}s}\frac{\dd z''}{z''}\int_{1/z''s}^{\min[x^2_{10},x^2_{21}z'/z'']} \frac{\dd x^2_{32}}{x^2_{32}} \ \as\!\! \left( \frac{1}{x_{32}^2} \right) \, \left[3 \, {\widetilde G} (x^2_{32},z''s) \right. \\
    &\hspace*{2cm}\left. + \; {\widetilde \Gamma}(x^2_{10},x^2_{32},z''s) + 2 \, G_2(x^2_{32},z''s)  +  \left(2 - \frac{N_f}{2N_c}\right) \Gamma_2(x^2_{10},x^2_{32},z''s) - \frac{1}{4N_c} \sum_q \overline{\Gamma}_q (x^2_{10},x^2_{32},z''s)  \right] \notag \\
    &\hspace*{1.75cm}- \frac{1}{8\pi}  \int_{\Lambda^2/s}^{z'x^2_{21}/x^2_{10}} \frac{\dd z''}{z''}\int_{\max[x^2_{10},\,1/z''s]}^{\min \left[ x^2_{21}z'/z'', 1/\Lambda^2 \right]} \frac{\dd x^2_{32}}{x^2_{32}} \ \as\!\! \left( \frac{1}{x_{32}^2} \right) \, \left[   \sum_q \, Q_q (x^2_{32},z''s) +  2  \,  N_f \, G_2(x^2_{32},z''s)  \right] , \notag \\[0.5cm]
    & G_2(x_{10}^2, z s)  =  G_2^{(0)} (x_{10}^2, z s) + \frac{N_c}{\pi} \, \int\limits_{\Lambda^2/s}^z \frac{\dd z'}{z'} \, \int\limits_{\max \left[ x_{10}^2 , \frac{1}{z' s} \right]}^{\min \left[ \frac{z}{z'} x_{10}^2, \frac{1}{\Lambda^2} \right] } \frac{\dd x^2_{21}}{x_{21}^2} \ \as\!\! \left( \frac{1}{x_{21}^2} \right) \, \left[ {\widetilde G} (x^2_{21} , z' s) + 2 \, G_2 (x_{21}^2, z' s)  \right] , \\[0.5cm]
    & \Gamma_2 (x_{10}^2, x_{21}^2, z' s)  =  G_2^{(0)} (x_{10}^2, z' s) + \frac{N_c}{\pi} \!\! \int\limits_{\Lambda^2/s}^{z' \frac{x_{21}^2}{x_{10}^2}} \frac{\dd z''}{z''} \!\!\!\!\!\!\!\!\! \int\limits_{\max \left[ x_{10}^2 , \frac{1}{z'' s} \right]}^{\min \left[  \frac{z'}{z''} x_{21}^2, \frac{1}{\Lambda^2} \right] } \!\! \frac{\dd x^2_{32}}{x_{32}^2} \ \as\!\! \left( \frac{1}{x_{32}^2} \right) \! \left[ {\widetilde G} (x^2_{32} , z'' s) + 2 \, G_2(x_{32}^2, z'' s)  \right] . 
\end{align}
\end{subequations}

The running coupling in Eqs.~\eqref{eq_LargeNcNf} is given by the standard one-loop expression, 
\begin{equation}\label{rc_1loop}
    \alpha_s(Q^2) = \frac{12\pi}{11N_c-2N_f}\frac{1}{\ln (Q^2 /\Lambda_\textrm{QCD}^2)},
\end{equation}
with $\Lambda_\textrm{QCD}$ the QCD confinement scale. We have also modified Eqs.~\eqref{eq_LargeNcNf} compared to Eqs.~(155) in Ref.~\cite{Cougoulic:2022gbk} in two additional ways:~first, we are  now treating the momentum scale $\Lambda$ as the infrared cutoff (assuming that $\Lambda > \Lambda_\textrm{QCD}$); second, since the amplitude $Q_q$ is now flavor dependent, we replaced the $N_f$ factors from Ref.~\cite{Cougoulic:2022gbk} by flavor sums~$\left( \sum_q \right)$. Eqs.~\eqref{eq_LargeNcNf} also include the dipole amplitude ${\widetilde G}$, which is defined in Ref.~\cite{Cougoulic:2022gbk}: as one can see from Eqs.~\eqref{g1_Dq}, \eqref{q+} and \eqref{JM_DeltaG}, the $g_1$ structure function and hPDFs do not depend on this dipole amplitude: this will affect our analysis below. Following Refs.~\cite{Kovchegov:2015pbl,Kovchegov:2016zex,Kovchegov:2017lsr,Kovchegov:2018znm,Cougoulic:2019aja,Cougoulic:2022gbk} we have introduced the impact-parameter integrated ``neighbor dipole amplitudes'' $\overline{\Gamma}_q (x^2_{10},x^2_{32},zs)$, ${\widetilde \Gamma} (x^2_{10},x^2_{32},zs)$ and $\Gamma_2(x^2_{10},x^2_{32},zs)$ for the amplitudes $Q_q$, ${\widetilde G}$ and $G_2$,  respectively, with physical dipole transverse size $x_{10}$ and lifetime $\sim x^2_{32}z$. This lifetime for the neighbor dipole amplitudes depends on the transverse size of another (adjacent) dipole, giving rise to the ``neighbor" amplitude name.

The inhomogeneous terms (initial conditions) in Eqs.~\eqref{eq_LargeNcNf} can be calculated at the Born level for a longitudinally polarized massless quark target instead of the proton. This gives \cite{Kovchegov:2015pbl,Kovchegov:2016zex, Cougoulic:2022gbk, Kovchegov:2017lsr}
\begin{equation}\label{BornIC}
    \widetilde{G}^{(0)}(x_{10}^2,zs) = Q_q^{(0)}(x_{10}^2,zs) = \frac{\alpha_s^2C_F}{2N_c}\pi\Bigl[C_F\ln{\frac{zs}{\Lambda^2}}-2\ln{(zsx_{10}^2)}\Bigr],\qquad G_2^{(0)}(x_{10}^2,zs) = \frac{\alpha_s^2C_F}{N_c}\pi\ln{\frac{1}{x_{10}\Lambda}}\, ,
\end{equation}
where $C_F = (N_c^2 -1)/(2 N_c)$ is the Casimir operator in the fundamental representation of SU($N_c$). These expressions will motivate our choice of the initial conditions for our phenomenological analysis. (While strictly-speaking we should have included running coupling corrections into the expressions \eqref{BornIC} as well, the fixed-coupling form has a sufficient variety of dependence on the relevant variables $zs$ and $x_{10}$ to motivate a fairly broad class of initial conditions we will implement below.)

%
\subsection{Flavor nonsinglet evolution at small $x$}
\label{sec:Non-sing-evol}
%

As one can see from \eq{g1_Dq} in the previous subsection, measurements of the $g_1$ structure function in DIS off a nucleon are only sensitive to a specific linear combination of $\Delta q^+ (x, Q^2)$. Such DIS measurements were the topic of our previous study~\cite{Adamiak:2021ppq}.  However, the polarized SIDIS  process, as we will see below, provides information on the individual flavor  hPDFs $\Delta q (x, Q^2)$, or, equivalently, on both $\Delta q^+ (x, Q^2)$ and
$\Delta q^- (x, Q^2) \equiv \Delta q (x, Q^2) - \Delta \bar{q}(x, Q^2)$. The above evolution equations \eqref{eq_LargeNcNf} only allow us to calculate $\Delta q^+ (x, Q^2)$. To perform the polarized SIDIS data analysis we need to supplement them with the small-$x$ helicity evolution in the flavor nonsinglet channel.

A closed evolution equation at small $x$ yielding $\Delta q^- (x, Q^2)$ in the LCOT framework can be obtained in the large-$N_c$ limit, which is equivalent to the large-$N_c\&N_f$ limit for the flavor nonsinglet helicity evolution in DLA. (In the DLA, the flavor nonsinglet evolution is $N_f$-independent, since virtual quark bubbles do not contribute.  Thus, the large-$N_c$ and large-$N_c \& N_f$ limits are identical for flavor nonsinglet evolution.) Employing Eq.~(54b) of \cite{Kovchegov:2016zex} we write in the DLA
\begin{equation}\label{dq-}
    \Delta q^- (x, Q^2) \equiv \Delta q (x, Q^2) - \Delta \bar{q} (x, Q^2) = \frac{N_c}{2\pi^3}\int\limits_{{\Lambda^2}/{s}}^1\frac{\dd z}{z}\int\limits_{{1}/{zs}}^{\min \left[ {1}/{z Q^2} , {1}/{\Lambda^2} \right]} \frac{\dd x_{10}^2}{x_{10}^2} \, G_q^\textrm{NS}(x_{10}^2,z s)\,. 
\end{equation}
We see that $\Delta q^- (x, Q^2)$ only depends on one (impact-parameter integrated) polarized dipole amplitude, $G^\textrm{NS}_q (x_{10}^2,z s)$, for each flavor $q = u, d, s$. The definition of this dipole amplitude can be found in Eqs.~(55) of Ref.~\cite{Kovchegov:2016zex}. Just as in the flavor singlet case, the nonsinglet dipole amplitude can be determined by solving the small-$x$ evolution equation, 
which reads \cite{Kovchegov:2016zex}
\begin{align}\label{NSeq_LargeNcNf}
    G_q^\textrm{NS}(x_{10}^2,z) = G_q^{\textrm{NS} \,(0)}(x_{10}^2, z) + \frac{N_c}{4\pi}\int\limits_{{\Lambda^2}/{s}}^z\frac{\dd z'}{z'}\int\limits_{{1}/{z's}}^{\min \left[ x_{10}^2{z}/{z'}, {1}/{\Lambda^2} \right]}\frac{\dd x_{21}^2}{x_{21}^2} \ \as\! \left( \frac{1}{x_{21}^2} \right) \, G_q^\textrm{NS}(x_{21}^2,z').
\end{align}
To be consistent with the flavor-singlet evolution, we have also inserted a running coupling into \eq{NSeq_LargeNcNf}, modifying it slightly compared to the fixed-coupling flavor nonsinglet evolution equation derived in Ref.~\cite{Kovchegov:2016zex}. The inhomogeneous term in \eq{NSeq_LargeNcNf} can also be calculated at Born level for a quark target~\cite{Kovchegov:2016zex}:
\begin{equation}\label{inhom_nonsing}
    G_q^{\textrm{NS} \,(0)}(x_{10}^2,zs) = \frac{\alpha_s^2C_F^2}{N_c}\pi\ln{\frac{zs}{\Lambda^2}}.
\end{equation}
This expression will again motivate our choice of the flavor nonsinglet initial conditions in phenomenology. 

%
\subsection{Numerical implementation of the flavor singlet and nonsinglet evolution}
\label{sec:sing_num_impl}
%

Similar to our previous works \cite{Kovchegov:2016weo,Kovchegov:2020hgb,Cougoulic:2022gbk,Adamiak:2023okq}, small-$x$ helicity evolution equations simplify if one performs the following change of variables,
\begin{equation}\label{varchange}
    \eta^{(n)} = \sqrt{\frac{N_c}{2\pi}}\, \ln{\frac{z^{(n)}s}{\Lambda^2}},\qquad s_{ij} = \sqrt{\frac{N_c}{2\pi}}\, \ln{\frac{1}{x_{ij}^2\Lambda^2}}.
\end{equation}
Here $z^{(n)} = z, z', z'', \ldots$, while $\eta^{(n)} = \eta, \eta', \eta'', \ldots$.
Note that this form, in contrast to the earlier works, removes the factor $\sqrt{\alpha_s}$ from the definition of the variables $\eta$ and $s_{ij}$, so that the one-loop running of the coupling can be implemented via (cf. \eq{rc_1loop})
\begin{equation}\label{rc}
    \alpha_s(s_{21}) = \sqrt{\frac{N_c}{2\pi}}\frac{12\pi}{(11N_c-2N_f)}\frac{1}{(s_{21}+s_0)}\,,\qquad s_0 = \sqrt{\frac{N_c}{2\pi}}\ln{\frac{\Lambda^2}{\Lambda_\textrm{QCD}^2}}\,.
\end{equation}
Since we assume that $\Lambda > \Lambda_\textrm{QCD}$, we have $s_0 >0$. As all our dipole sizes are smaller than $1/\Lambda$, we see that $s_{21} >0$, thus avoiding the Landau pole at $s_{21} = - s_0 < 0$ in the coupling. (In general, having an IR cutoff for the dipole sizes, $x_{ij} < 1/\Lambda$, implies that all $s_{ij} >0$.)

Before discretizing our evolution equations, we need to impose the starting value of $x$ for our evolution (cf. Ref.~\cite{Adamiak:2021ppq}). For $z=1$ and $x_{10} = 1/Q$, we have the ``rapidity" variable $y \equiv \eta - s_{10} = \sqrt{\frac{N_c}{2\pi}}\ln{\frac{1}{x}}$. Hence, if our evolution starts at some value of $x$ labeled by $x_0$, then the $x<x_0$ condition implies that $\eta - s_{10} > \sqrt{\frac{N_c}{2\pi}}\ln{\frac{1}{x_0}} \equiv y_0$. Regarding the value of $x_0$, it was observed in Ref.~\cite{Adamiak:2021ppq}, using the older (KPS) version of our helicity evolution, that good-$\chi^2$ fits of the polarized DIS data can be obtained with $x_0 = 0.1$ (and even for a slightly higher values of $x_0$). This is in contrast to the $x_0 = 0.01$ starting point of the evolution \cite{Balitsky:1995ub,Balitsky:1998ya,Kovchegov:1999yj,Kovchegov:1999ua,Jalilian-Marian:1997dw,Jalilian-Marian:1997gr,Weigert:2000gi,Iancu:2001ad,Iancu:2000hn,Ferreiro:2001qy} for phenomenological analyses of the unpolarized observables (see, {\it e.g.}, Refs.~\cite{Albacete:2010sy,Albacete:2009fh}). As discussed in Sec.~\ref{sec:results_datavtheory} below, it was speculated in Ref.~\cite{Adamiak:2021ppq} that such a discrepancy could be attributed to the helicity evolution resumming the double-logarithmic parameter $\as \, \ln^2 (1/x)$ while the unpolarized evolution \cite{Kuraev:1977fs, Balitsky:1978ic, Kovchegov:1999yj, Kovchegov:1999ua, Jalilian-Marian:1997dw, Jalilian-Marian:1997gr, Weigert:2000gi, Iancu:2001ad, Iancu:2000hn, Ferreiro:2001qy} resums single logarithms $\as \, \ln (1/x)$. This way, the resummation parameter for helicity evolution is larger at small $x$, making the helicity evolution start at larger $x$ values. We thus put $x_0 = 0.1$ in all our analyses below.
\footnote{Note that the $x< x_0$ condition is applied only to our small-$x$ helicity evolution equations. The expressions for the $g_1$ structure function~\eqref{g1_Dq} and the quark \eqref{q+} and gluon \eqref{JM_DeltaG} hPDFs remain as shown above: for $x>x_0$ they are driven by the initial conditions/inhomogeneous terms for our evolution (cf.~Ref.~\cite{Adamiak:2021ppq}). The coupling in \eq{JM_DeltaG} runs with $Q^2$.} 

The full process of discretizing our flavor singlet and nonsinglet evolution equations with running coupling is detailed in Appendix~\ref{Disc_apdx}. In the end, the discretized version of Eqs.~\eqref{eq_LargeNcNf} written in terms of the variables \eqref{varchange} reads
\begin{subequations}\label{DiscreteEvol}
    \begin{align}
        Q_q[i,j] & = Q_q[i,j-1] + Q_q^{(0)}[i,j] -Q_q^{(0)}[i,j-1] \label{DiscreteEvol_Q} \\
        &\;\;\;\;\;+ \Delta^2\sum_{i'=i}^{j-2-y_0}\,\alpha_s[i']\Bigl[\frac{3}{2}Q_q[i',j-1]+2\widetilde{G}[i',j-1]+2\widetilde{\Gamma}[i,i',j-1] \notag \\
        &\;\;\;\;\;\;\;\;\;\;-\overline{\Gamma}_q[i,i',j-1]+3G_2[i',j-1]+2\Gamma_2[i,i',j-1]\Bigr] \notag  \\
        &\;\;\;\;\;+ \frac{1}{2}\Delta^2\sum_{j'=j-1-i}^{j-2}\,\alpha_s[i+j'-j+1]\Bigl[Q_q[i+j'-j+1,j'] + 2G_2[i+j'-j+1,j']\Bigr] \, , \notag  \\[0.5cm]
        \overline{\Gamma}_q[i,k,j] &= \overline{\Gamma}_q[i,k-1,j-1]+Q_q^{(0)}[i,j] - Q_q^{(0)}[i,j-1] \label{DiscreteEvol_Gmb} \\
        &\;\;\;\;\;+ \Delta^2\sum_{i'=k-1}^{j-2-y_0}\,\alpha_s[i']\Bigl[\frac{3}{2}Q_q[i',j-1]+2\widetilde{G}[i',j-1]+2\widetilde{\Gamma}[i,i',j-1] \notag \\
        &\;\;\;\;\;\;\;\;\;\;-\overline{\Gamma}_q[i,i',j-1]+3G_2[i',j-1]+2\Gamma_2[i,i',j-1]\Bigr] \, , \notag  \\[0.5cm]
        \widetilde{G}[i,j] & = \widetilde{G}[i,j-1] + \widetilde{G}^{(0)}[i,j] - \widetilde{G}^{(0)}[i,j-1] \label{DiscreteEvol_Gt} \\ 
        &\;\;\;\;\;+\Delta^2\sum_{i'=i}^{j-2-y_0}\,\alpha_s[i']\Bigl[3\widetilde{G}[i',j-1]+\widetilde{\Gamma}[i,i',j-1] \notag \\
        &\;\;\;\;\;\;\;\;\;\;+2G_2[i',j-1]+\bigl(2-\frac{N_f}{2N_c}\bigr)\Gamma_2[i,i',j-1]-\frac{1}{4N_c}\sum_q\overline{\Gamma}_q[i,i',j-1]\Bigr] \notag  \\
        &\;\;\;\;\;- \Delta^2\frac{1}{4N_c}\sum_{j'=j-1-i}^{j-2}\,\alpha_s[i+j'-j+1]\Bigl[\sum_qQ_q[i+j'-j+1,j'] + 2N_fG_2[i+j'-j+1,j']\Bigr] \, , \notag  \\[0.5cm]
        \widetilde{\Gamma}[i,k,j] &= \widetilde{\Gamma}[i,k-1,j-1]+ \widetilde{G}^{(0)}[i,j] - \widetilde{G}^{(0)}[i,j-1] \label{DiscreteEvol_Gmt} \\
        &\;\;\;\;\;+ \Delta^2\sum_{i'=k-1}^{j-2-y_0}\,\alpha_s[i']\Bigl[3\widetilde{G}[i',j-1]+\widetilde{\Gamma}[i,i',j-1] \notag \\
        &\;\;\;\;\;\;\;\;\;\;+2G_2[i',j-1]+\bigl(2-\frac{N_f}{2N_c}\bigr)\Gamma_2[i,i',j-1] - \frac{1}{4N_c}\sum_q\overline{\Gamma}_q[i,i',j-1]\Bigr] \, , \notag  \\[0.5cm]
        G_2[i,j] &= G_2[i,j-1] + G_2^{(0)}[i,j]-G_2^{(0)}[i,j-1]  \label{DiscreteEvol_G2} \\
        &\;\;\;\;\;+2\Delta^2\sum_{j'=j-1-i}^{j-2}\,\alpha_s[i+j'-j+1]\Bigl[\widetilde{G}[i+j'-j+1,j']+2G_2[i+j'-j+1,j']\Bigr] \, , \notag  \\[0.5cm]
    \Gamma_2[i,k,j] &= \Gamma_2[i,k-1,j-1] + G_2^{(0)}[i,j] - G_2^{(0)}[i,j-1]   \, , \label{DiscreteEvol_Gm2}
\end{align}
\end{subequations}
where the numerical step sizes are chosen such that $\Delta \eta = \Delta s_{10} = \Delta s_{21} \equiv \Delta$, and the indices are defined by $\{\eta,s_{10},s_{21}\}\to \{j,i,k\}\cdot\Delta$. Eqs.~\eqref{DiscreteEvol} allow us to compute the numerical solution for the flavor singlet evolution equations \eqref{eq_LargeNcNf}. Note that it is only necessary to loop over the ranges dictated by our physical assumptions, $0\leq i\leq k\leq j\leq j_{max}$ and $i<j$. Furthermore, it is useful to notice that the neighbor dipole amplitudes reduce to their dipole-amplitude counterparts when $k=i$, that is,
\begin{subequations}\label{GmG}
    \begin{align}
        \overline{\Gamma}_q[i,k=i,j] &= Q_q[i,j]\,, \\
    \widetilde{\Gamma}[i,k=i,j] &= \widetilde{G}[i,j]\,,\\ 
    \Gamma_2[i,k=i,j] &= G_2[i,j]\,.
    \end{align}
\end{subequations}

We can continue this convention and write the quark and gluon hPDFs from Eqs.~\eqref{q+} and \eqref{JM_DeltaG} in the new variables,
\begin{align}\label{q+2}
    \Delta q^+ (x, Q^2) = - \frac{1}{\pi^2} \, \int\limits_0^{\sqrt{\frac{N_c}{2 \pi}} \, \ln \frac{Q^2}{x \, \Lambda^2}} \dd \eta \, \int\limits^\eta_{\max \left[ 0, \eta - \sqrt{\frac{N_c}{2 \pi}} \ln \frac{1}{x} \right] } \dd s_{10} \, \left[ Q_q (s_{10}, \eta) + 2 \, G_2 (s_{10}, \eta) \right] ,
\end{align}
and 
\begin{align}\label{JM_DeltaG2}
\Delta G (x, Q^2) = \frac{2 N_c}{\as (Q^2) \, \pi^2} \, G_2 \!\left(\!  s_{10} = \sqrt{\frac{N_c}{2 \pi}} \ln \frac{Q^2}{\Lambda^2} , \, \eta = \sqrt{\frac{N_c}{2 \pi}} \, \ln \frac{Q^2}{x \Lambda^2} \right) ,
\end{align}
where the only difference compared to $\Delta G$ from \eq{JM_DeltaG} is the running coupling. 

    \begin{figure}[t!]
    \begin{centering}
    \includegraphics[width= 0.85 \textwidth] 
    {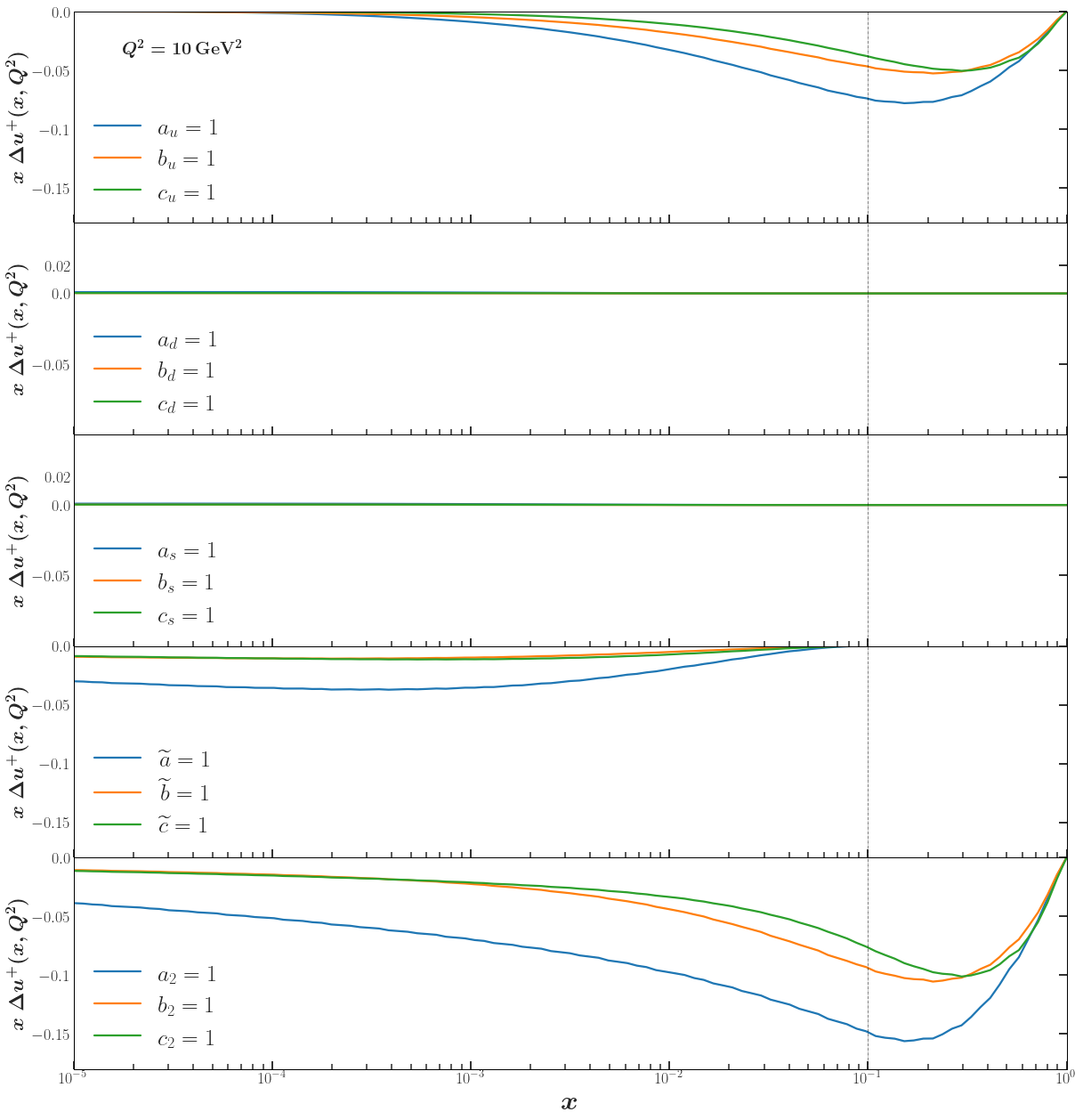}
    \caption{The $u$-quark hPDF, $x\Delta u^+(x)$, constructed solely out of each basis function in the range $x\in[10^{-5}, 1]$. The legend in each panel shows which basis function was used for which curve. For example, the blue curve in the top panel corresponds to $x\Delta u^+(x)$ constructed from the initial conditions $Q_u^{(0)}=\eta$ and $Q_q^{(0)}=\widetilde{G}^{(0)}=G_2^{(0)}=0$ for $q\in\{d,s\}$. The evolution begins at $x_0=0.1$, and the coupling constant runs with the daughter-dipole prescription specified in Eqs.~\eqref{setaint}. 
        \label{xhpdfu_medx}
    }
    \end{centering}
    \end{figure}
    
The last pieces to consider are the inhomogeneous terms. According to the Born-level initial conditions \eqref{BornIC}, they can be re-written using our new logarithmic variables as
\begin{subequations}\label{BornICdisc}
\begin{align}
    Q^{(0)}_q(s_{10},\eta) &= \widetilde{G}^{(0)}(s_{10},\eta)   = \frac{\alpha_s^2C_F\pi}{2N_c}\sqrt{\frac{2\pi}{N_c}}\Bigl[(C_F-2)\eta+2s_{10}\Bigr],\\
    G_2^{(0)}(s_{10},\eta)  &= \frac{\alpha_s^2C_F\pi}{2N_c}\sqrt{\frac{2\pi}{N_c}}\,s_{10}\,.
\end{align}
\end{subequations}
Since Eqs.~\eqref{BornICdisc} are linear in $\eta$ and $s_{10}$, we follow Ref.~\cite{Adamiak:2021ppq} and employ the linear-expansion ansatz, {\it i.e.},
\begin{subequations}\label{LinICdisc}
\begin{align}
    Q^{(0)}_q(s_{10},\eta) &= a_q \, \eta + b_q \, s_{10} + c_q \, ,  \\
    \widetilde{G}^{(0)}(s_{10},\eta)   &= \widetilde{a} \, \eta + \widetilde{b} \, s_{10} + \widetilde{c}\, ,\\
    G_2^{(0)}(s_{10},\eta)  &= a_2 \, \eta + b_2 \, s_{10} + c_2 \, .
\end{align}
\end{subequations}
Thus, for the three light flavors we consider, $q=u,d,s$, the full set of initial conditions for the flavor singlet evolution depends on 15 parameters $a_u, b_u, c_u, a_d, \ldots , c_2$ which we will fit to the data. Moreover, because the evolution equations we are solving are linear, their solution can be written as a linear combination of 15 ``basis'' dipole amplitudes, each of which is constructed by performing the iterative calculation outlined above while setting one parameter (from all the $a$'s, $b$'s and $c$'s) in Eqs.~\eqref{LinICdisc} to be 1 and all the other parameters to 0. Furthermore, since all hPDFs and the $g_1$ structure function depend linearly on the polarized dipole amplitudes, they are also linear combinations of their corresponding basis functions as well. 

For example, $\Delta u^+(x)$ can be expressed as a linear combination of the 15 ``basis hPDFs'' shown in \fig{xhpdfu_medx}. Since $\Delta u^+(x)$ depends directly on the linear combination $Q_u + 2 G_2$ (see \eq{q+}), one may expect that $Q_u$ and $G_2$ have the largest contributions to $\Delta u^+(x)$ at moderate $x$. This is indeed the case, with the top and bottom panels in \fig{xhpdfu_medx} having the largest-magnitude contributions to $\Delta u^+(x)$. Some of the other amplitudes contribute more significantly at lower $x$'s, as their magnitudes begin to influence those of $Q_u$ and/or $G_2$ through evolution. At the smallest values of $x$ in \fig{xhpdfu_medx}, the largest contributor is $G_2$, followed by $\widetilde{G}$, while the contributions from $Q_d$ and $Q_s$ remain small for all values of $x$. 

A consequence of this observation, that we will return to later, is that the sign of the $g_1$ structure function is influenced mainly by the sign of $G_2$ (or, equivalently, the sign of $\Delta G$) and the sign of $\widetilde G$. A challenge for phenomenology presents itself: $\widetilde G$ is slow to grow and hence less sensitive to available data near $x=x_0$, but it has a potentially large effect on the small-$x$ asymptotics. Unless we have sufficient data from an observable that is directly sensitive to $\widetilde G$, constraining that amplitude will be difficult.

Similar to the singlet evolution, the discretization of the nonsinglet evolution equation (\ref{NSeq_LargeNcNf}) reads (again, see Appendix~\ref{Disc_apdx} for details)
\begin{align}\label{eq_nonsinglet_numerical_solution}
        G^\textrm{NS}[i,j] & = G^\textrm{NS}[i,j-1]+G^{\textrm{NS} \,(0)}[i,j] - G^{\textrm{NS} \,(0)}[i,j-1] \\
        &+ \frac{1}{2}\Delta^2\Biggl[\sum\limits_{i'=i}^{j-2-y_0}\alpha_s[i'] \, G^\textrm{NS}[i',j-1]+\sum\limits_{j'=j-1-i}^{j-2} \as[i-j+1+j'] \, G^\textrm{NS}[i-j+1+j',j']\Biggr] . \notag
\end{align}
The corresponding flavor nonsinglet quark hPDF is given by
\begin{align}\label{q-2}
    \Delta q^- (x, Q^2) = - \frac{1}{\pi^2} \, \int\limits_0^{\sqrt{\frac{N_c}{2 \pi}} \, \ln \frac{Q^2}{x \, \Lambda^2}} \dd \eta \, \int\limits^\eta_{\max \left[ 0, \eta - \sqrt{\frac{N_c}{2 \pi}} \ln \frac{1}{x} \right] } \dd s_{10} \, G^\textrm{NS}(s_{10},\eta)\, ,
\end{align}
with the integrals also discretized and evaluated numerically. Interested readers are directed to Appendix \ref{Convergence} for a discussion about convergence testing the numerical solutions of the flavor (non-)singlet evolution equations and the discretized versions of the hPDFs.

The Born-level approximation \eqref{inhom_nonsing} is linear in the logarithmic variables \eqref{varchange}, so we make a linear expansion ansatz for the inhomogeneous term in the flavor nonsinglet evolution, 
\begin{align}\label{abcNS}
    G_q^{\textrm{NS} \,(0)} = a_q^\textrm{NS}\,\eta + b_q^\textrm{NS}\,s_{10}+c_q^\textrm{NS} ,
\end{align}
for each of the three light flavors, $q=u,d,s$. This means that flavor nonsinglet hPDFs can be reconstructed as a linear combination of 9 flavor nonsinglet basis functions, generated by putting one of the 9 parameters  ($a_u^\textrm{NS}, b_u^\textrm{NS}, \ldots , c_s^\textrm{NS}$) to $1$, while setting all others equal to $0$. Combining this with the 15 parameters from Eqs.~\eqref{LinICdisc} describing the inhomogeneous terms for the flavor singlet dipole amplitudes, we have 24 parameters  (and associated basis functions) for the eight amplitudes ($Q_u, Q_d, Q_s, {\widetilde G}, G_2, G_u^\textrm{NS}, G_d^\textrm{NS}$ and $G_s^\textrm{NS}$), which we will fit to describe the world polarized DIS and SIDIS experimental data at low $x$.

\subsection{SIDIS cross section at small $x$}
%

We will now derive a formula for the SIDIS structure function $g_1^h(x,z)$ at small $x$.  Using the notation of Ref.~\cite{Cougoulic:2022gbk}, we start with the DIS  structure function $g_1(x)$ and write it as
\begin{align}\label{g1_xsect}
    g_1 (x, Q^2) = - \frac{Q^2}{16 \pi^2 \alpha_{\rm em} \, x} \, \sum_{\lambda = \pm} \lambda \, \sigma^{\vec{\gamma}^* + \vec{p} \to X} (\lambda, +)\,,
\end{align}
where $\sigma^{\vec{\gamma}^* + \vec{p} \to X} (\lambda, \Sigma)$ is the total virtual photon--proton cross section for the proton with helicity $\Sigma$ and for the transversely polarized virtual photon with polarization $\lambda$, and $\alpha_{\rm em}$ is the fine structure constant. The virtual photon--proton cross section is always inelastic at this order in $\alpha_{\rm em}$, as the virtual photon has to decay into a quark--anti-quark pair, with the quark and anti-quark fragmenting into hadrons in the final state. 

\begin{figure}[ht]
    \begin{center}
    \includegraphics[width=0.4 \textwidth]{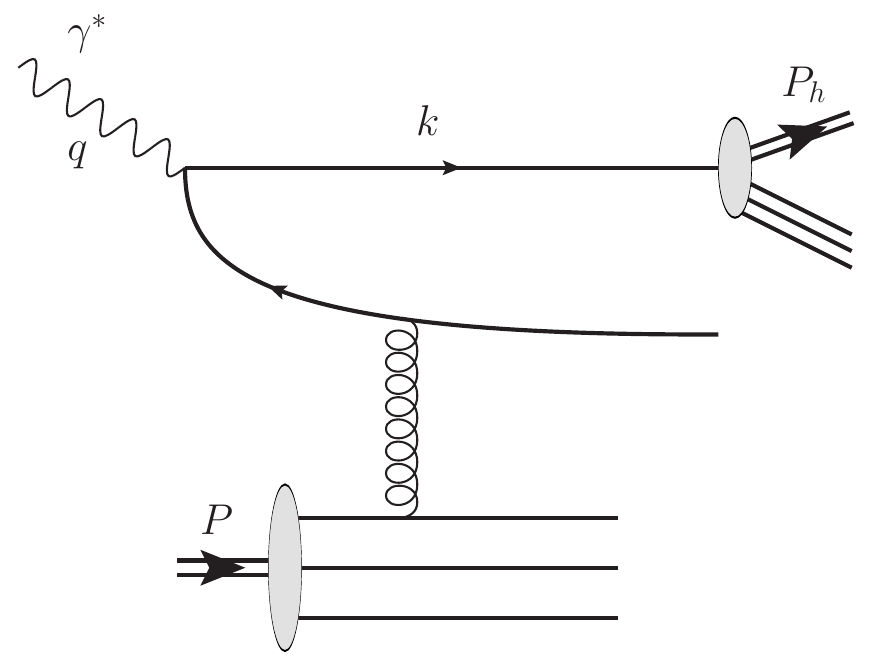} 
    \caption{The SIDIS process at small $x$.  An incoming virtual photon with momentum $q$ decays into a quark--antiquark pair which interacts with the target proton carrying momentum $P$.  The quark and antiquark then fragment into hadrons, and one of these hadrons is detected with momentum $P_h$.}
    \label{fig:SIDIS}
    \end{center}
    \end{figure}

Consider producing a hadron with a fixed value of $z \equiv P \cdot P_h/P \cdot q$, where $P$ and $q$ are the 4-momenta of the proton and virtual photon, respectively, while $P_h$ is the momentum of the detected hadron, as shown in \fig{fig:SIDIS}. At high energy/small $x$ we can work in the frame where the proton has a large $P^+$ momentum component, while the virtual photon has a large $q^-$ momentum component. Then $z \approx P^-_h/q^-$ is the fraction of the virtual photon's minus momentum carried by the produced hadron. All other components of the hadron's momentum are integrated over. 

We then write, by analogy to \eq{g1_xsect}, in the collinear approximation~\cite{deFlorian:1995fd,Bacchetta:2006tn, Signori:2013mda}
\begin{align}\label{g1h_xsect}
    g_1^h (x, z, Q^2) = - \frac{Q^2}{16 \pi^2 \alpha_{\rm em} \, x} \, \sum_{\lambda = \pm} \lambda \, \int \dd^2 k_\perp \, \dd^2 P_{h\perp} \, \delta^{(2)} \!\left( z \, {\bf k}_\perp - {\bf P}_{h\perp} \right) \, \sum_{q , {\bar q}}  \frac{\dd \sigma^{\vec{\gamma}^* + \vec{p} \to q + X}}{\dd^2 k_\perp} (\lambda, +) \,  D_1^{h/q}(z,Q^2)\,,
\end{align}
where ${\bf k}_\perp$ and ${\bf P}_{h\perp}$ are the transverse momentum vectors for the quark and produced hadron in \fig{fig:SIDIS}, while $ D_1^{h/q}(z,Q^2)$ is the collinear fragmentation function. The sum $\sum_{q , {\bar q}}$ goes over the produced quarks and antiquarks. While only quark fragmentation is depicted in \fig{fig:SIDIS}, an antiquark could instead fragment there, by reverting the particle number flow direction on the quark line in the diagram. 

In arriving at \eq{g1h_xsect} we have employed the aligned jet configuration, dominant in DLA \cite{Kovchegov:2015pbl,Cougoulic:2022gbk}, in which $k^- \approx q^-$, such that the produced hadron carries the fraction $P^-_h/k^- \approx P^-_h/q^- =z$ of the quark's momentum. Consequently, we assume that $z$ is not very small, such that the hadron is produced in the forward (virtual photon) direction/current fragmentation region and arises from the fragmentation of the forward-moving quark with 4-momentum $k$ in \fig{fig:SIDIS}, and not from the fragmentation of the antiquark, which is separated from the quark by a large rapidity interval. This is similar to the hybrid factorization approach to particle production \cite{Dumitru:2005gt,Chirilli:2011km,Chirilli:2012jd}. (The fragmentation of the antiquark in \fig{fig:SIDIS} would contribute to small-$z$ hadron production, and is neglected here since we are interested in order-one values of $z$.) In addition, the scale in the argument of the fragmentation function could be chosen to be $k_\perp^2$. However, in our small-$x$ kinematics, the typical value of $k_\perp^2$ is not too far from $Q^2$, allowing us to use $Q^2$ in the argument of $ D_1^{h/q}(z,Q^2)$. 

Integrating \eq{g1h_xsect} over ${\bf k}_\perp$ and ${\bf P}_{h\perp}$ we obtain
\begin{align}\label{g1h_xsect2}
    g_1^h (x, z, Q^2) = - \frac{Q^2}{16 \pi^2 \alpha_{\rm em} \, x} \, \sum_{\lambda = \pm} \lambda \, \sum_{q , {\bar q}}  \sigma^{\vec{\gamma}^* + \vec{p} \to q + X} (\lambda, +) \,  D_1^{h/q}(z,Q^2)\,.
\end{align}
Comparing this to Eqs.~\eqref{g1_xsect} and \eqref{g1_Dq}, we arrive at
\begin{align}\label{g1h_final}
	g_1^h(x,z,Q^2) = \frac{1}{2} \sum_{q, {\bar q}} \, e_q^2 \, \Delta q (x,Q^2) \, D_1^{h/q}(z,Q^2)\,,
\end{align}
reproducing the result in Eq.~(2) of Ref.~\cite{Ethier:2017zbq} (see also Refs.~\cite{Frankfurt:1989wq,deFlorian:1995fd,deFlorian:1996vg}), derived in the collinear factorization framework. (As we mentioned above, since quarks and antiquarks have different fragmentation functions, the right-hand-side of \eq{g1h_final} cannot be expressed solely in terms of the $\Delta q^+$ linear combinations of hPDFs, and the $\Delta q^-$ functions will enter as well.) We conclude that the expression \eqref{g1h_final} for the polarized SIDIS structure function is the same in the collinear and small-$x$ formalisms for large $z$. However, we emphasize that a similar discussion as that surrounding Eqs.~\eqref{g1_Dq} and \eqref{g1_coef_ftns} applies to Eq.~\eqref{g1h_final} regarding its interpretation  in the LCOT framework as implicitly including higher-order $\alpha_s$ corrections. \footnote{Strictly speaking, for consistency the fragmentation functions $D_1^{h/q} (z, Q^2)$ should also be taken in the polarized DIS scheme, but since the only presently available fragmentation functions are given in the $\overline{\mathrm{MS}}$ scheme, we make use of the existing extractions.}

%
\subsection{Global analysis}
%

\label{sec:global}

Our goal is to describe the world data on the longitudinal double-spin asymmetries in DIS and SIDIS at low $x$ using small-$x$ helicity evolution. We start with the longitudinal DIS asymmetry, $A_{\parallel}$~(see, {\it e.g.}, Refs.~\cite{Sato:2016tuz,COMPASS:2007qxf}),  
\begin{align}\label{eq: DSA}
	A_{\parallel} &=\frac{\sigma^{\downarrow \Uparrow}-\sigma^{\uparrow \Uparrow}}{\sigma^{\downarrow \Uparrow}+\sigma^{\uparrow \Uparrow}}
	=  D(A_1 + \eta A_2)\,, 
\end{align}
where the arrow $\uparrow(\downarrow)$ denotes the lepton spin along (opposite to) the beam direction, and the arrow $\Uparrow$ denotes the target polarization along the beam axis. The kinematic variables are given by
\begin{align}
    D &= \frac{y(2-y)(2+\gamma^2 y)}{2(1+\gamma^2)y^2+(4(1-y)-\gamma^2y^2)(1+R)},
    \ \ \ 
    \eta = \gamma \frac{4(1-y)-\gamma^2 y^2}{(2-y)(2+\gamma^2y)}\,,
\end{align}
where $y=\nu/E$ is fractional energy transfer of the lepton in the target rest frame, $\gamma^2 = 4 M^2x^2/Q^2$, and $R=\sigma_L/\sigma_T$ is the ratio of the longitudinal to transverse virtual photoproduction cross sections. When $4M^2x^2\ll Q^2 ~(\gamma^2\ll 1)$, we have $\eta \ll 1
$ and the virtual photon--target asymmetries are
\begin{align}\label{A1_fin}
    A_1 &= \frac{g_1 -\gamma^2 g_2}{F_1} \approx \frac{g_1}{F_1}, \ \ \ A_2 = \gamma \, \frac{g_1 + g_2}{F_1} \ll 1\,,
\end{align}
implying
\begin{align}\label{Apar_fin}
    A_\parallel \approx D \, A_1 . 
\end{align}
Similarly, in polarized SIDIS for the production of a hadron $h$, the asymmetry $A_1^h$ can be expressed as (see, {\it e.g.}, Refs.~\cite{Ethier:2017zbq, Leader:2010rb})
\begin{align}\label{A1h}
    A_1^h = \frac{g_1^h-\gamma^2 g_{2}^h}{F_1^h} \approx \frac{g_1^h}{F_1^h}\,.
\end{align}

In principle there is another observable in the DIS/SIDIS family that could help constrain hPDFs:~parity-violating DIS. This process is sensitive to the $g_1^{\gamma Z}$ structure function which is approximately proportional to $\Delta \Sigma$ \cite{Hobbs:2008mm, Zhao:2016rfu}. Unfortunately there is little to no data for $g_1^{\gamma Z}$ in the small-$x$ ($x<0.1$) region (see, {\it e.g.}, Ref.~\cite{Wang:2013iiv}), not allowing us to employ this observable in our analysis.

Between the two scattering processes, we have ten unique observables:~two in DIS (proton or deuteron/${}^3 \mathrm{He}$ target) and eight in SIDIS (proton or deuteron/${}^3 \mathrm{He}$ target with charged pion or kaon final states) from which in principle we can constrain the eight polarized dipole amplitudes (five associated with the $C$-even and flavor singlet hPDFs ($Q_u, Q_d, Q_s, {\widetilde G}, G_2$),  and three with the flavor nonsinglet hPDFs ($G_u^\textrm{NS}, G_d^\textrm{NS}$ and $G_s^\textrm{NS}$)). In our formalism, the $g_1$ and $g_1^h$  structure functions are calculated in terms of hPDFs using Eqs.~(\ref{g1_Dq}), (\ref{g1h_final}), respectively. (Note that $ \Delta q = ( \Delta q^+ + \Delta q^-)/2$ and $\Delta \bar{q} = (\Delta q^+ - \Delta q^-)/2$.) This is the bridge connecting small-$x$ helicity evolution to the experimental data. Fitting the hPDFs to $A_\parallel, A_1$ and $A_1^h$ at moderate $x\lesssim 0.1$ allows us to determine the initial conditions of the polarized dipole amplitudes \eqref{LinICdisc}, \eqref{abcNS}. We then evolve the polarized dipole amplitudes toward lower values of $x$ using Eqs.~\eqref{eq_LargeNcNf} and \eqref{NSeq_LargeNcNf} to obtain hPDFs in that region, and compare with  existing data as well as make predictions at smaller~$x$. We mention that the structure functions $F_1$ and $F_1^h$ involve the unpolarized PDF $q (x,Q^2)$ and, for the latter, the unpolarized fragmentation function (FF) $D_1^{h/q}(z,Q^2)$.  We compute $F_1$ and $F_1^h$ up to next-to-leading order using collinear factorization and DGLAP evolution, based on the JAM analysis in Ref.~\cite{Cocuzza:2022jye}. (To be consistent, strictly speaking one should include small-$x$ evolution also for $F_1$ and $F_1^h$.  However, for us the results of Ref.~\cite{Cocuzza:2022jye} serve as a faithful proxy of the experimental data for these structure functions. A more comprehensive analysis that also utilizes small-$x$ evolution for $F_1$ and $F_1^h$ is left for future work.)

Let us present a short discussion about our ability to constrain $G_2$ and $\widetilde{G}$, which are two important polarized dipole amplitudes driving the small-$x$ evolution of the hPDFs.  The polarized dipole amplitude $G_2$ is directly related to the gluon hPDF, per \eq{JM_DeltaG2}. However, the observables we consider here do not directly couple to the gluon hPDF. Instead, as we saw above, they couple only to quark hPDFs. The dipole amplitude $G_2$ enters the quark hPDFs $\Delta q^+$ along with the dipole amplitude $Q_q$. Moreover, they always enter in the same linear combination, $Q_q + 2 \, G_2$ for $q=u,d,s$ (see \eq{q+2}). We see that while $G_2$ and $Q_q$ couple directly to the spin-dependent structure functions for DIS and SIDIS, we do not have an observable (or a linear combination of observables) in this analysis which separately couples only to $G_2$  or only to $Q_q$. 

What may help us to separate $G_2$ and $Q_q$ is the fact that these dipole amplitudes have a different pre-asymptotic form. While it is established numerically that at asymptotically small $x$, both polarized dipole amplitudes $G_2$ and $Q_q$ are proportional to the same power of $x$ with the same intercept \cite{Adamiak:2023okq} and are, therefore, probably hard to distinguish, in the pre-asymptotic region where the asymptotic form has not yet been reached, their contributions to the quark hPDFs may be quite different. This can be studied by comparing the $Q_u$ and $G_2$ basis functions for $\Delta u^+$ in \fig{xhpdfu_medx}, shown in the top and bottom panels of that figure, respectively. If these functions were identical, they could be freely interchanged against each other while still producing the same structure functions:~in such a case it would be impossible to separate $G_2$ and $Q_u$ from the data. Since the contributions of different amplitudes to quark hPDFs differ from each other, as follows from \fig{xhpdfu_medx}, these basis contributions cannot be adjusted at one value of $x$ while maintaining the same value for the observables at all other $x$. Therefore, we may be able to separate $G_2$ and $Q_u$ using the polarized DIS and SIDIS data. However, since the $Q_u$ and $G_2$ basis functions have similar shapes, per \fig{xhpdfu_medx}, it might be the case that the uncertainties in the resulting extractions of $Q_u$ and $G_2$ will be large.

The polarized dipole amplitude $\widetilde{G}$, on the other hand, does not couple to any of the polarized DIS or SIDIS observables we consider here. Rather, it mixes with other polarized dipole amplitudes only through evolution (see Eqs.~\eqref{eq_LargeNcNf}). This is why the $\widetilde{G}$ basis function of $\Delta u^+$ (second from the bottom panel in \fig{xhpdfu_medx}) appears to be vanishingly small above $x>x_0$. The consequence of this is that in the region of $x$ where the polarized DIS and SIDIS data exist, $5\times 10^{-3}<x<0.1$, the $\widetilde{G}$ amplitude is very small and is, therefore, much less constrained by the data than the $Q_q$ and $G_2$ dipole amplitudes. At small $x$, however, the $\widetilde{G}$ amplitude is quite large, second only to $G_2$ (see \fig{xhpdfu_medx}). As we will see below, $\widetilde{G}$, unconstrained by the existing polarized DIS and SIDIS data, will dominate over the other polarized dipole amplitudes at small $x$, adversely affecting our ability to make precise predictions at even smaller~$x$. Nevertheless, it is possible that $\widetilde{G}$ might be constrained with slightly more leverage in $x$. We will discuss this  in \sec{sec:results-EICimpact} when we explore the impact of the future EIC data on our uncertainties.

In our global analysis we use the JAM Bayesian Monte Carlo framework~(see, {\it e.g.}, \cite{Sato:2016tuz,Sato:2019yez,Moffat:2021dji}) to
randomly sample (roughly 500 times) the space of 24 parameters $a, b, c$ from Eqs.~\eqref{LinICdisc} and \eqref{abcNS}, namely $a_u, b_u, c_u, a_d, \ldots , c_s^\textrm{NS}$. For each combination of these parameters, we solve our evolution equations~\eqref{eq_LargeNcNf} and \eqref{NSeq_LargeNcNf} to determine the polarized dipole amplitudes $Q_u, Q_d, Q_s, {\widetilde G}, G_2, G_u^\textrm{NS}, G_d^\textrm{NS}$ and $G_s^\textrm{NS}$. 
(The actual numerical solution is facilitated by the basis functions introduced above.) 
Next, using Eqs.~\eqref{q+2} and \eqref{q-2}, we calculate the quark hPDFs at small $x$, which, via Eqs.~\eqref{g1_Dq} and \eqref{g1h_final}, 
can be used to determine the structure functions $g_1$ and $g_1^h$ that enter the numerator of the asymmetries $A_\parallel,A_1$ (Eqs.~\eqref{A1_fin}, \eqref{Apar_fin}) and $A_1^h$ (Eq.~\eqref{A1h}), respectively. 
The $\chi^2$-minimization procedure allows us to construct the posterior distributions of the parameters, and the corresponding solutions of our evolution equations then allow us to infer the quark and gluon hPDFs 
(the latter via \eq{JM_DeltaG2}). We confirmed that the posterior distributions of the parameters are distributed more narrowly than the initial flat sampling and are approximately Gaussian, indicating a convergence in their values. These extracted quark and gluon hPDFs, and the quantities that can be computed from them, are the main results of our work, which we  present below.

%
\section{Results}
%
\label{sec:results}

In this section we present the results of our numerical analysis. We will concentrate on the proton $g_1$ structure function, and the quark and gluon hPDFs (along with quantities, such as net spin, that can be computed from them).  

%
\subsection{Data versus theory} \label{sec:results_datavtheory}
%

    \begin{figure}[h!]
            \begin{centering}
            \includegraphics[width=500pt]{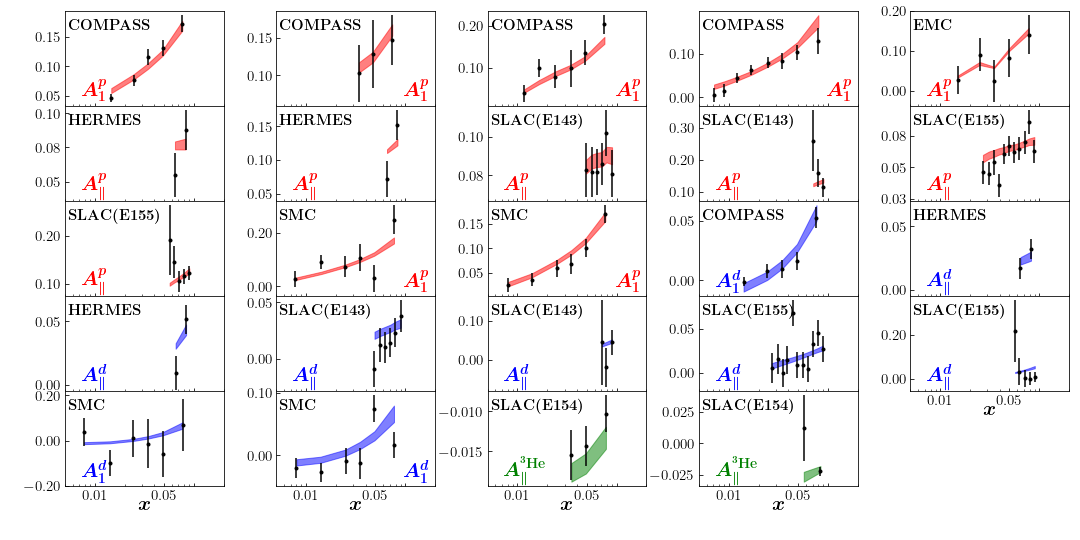}
            \vspace{-0.35cm}
            \caption{Comparison of the experimental data and the fit based on our small-$x$ theory  for the double-spin asymmetries $A_1$ and $A_\parallel$ in polarized DIS on a proton (red), deuteron (blue) and ${}^3 \mathrm{He}$ (green) target.
                \label{Plot_pidis_data}
                \vspace{-0.1cm}
            }
            \end{centering}
            \end{figure}
    \begin{figure}[h!]
            \begin{centering}
            \includegraphics[width=500pt]{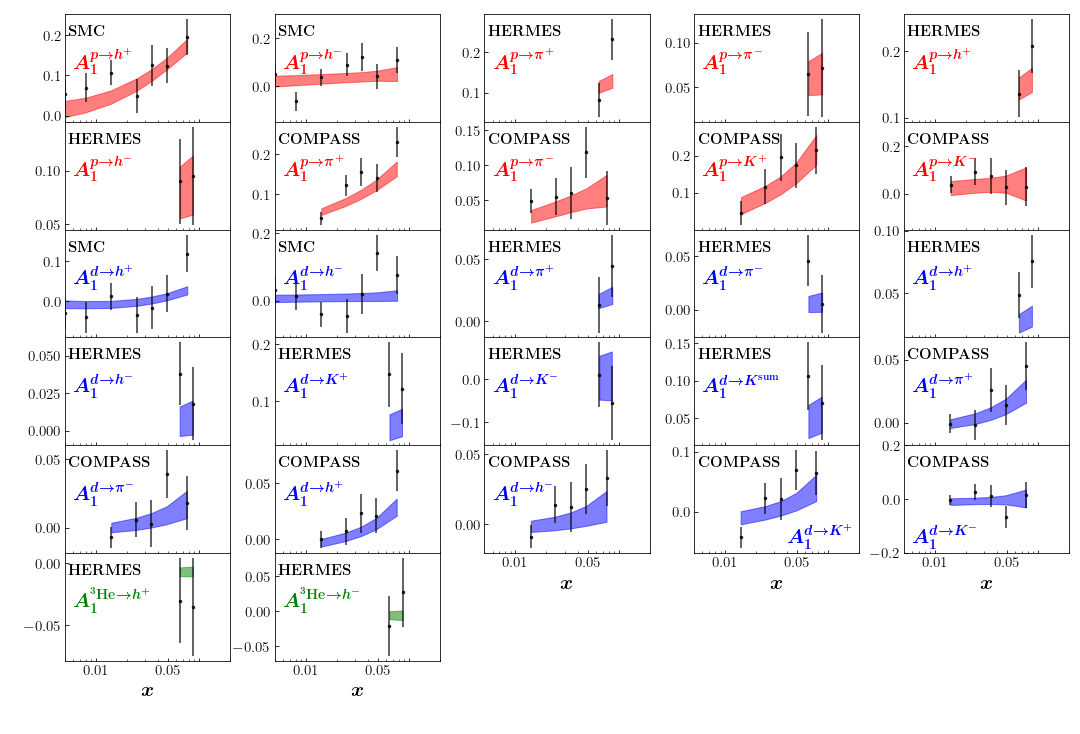}
            \vspace{-0.35cm}
            \caption{Comparison of experimental data and fit based on our small-$x$ theory  for the double-spin asymmetry $A_1^h$ in polarized SIDIS on a proton (red), deuteron (blue) and ${}^3 \mathrm{He}$ (green) target for charged pion, kaon and unidentified hadron  final states.
                \label{Plot_psidis_data}
                \vspace{-0.1cm}
            }
            \end{centering}
    \end{figure}

Our analysis (JAMsmallx) of the world polarized DIS and SIDIS data at low $x$ utilizes measurements from SLAC~\cite{Anthony:1996mw, Abe:1997cx, Abe:1998wq, Anthony:1999rm, Anthony:2000fn}, EMC~\cite{Ashman:1989ig}, SMC~\cite{Adeva:1998vv, SpinMuon:1998eqa, Adeva:1999pa}, COMPASS~\cite{Alekseev:2010hc, Adolph:2015saz,Adolph:2016myg}, and HERMES~\cite{Ackerstaff:1997ws, Airapetian:2007mh} for DIS, and SMC~\cite{ SpinMuon:1997yns}, COMPASS~\cite{COMPASS:2010hwr, COMPASS:2009kiy}, and HERMES~\cite{HERMES:2004zsh, HERMES:1999uyx} for SIDIS. The data of interest falls in the Bjorken-$x$ range of $5\times 10^{-3}<x<0.1 \equiv x_0$, and the $Q^2$ range is $1.69 ~\text{GeV}^2 <Q^2< 10.4 ~\text{GeV}^2$. Since $x \approx Q^2 / s$, the minimum cut on $Q^2$ determines the minimum accessible $x$ in the data set (for a given experimental center-of-mass energy), and conversely the maximum cut on $x$ determines the maximum $Q^2$. The upper limit on $x$ (denoted by $x_0$) was chosen based on our previous (DIS-only) work \cite{Adamiak:2021ppq}, as (almost) the highest value of $x$ which gave a good-$\chi^2$ fit. This $x_0$ is the point where we start the small-$x$ helicity evolution. The fact that our small-$x$ approach was able to describe data up to such a high value of $x$ could be due to the fact that, unlike the unpolarized Balitsky--Fadin--Kuraev--Lipatov (BFKL) \cite{Kuraev:1977fs,Balitsky:1978ic}, Balitsky--Kovchegov (BK) \cite{Balitsky:1995ub,Balitsky:1998ya,Kovchegov:1999yj,Kovchegov:1999ua} and Jalilian-Marian--Iancu--McLerran--Weigert--Leonidov--Kovner
(JIMWLK)\cite{Jalilian-Marian:1997dw,Jalilian-Marian:1997gr,Weigert:2000gi,Iancu:2001ad,Iancu:2000hn,Ferreiro:2001qy} small-$x$ evolution, which resums powers of $\as \, \ln (1/x)$ at the leading order, our helicity evolution has a different (larger) resummation parameter,  $\alpha_s \ln^2(1/x)$. For $\as \approx 0.25$, our resumation parameter becomes of order 1 for $x \approx 0.1$, potentially justifying our use of $x_0 = 0.1$ as the starting point for our evolution. Note that the value of our resummation parameter $\alpha_s \ln^2(1/x)$ at $x=x_0 = 0.1$ is comparable to (and even slightly larger than) the value of the resummation parameter $\as \, \ln (1/x)$ for the unpolarized small-$x$ evolution at $x=0.01$, which is where the latter evolution is usually initiated in phenomenological analyses~\cite{Albacete:2009fh,Albacete:2010sy}. The lower limit of $Q^2$ is set by the charm quark mass, $m_c^2 = 1.69\, \mathrm{GeV}^2$.  This is also the cut placed by the JAM FF set we use~\cite{Cocuzza:2022jye}, which has independent functions for $\pi^+,K^+,h^+$ ($\pi^-,K^-,h^-$ are found through charge conjugation) that we evolve through the DGLAP equations.  By analogy to \cite{Adamiak:2021ppq}, we choose our IR cutoff to be $\Lambda = 1$~GeV. Also, in the $Q^2$ range specified above, the strong coupling in \eq{rc} is taken with $N_f =3$ (and $N_c =3$). 

\begin{table}[t!]
  \caption{Summary of polarized DIS data included in the fit, separated into $A_1$  (left) and $A_{\parallel}$ (right), along with the  $\chi^2/N_{\rm pts}$ for each data set.\\}
    \begin{tabular}{l|c|c|c} 
    \hline
    {\bf Data set ($\boldsymbol{A_1}$)} & 
        ~{\bf Target}~ &
        ~$\boldsymbol{N_\mathrm{pts}}$~ & 
        ~$\boldsymbol{\chi^2 / N_\mathrm{pts}}$~  
        \\ \hline     
    SLAC (E142) \cite{Anthony:1996mw} &
        ${}^3 \mathrm{He}$ &
        $1$       & 
        $0.60$    \\ \hline
    EMC  \cite{Ashman:1989ig} &
        $p$ &
        $5$ &
        $0.20$   \\ \hline
    SMC  \cite{Adeva:1998vv, Adeva:1999pa} &
        $p$ &
        $6$ &
        $1.29$  \\ 
        &
        $p$ &
        $6$ &
        $0.53$  \\
        &
        $d$ &
        $6$ &
        $0.67$  \\
        &
        $d$ &
        $6$ &
        $2.26$  	\\ \hline
    COMPASS   \cite{Alekseev:2010hc} &
        $p$ &
        $5$ &
        $1.02$  \\ \hline
    COMPASS   \cite{Adolph:2015saz} &
        $p$ &
        $17$ &
        $0.74$   \\ \hline
    COMPASS  \cite{Adolph:2016myg} &
        $d$ &
        $5$ &
        $0.88$  \\ \hline
    HERMES   \cite{Ackerstaff:1997ws} &
        $n$ &
        $2$ &
        $0.73$ \\ \hline\hline
    {\bf Total} &    & 59 & 0.91 \\ \hline
   \end{tabular}
    \qquad\qquad\qquad
    \begin{tabular}{l|c|c|c} 
    \hline
    {\bf Data set ($\boldsymbol{A_{\parallel}}$)}& 
        ~{\bf Target}~ &
        ~$\boldsymbol{N_\mathrm{pts}}$~ & 
        ~$\boldsymbol{\chi^2 / N_\mathrm{pts}}$~  
        \\ \hline
    SLAC(E155) \cite{Anthony:1999rm} &
        $p$ &
        $16$ &
        $1.28$ 	\\ 
        &
        $d$ &
        $16$ &
        $1.62$      \\ \hline
    SLAC (E143) \cite{Abe:1998wq} &
        $p$ &
        $9$ &
        $0.56$    \\
        &
        $d$ &
        $9$ &
        $0.92$  	\\ \hline
    SLAC (E154) \cite{Abe:1997cx} &
        ${}^3 \mathrm{He}$ &
        $5$       &
        $1.09$     \\ \hline
    HERMES    \cite{Airapetian:2007mh} &
        $p$ &
        $4$ &
        $1.54$  \\ 
        &
         $d$ &
         $4$ &
         $0.98$  \\ \hline\hline
    {\bf Total} &   & 63 & 1.19 \\ \hline
    \end{tabular}
  \label{t:Chi2_DIS_Apa}
\end{table}

\begin{table}[h!]
    \begin{center}
    \caption{Summary of the polarized  SIDIS data on $A_1^h$ included in the fit, along with the $\chi^2/N_{\rm pts}$ for each data set.\\}
    \label{t:Chi2_SIDIS}
    \vspace{0.3cm}
    \begin{tabular}{l|c|c|c|c} 
    \hline
    {\bf Dataset ($\boldsymbol{A_1^h}$})& 
        ~{\bf Target}~ &
        ~{\bf Tagged Hadron}~ &
        ~$\boldsymbol{N_\mathrm{pts}}$~ & 
        ~$\boldsymbol{\chi^2 / N_\mathrm{pts}}$~  
        \\ \hline
    SMC  \cite{SpinMuon:1998eqa} &
        $p$ &
        $h^+$ &
        $7$ &
        $1.03$  \\ 
        &
        $p$ &
        $h^-$ &
        $7$ &
        $1.45$  		\\ 
        &
        $d$ &
        $h^+$ &
        $7$ &
        $0.82$  \\ 
        &
        $d$ &
        $h^-$ &
        $7$ &
        $1.49$   \\ \hline
    HERMES  \cite{HERMES:2004zsh} &
        $p$ &
        $\pi^+$ &
        $2$ &
        $2.39$ \\ 
        &
        $p$ &
        $\pi^-$ &
        $2$  &
        $0.01$    \\ 
        &
        $p$ &
        $h^+$ &
        $2$ &
        $0.79$  \\
        &
        $p$ &
        $h^-$ &
        $2$  &
        $0.05$  \\ 
        &
        $d$  &
        $\pi^+$ &
        $2$ &
        $0.47$ 	 \\
        &
        $d$ &
        $\pi^-$ &
        $2$ &
        $1.40$ 	 \\
        &
        $d$ &
        $h^+$  &
        $2$  &
        $2.84$   \\ 
        &
        $d$  &
        $h^-$ &
        $2$  &
        $1.22$ 	 \\ 
        &
        $d$ &
        $K^+$ &
        $2$  &
        $1.81$ 	 \\ 
        &
        $d$  &
        $K^-$ &
        $2$  &
        $0.27$ 	\\ 
        &
        $d$  &
        $K^+ + K^-$ &
        $2$  &
        $0.97$ 	\\ \hline	
    HERMES \cite{HERMES:1999uyx} &
        ${}^3 \mathrm{He}$  &
        $h^+$  &
        $2$ &
        $0.49$ 	\\ 
        &
        ${}^3 \mathrm{He}$ &
        $h^-$  &
        $2$   &
        $0.29$ 	\\ \hline
    COMPASS \cite{COMPASS:2010hwr} &
        $p$ &
        $\pi^+$ &
        $5$ &
        $1.88$ 		\\ 
        &
        $p$   &
        $\pi^-$ &
        $5$ &
        $1.10$ \\ 
        &
        $p$  &
        $K^+$ &
        $5$   &
        $0.42$ 	\\
        &
        $p$ &
        $K^-$ &
        $5$   &
        $0.31$ 		\\ \hline
    COMPASS \cite{COMPASS:2009kiy}  
        &
        $d$ &
        $\pi^+$ &
        $5$   &
        $0.50$ 	\\ 
        &
        $d$  &
        $\pi^-$ &
        $5$   &
        $0.78$ \\
        &
        $d$   &
        $h^+$ &
        $5$ &
        $0.90$ \\ 
        &
        $d$ &
        $h^-$  &
        $5$  &
        $0.86$ 		\\ 
        &
        $d$     &
        $K^+$ &
        $5$  &
        $1.50$ 		\\
        &
        $d$ &
        $K^-$ &
        $5$ &
        $0.78$ 	\\ \hline\hline
    {\bf Total} &    &    & 104 & 1.01 \\ \hline
    \end{tabular}
    \end{center}
\end{table}

The range of the outgoing hadron momentum fraction $z$ in polarized SIDIS is $0.2<z<1.0$, and we do not place any explicit cut on this variable. In practice, the data (after all the appropriate cuts) generally has values of $0.4<z<0.6\,$; some data sets integrate $z\in[0.2, 1]$ while others cover   $z\in[0.2, 0.85]$. After all the cuts we are left with 122 polarized DIS data points and 104 polarized SIDIS data points, for a total $N_{\rm pts}=226$. The overall $\chi^2/N_{\rm pts}$ of our fit, based on the central theory curves, is 1.03. (We have also performed fits with cutoffs of $x_0=0.08$ and $x_0=0.05$, which produced no significant change in $\chi^2/N_{\rm pts}$.) The breakdown of the data by experiment, along with our $\chi^2/N_{\rm pts}$ for those individual data sets, is shown in Table~\ref{t:Chi2_DIS_Apa} for DIS and Table~\ref{t:Chi2_SIDIS} for SIDIS. The plots of the experimental data versus our JAMsmallx theory are shown in \fig{Plot_pidis_data} for polarized DIS and \fig{Plot_psidis_data} for polarized SIDIS. Overall, our results demonstrate very good agreement with the existing world data.

\subsection{Proton $g_1$ structure function}

    \begin{figure}[t!]
            \begin{centering}
            \includegraphics[width=345pt]
            {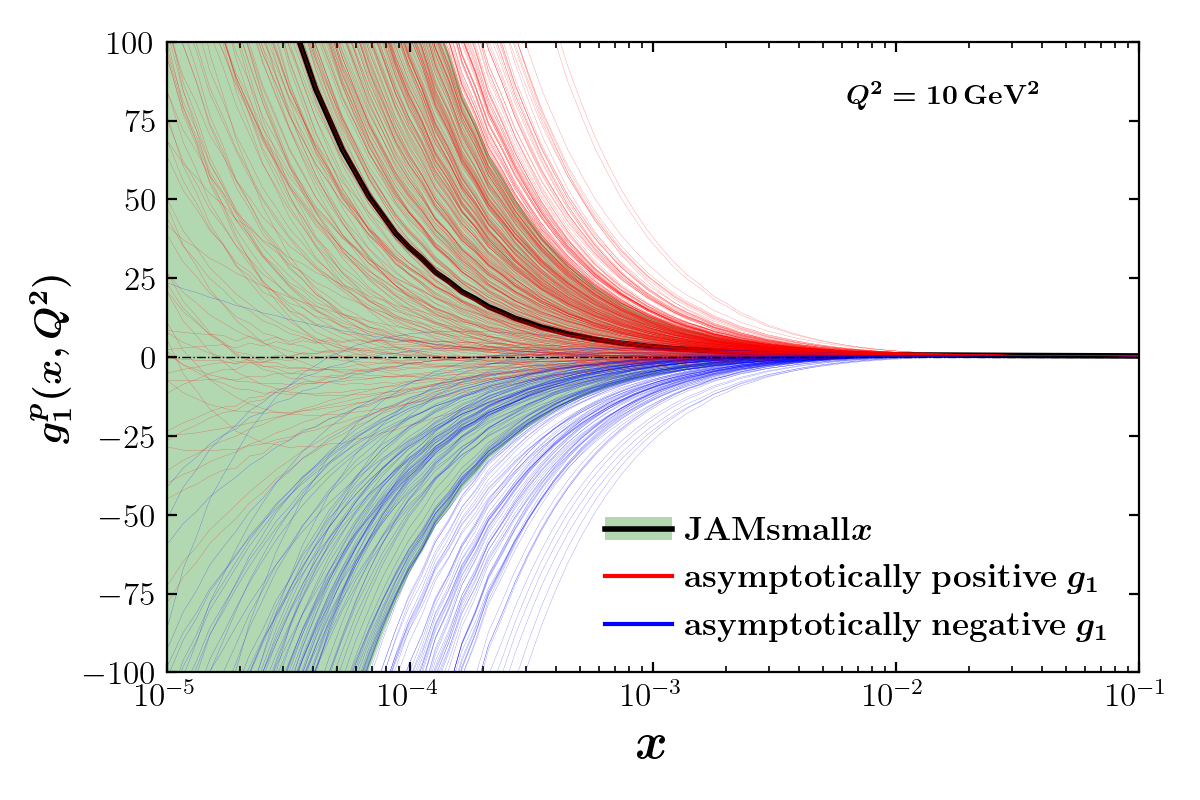}
            \caption{The small-$x$ calculation of the $g_1$ structure function of the proton. The black curve is the mean of all the replicas with the green band giving the 1$\sigma$ uncertainty. Red (blue) curves are solutions that are asymptotically positive (negative).
                \label{Plot_g1}
            }
            \end{centering}
    \end{figure}

We now examine our result for the $g_1$ structure function of the proton to analyze the predictive capability of our formalism. Our calculation of $g_1^p$ for all  replicas is given in \fig{Plot_g1}. This is the result of 500 individual fits of the experimental data where the (quark and gluon) hPDFs were extracted and then (the quark ones) used to compute $g_1^p$.  We color code each replica by its asymptotic sign at small $x$ in order to clarify the structure of the plot as well as to help establish correlations with the hPDFs below. While $g_1^p$ is well constrained in the region where there is experimental data ($5\times 10^{-3}<x<10^{-1}$), it is largely unconstrained at smaller $x$.
The major difficulty in constraining $g_1^p$ is caused by the insensitivity of the data to the $G_2$ and $\widetilde G$ amplitudes described above.

That being said, the asymptotic solution  of the large-$N_c \& N_f$ evolution equations~\cite{Adamiak:2023okq} guarantees that the small $x$ behavior of $g_1^p$ must be exponential in $\ln (1/x)$. This implies that it has to pick a sign (positive or negative) when $x\to 0$. Our results indicate (see \fig{Plot_g1_ambiguities}) that, given the existing experimental data constraining our formalism, the asymptotic sign is likely to be picked by $x=3.5\times10^{-4}$ with 10\% uncertainty, with the uncertainty decreasing to 5\% at approximately $x=2.5\times10^{-5}$. Currently, $70\%$ of the replicas are asymptotically positive and $30\%$ are asymptotically negative. These percentages are stable as the number of replicas increases. The primary source of uncertainty is how low in $x$ one must go to determine the sign, as some replicas that appear positive may undergo a sign change at smaller $x$. Interestingly, our observation of a preference for $g_1^p$ to be positive at small $x$ agrees with the recent papers analyzing (unpolarized and polarized) DIS structure functions using the anti--de Sitter space/Conformal Field Theory (AdS/CFT) correspondence~\cite{Kovensky:2018xxa,Jorrin:2022lua,Borsa:2023tqr} that make an even stronger statement that $g_1^p$ clearly grows positive at small~$x$.  This behavior also has implications for the net parton spin expected at small~$x$, as we discuss in Sec.~\ref{s:spin}.

\subsubsection{Sign of $g_1^p$ and quantifying numerical ambiguity} \label{subsubsec: Sign_Ambiguity}

From Fig.~\ref{Plot_g1} alone, one can make the qualitative observation that indeed each replica of $g_1^p$ grows exponentially with $\ln (1/x)$ as we suggested earlier, and the color  indicates the asymptotic sign of $g_1^p$ for that given replica. We mentioned in the previous section that the exponential behavior of helicity functions in our theory makes it difficult for a given replica to maintain a near-zero value, and thus it must eventually choose to (rapidly) increase in magnitude towards positive or negative values. Given the numerical nature of our global analysis, we cannot compute each fitted replica down to $x=0$ (corresponding to $\ln x \to - \infty$), so the color-coding and sign assignment is determined by the slope of a replica at the lowest-computed value of $x$:~if the slope increases (decreases) as $x$ goes to zero, then it is considered ``asymptotically" positive (negative). To balance our time and computational resources, the results discussed in this section use replica data computed down to $x_{\mathrm{asymp}}=10^{-7.5}$. One may realize potential issues with this system: a given replica may have multiple different ``asymptotic'' signs depending on the lowest computed value of $x$.

\begin{figure}[t] 
    \begin{centering}
    \includegraphics[width=355pt]{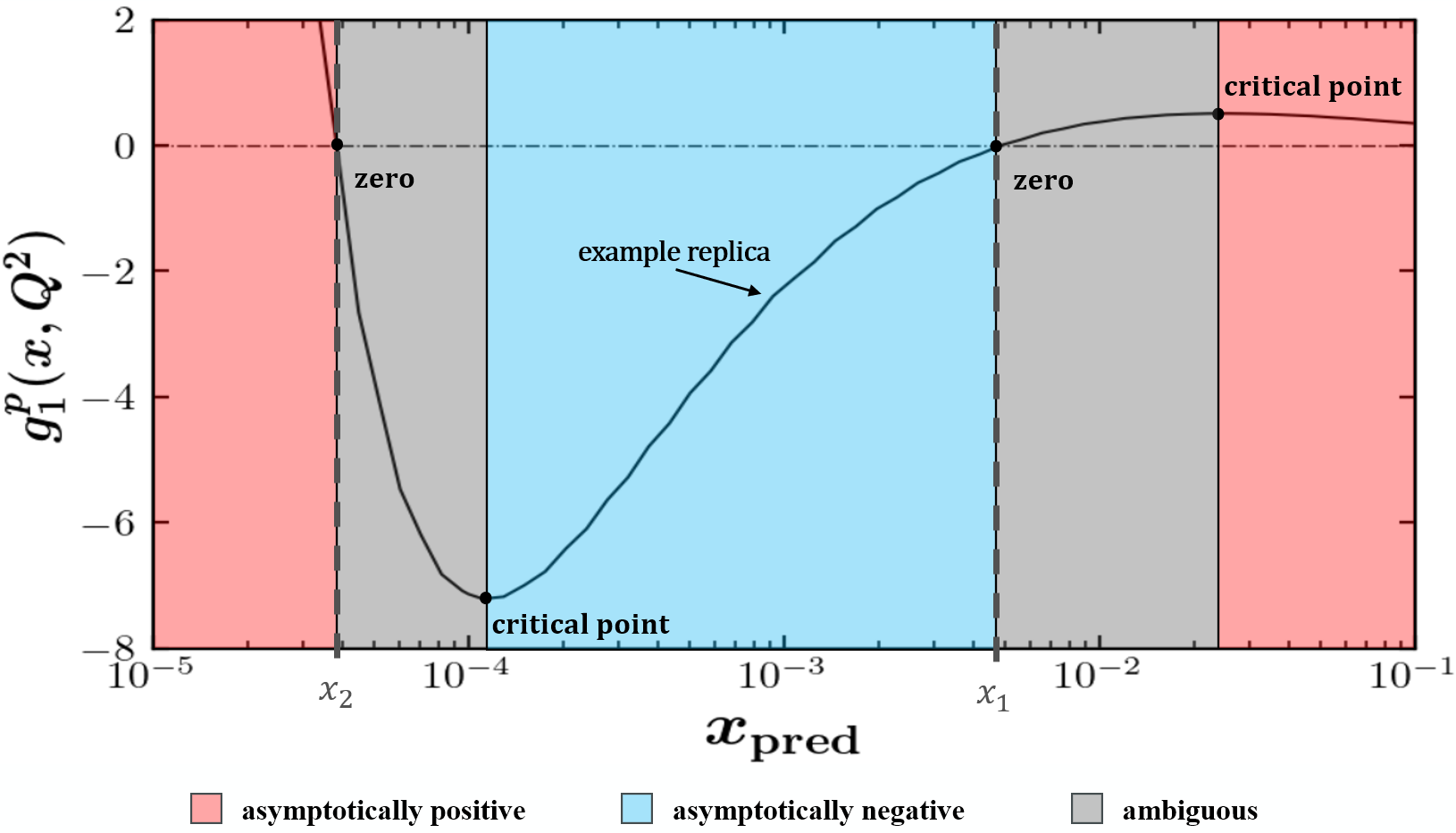}
    \caption{An example replica of $g_1^p(x, Q^2=10\, {\rm GeV^2})$ that demonstrates how the asymptotic sign is dependent on $x_{\mathrm{pred}}$. If $x_{\mathrm{pred}}$ resides in the red (blue) region then the replica will be considered asymptotically positive (negative) according to the sign of the first derivative (for decreasing $x$), and its agreement with the sign of the magnitude. If $x_{\mathrm{pred}}$ resides in either gray region then the asymptotic sign is ambiguous due to a contradiction between the sign of the slope and the sign of the magnitude.
        \label{Ambiguity_Ex}
    }
    \end{centering}
    \end{figure}

Any given replica is defined by its specific combination of basis functions, and since our Bayesian analysis samples parameters (Eq.~\eqref{BornICdisc}) that may be either positive or negative, competition between basis functions can result in nodes. Replicas with two nodes in $g_1^p(x)$, such as the one illustrated in \fig{Ambiguity_Ex}, can occur for linear combinations of similar basis functions with opposite signs, as in the top/bottom panels of \fig{xhpdfu_medx}. These changes in sign can occur at various values of $x$ depending on the initial conditions, making the prediction of the asymptotic sign dependent on what $x$ value is used to make the prediction.

Careful readers may have already noticed this from \fig{Plot_g1}, where there are a few red-coded replicas that appear to be growing negative (and a blue-coded replica that appears to be growing positive) at $x=10^{-5}$.  This is due to each of these replicas having a delayed critical point ($\tfrac{\dd\,g_1^p(x)}{\dd\,x}=0$) that occurs at $x < 10^{-5}$, where a different basis function takes over the growth and the replica changes the sign of its slope. 
These critical points also are connected to the issue of \textit{ambiguity}, where at a specific value of $x$ we may be able to measure that a replica is \textit{growing} positive (or negative) but has a magnitude that is actively negative (or positive), leaving its asymptotic sign unconfirmed. Luckily, investigations of these incidents show that they occur in a statistically small portion of replicas from the perspective of our considerably small $x_{\mathrm{asymp}}$.

Since our goal is predictability at small $x$, we decided to quantify the amount of ambiguity by its probability density in $x$. That is, for each replica we count the smallest-$x$ instance of ambiguity, and take note of where in $x$ it occurred. For example, \fig{Ambiguity_Ex} shows a replica that begins positive (true for all replicas) and evolution drives it more positive until it reaches a critical point, after which the replica then grows negative. After the critical point (in the gray region), the replica will be considered \textit{ambiguous} until it crosses $g_1^p(x_1)=0$, and then it is considered asymptotically negative (in the blue region). Only when the sign of $g_1^p$ and the sign of its first derivative (as $x$ decreases) agree can the replica be considered asymptotically positive or negative. If we wanted to predict the asymptotic sign of the replica based on an observation at $x=x_{\mathrm{pred}}$ that resides in this (blue) region, then we would predict that this replica is ``asymptotically negative" as $x\to0.$ However, this same replica has a small-$x$ critical point (around $x=10^{-4}$) that causes the sign of its slope to change; the replica observed in the (gray) region (on the left) between the critical point and $g_1^p(x_2)=0$ would be considered ambiguous again. After crossing zero a second time, a prediction made at $x_{\mathrm{pred}} < x_2$ would therefore designate the replica to be ``asymptotically positive." The smallest-$x$ instance of ambiguity is thus counted in a bin at $x_2$. In this way, each replica is counted exactly once, and replicas that oscillate multiple times about the $g_1^p=0$ axis only have their most delayed ambiguity counted. We can define the number of replicas that have their smallest-$x$ instance of ambiguity in a particular bin of $x$ as $C_A(x)$ (the counts of ambiguities) and make a histogram.  The ambiguity count $C_A (x)$ is normalized such that it sums to the total number of replicas $N_{\mathrm{ambig}}$ containing at least one ambiguity:
\begin{align}   \label{e:ambigsum1}
    \sum_{x = x_{\mathrm{asymp}}}^{x_0} C_A (x) = N_{\mathrm{ambig}} \leq N_{\rm tot}  \: .
\end{align}
Because some replicas are always unambiguous across the entire range of $x$, the ambiguity count is less than the total number of replicas: $N_{\mathrm{ambig}} \leq N_{\rm tot}$.

\begin{figure}[t!]
    \begin{centering}
    \includegraphics[width=500pt]{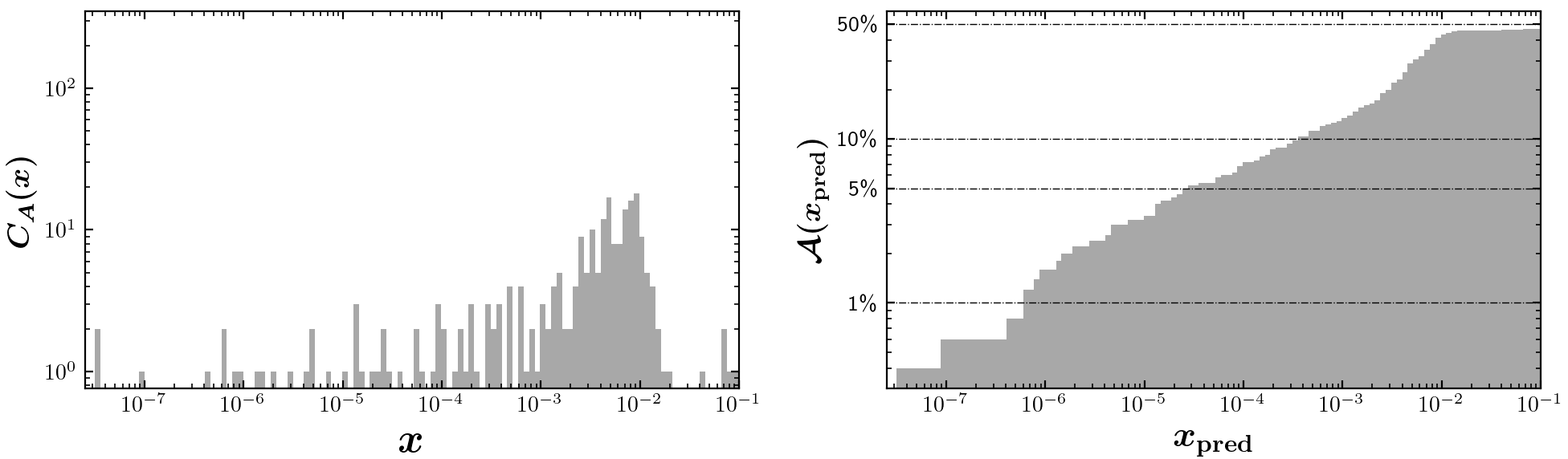}
    \caption{(Left) Histogram that counts the number of replicas with a smallest-$x$ ambiguity at a given value of $x$. (Right) The running sum of the ambiguity histogram, telling us what percentage of replicas have an ambiguity below a given value of~$x$.
        \label{Plot_g1_ambiguities}
    }
    \end{centering}
    \end{figure}

Now suppose we want to predict the asymptotic behavior of $g_1^p$ at small $x$ based on the behavior of the function at some value $x_{\mathrm{\mathrm{pred}}}$.  Knowledge of the ambiguity count $C_A (x)$ allows us to estimate the accuracy of this prediction by estimating the probability that an unobserved ambiguity remains at $x_{\mathrm{asymp}} < x < x_{\mathrm{\mathrm{pred}}}$.  This probability is given by a summation as in Eq.~\eqref{e:ambigsum1}, but over the truncated range in $x$:
\begin{align}   \label{e:ambigsum2}
    \mathcal{A}(x_{\mathrm{\mathrm{pred}}}) = \frac{1}{N_{\mathrm{rep}}}\sum\limits_{x=x_{\mathrm{\mathrm{asymp}}}}^{x_{\mathrm{\mathrm{pred}}}} C_A(x) \: .
\end{align}
From the normalization condition \eqref{e:ambigsum1}, we see that Eq.~\eqref{e:ambigsum2} implies the truncated moment is normalized at $x_{\mathrm{\mathrm{pred}}} = x_0$ to the total fraction of replicas containing at least one ambiguity:
\begin{align} \label{e:Ax0}
    \mathcal{A}(x_0) = \frac{N_{\mathrm{ambig}}}{N_\mathrm{rep}} \: .
\end{align}
From the left panel of \fig{Plot_g1_ambiguities} we see that the number of smallest-$x$ ambiguities decreases greatly as $x$ approaches zero. The right panel shows we must go down to approximately $x=3.5\times10^{-4},\,2.5\times10^{-5},\,\mathrm{and}\,6\times10^{-7}$ to capture the asymptotic sign with 10\%, 5\%, and 1\% uncertainty, respectively. This is strong justification that $x_{\mathrm{\mathrm{asymp}}}=10^{-7.5}$ is reasonably low enough to capture the asymptotic sign of our replicas with low uncertainty. Due to Eq.~\eqref{e:Ax0} we also know how many replicas are completely unambiguous; since we impose our evolution to begin at $x_0=0.1$, the running integral at that point quantifies the total ratio of replicas that have at least one ambiguity. According to the right panel of \fig{Plot_g1_ambiguities}, approximately 50\% of replicas choose their asymptotic sign immediately as evolution begins.  Note that the data constrains the initial condition for $g_1^p$ to be positive, so all completely unambiguous replicas are asymptotically positive.

Furthermore, splitting the replicas by their asymptotic sign (not shown in Fig.~\ref{Plot_g1_ambiguities}) allows us to also investigate how early (or late) the different solutions are chosen relative to each other. We gather that ambiguously negative replicas tend to choose their sign earlier than their positive counterparts, with the caveat that the majority of asymptotically positive replicas do not have any ambiguities at all. Approximately 75\% of asymptotically positive replicas are completely unambiguous, and the remaining 25\% are determined by $x\approx2\times10^{-5}$ with 5\% uncertainty. Though fewer in number, a still significant portion of replicas are asymptotically negative, 95\% of which are confirmed by $x\approx4.3\times10^{-4}$. This suggests that using a lower $x_{\mathrm{pred}}$ will affect the positive-identified and negative-identified solutions differently.  In particular, a lower $x_{\mathrm{pred}}$ is likely to identify a greater number of asymptotically positive solutions by correcting replicas that would have been misidentified as asymptotically negative at a higher $x_{\mathrm{pred}}$.  This asymmetric impact on positive-identified versus negative-identified solutions can be traced back to constraints from the data at large $x$, which strongly prefers $g_1^p > 0$.  The fact that this positive preference persists down to small $x$ suggests that the polarized dipole(s) which dominate the small-$x$ asymptotics are partially (but not fully) constrained by the large-$x$ data.  This will be discussed in detail in Sec.~\ref{subsubsec: origins}.

We performed a similar analysis of the smallest-$x$ critical points of each replica (rather than the ambiguities). On average, the smallest-$x$ critical point occurs 4\% earlier in $\ln(1/x)$ than its smallest-$x$ zero.  Since the ambiguous region of a replica is precisely the region in $x$ between its critical point and zero, this small $4\%$ difference indicates that any remaining ambiguities are quickly resolved at small $x$. 
Thus, we conclude that, from the perspective of Fig.~\ref{Plot_g1_ambiguities}, if we had data down to $x\approx10^{-5}$ we could determine the asymptotic sign of $g_1^p$ with high certainty ($> 95\%$).

\subsubsection{Asymptotic behavior of $g_1^p$}

Collectively utilizing the information in Figs.~\ref{Plot_g1} and \ref{Plot_g1_ambiguities} paints a curious picture:~there are many more $g_1^p$ replicas that adopt their asymptotic forms early than there are replicas that change their signs at small $x$. This results in some clustering behavior, {\it e.g.}, in the left panel of \fig{Plot_g1_ambiguities} there is a cluster of replicas around $x=5\times 10^{-3}$, implying that these replicas share similar critical points and rates of growth.  As mentioned previously, the majority of replicas have no ambiguities and adopt their asymptotic growth rather quickly, effectively clustering their critical points at $x=x_0\equiv0.1$ (not explicitly shown). This behavior supports the idea that early adoption of asymptotic growth is preferred, whereas replicas with late critical points are fewer in nature. Consequently, we expect that there should be a form of bimodality in $g_1^p$ between the rapidly growing \textit{positive} solutions versus the rapidly growing \textit{negative} solutions. This is a novel result, which we quantitatively analyze below.

    \begin{figure}[b!]
    \begin{centering}
    \includegraphics[width=500pt]{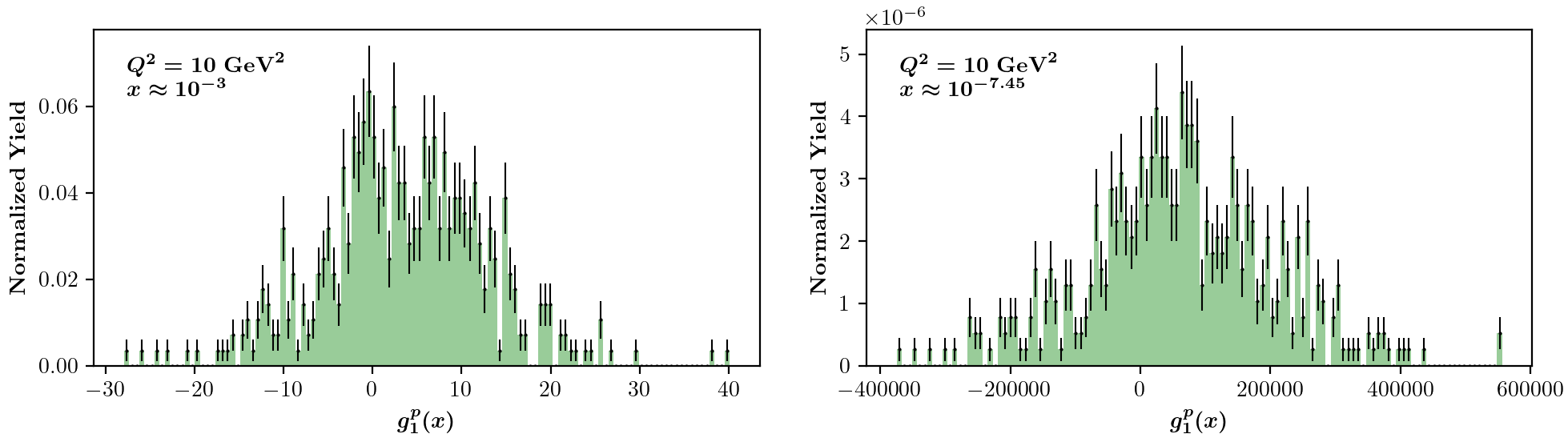}
    \caption{Histograms counting all values of $g_1^p$ at $x=10^{-3}$ (left) and $10^{-7.45}$ (right), displaying normal distributions centered slightly above zero.
        \label{g1_normdist}
    }
    \end{centering}
    \end{figure}

\begin{figure}[t]
    \begin{centering}
    \includegraphics[width=512pt]{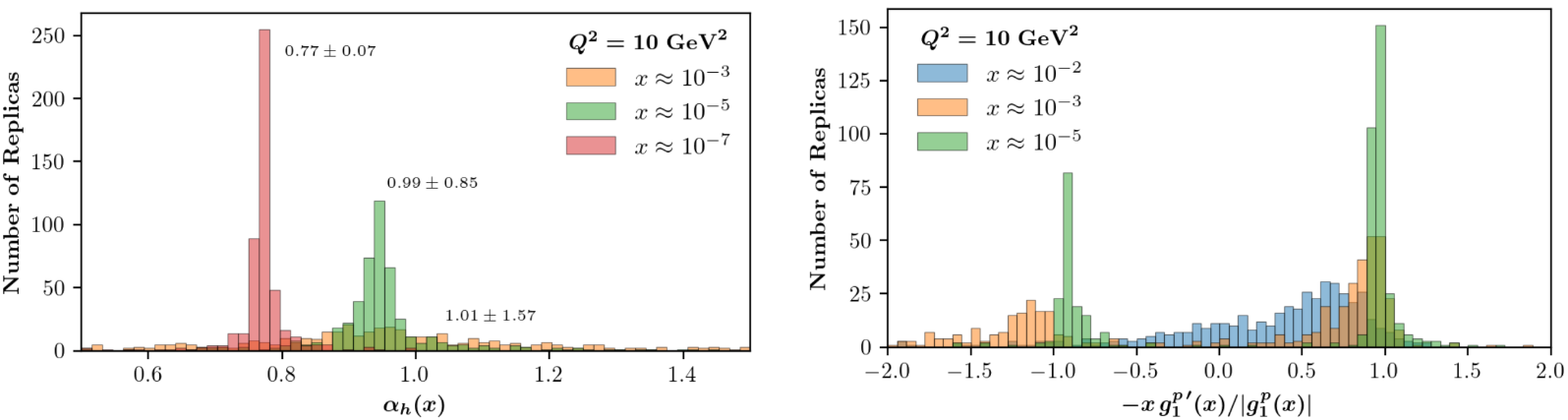}
    \caption{(Left) Histograms utilizing Eq.~\eqref{e:logderv1} showing that as $x$ decreases, the intercept $\alpha_h(x)$ becomes more constrained as a consequence of the small-$x$ evolution equations. (Right) Keeping information on the sign dependence by using Eq.~\eqref{e:logderv2} produces bimodal peaks at $\pm\alpha_h(x)$. At large $x$ there is no asymptotic behavior, and for smaller values of $x$ two refined peaks emerge.
        \label{Bimodal_Curvature}
    }
    \end{centering}
    \end{figure}

While Fig.~\ref{Plot_g1} may appear to show the anticipated bimodality (red versus blue curves), upon closer inspection the values of $g_1^p$ are normally distributed, both at small $x$ ($x=10^{-3}$) and very small $x$ ($x=10^{-7.45}$), as depicted in Fig.~\ref{g1_normdist}. To uncover the bimodal behavior it is necessary to construct a new observable related to the \textit{curvature} of $g_1^p$ which is sensitive to how quickly our evolution equations drive the $g_1^p$ replicas toward the asymptotic limit. The emphasis, therefore, is not so much on $g_1^p$ as on the exponent of its power-law behavior at small-$x$, {\it i.e.}, $g_1^p (x) \sim x^{-\alpha_h}$.  The generalized $x$-dependent exponent $\alpha_h (x)$ can be extracted through the logarithmic derivative of $g_1^p$:
\begin{align}   \label{e:logderv1}
    \lim_{x\to0}\,g_1^p(x) \equiv g_1^{p\,(0)} \, x^{-\alpha_h(x)}
    \qquad \therefore \qquad
    \alpha_h(x) \equiv \frac{1}{g_1^p(x)}\frac{\dd \,g_1^p (x)}{\dd \ln (1/x)}, 
\end{align}
where $g_1^{p\,(0)}=\mathrm{const.}$ Examining the distribution of $\alpha_h(x)$ across replicas can provide complementary information to the distribution of $g_1^p(x)$ itself.  Notably, the exponent provides a meaningful way to scale the solutions:~if they have the same $\alpha_h (x)$, they have the same \textit{curvature}, whether the magnitude of $g_1^p (x)$ is large or small.  To further capture the \textit{signed} behavior of $g_1^p(x)$ and distinguish between solutions trending positive or negative at small $x$, we can generalize the logarithmic derivative \eqref{e:logderv1} to reflect the sign of $g_1^p$ itself:
\begin{align}   \label{e:logderv2}
    \alpha_h(x) = \frac{1}{g_1^p(x)} \frac{\dd \,g_1^p (x)}{\dd \ln (1/x)}
    \qquad \Rightarrow \qquad
    \mathrm{Sign}\bigl[g_1^p (x)\bigr]\,\alpha_h (x) = \frac{1}{\big| g_1^p(x) \big|} \frac{\dd \,g_1^p (x)}{\dd \ln (1/x)}    \: .
\end{align}
Both the effective exponent $\alpha_h (x)$ \eqref{e:logderv1} and its signed generalization \eqref{e:logderv2} are shown in Fig.~\ref{Bimodal_Curvature} at varying values of $x$ (from the same global fit that produced \fig{Plot_g1}). (We remark that if a $g_1^p$ replica has a delayed critical point it will result in a delayed zero that may cause an artificially large ratio if $g_1^{p\,\prime}(x) \gg g_1^p(x)\approx 0$. In order to avoid these statistical outliers, any replica with a ratio value outside of 5$\sigma$ from the average are omitted from the results in Fig.~\ref{Bimodal_Curvature}.) The distribution in the right panel at $x=10^{-2}$ (blue histogram)  is skew-normal, which is expected since we are definitively outside of the asymptotic regime.  However, at $x=10^{-3}$ (yellow histogram)  we already see the formation of two separated peaks, one positive and one negative. As $x$ continues to decrease down to $x=10^{-5}$ (green histogram), the two peaks become more refined as the evolution equations predict specific curvature related to the intercept $\alpha_h$ (see Eq.~\eqref{e:logderv2}). Without the sign dependence, as displayed in the left panel of Fig.~\ref{Bimodal_Curvature}, as $x\to x_{\mathrm{asymp}}$, a single peak emerges that approaches the expected asymptotic value for $\alpha_h$. The decreasing uncertainties are a consequence of our small-$x$ evolution, where the predictive power constrains the value of $\alpha_h(x)$. 

From the perspective of the right panel of Fig.~\ref{Bimodal_Curvature} it appears that data sensitive to this curvature at $x$ as large as $x=10^{-3}$ may be enough to identify which bimodal peak $g_1^p$ belongs to. Unambiguously identifying this curvature will provide us the asymptotic sign of $g_1^p$ as well as the asymptotic sign of all the (flavor singlet and $C$-even) hPDFs, as will be discussed below. The fact that such a conclusion could be made  at $x\approx 10^{-3}$ by analyzing the {\it curvature} of $g_1^p(x)$, compared to $x\approx 10^{-5}$ by studying $g_1^p(x)$ itself (see the discussion around  Fig.~\ref{Plot_g1_ambiguities}), makes the idea of curvature a useful quantity to consider once future low-$x$ data is available from the EIC.

\subsubsection{Origins of asymptotic behavior}\label{subsubsec: origins}

To understand what differentiates the positively and negatively growing solutions for $g_1^p$ displayed in Fig.~\ref{Plot_g1}, we examine the polarized dipole amplitude parameters themselves, defined in Eqs.~\eqref{LinICdisc}. We note that the experimental data is only sensitive to the polarized dipole amplitudes as a whole, and not to any specific basis function. For example, combining Eqs.~\eqref{g1_Dq}, \eqref{q+}, and \eqref{A1_fin}  shows that $A_1$ is constructed from the dipole amplitudes $Q_q$ and $G_2$, and any combination of parameters that reconstructs the experimental data with good $\chi^2$ is equally valid. An appropriate change of variables can reorganize the basis hPDFs to increase the sensitivity to their overall sign.  We can then classify which of these parameters are most correlated with the asymptotic sign of $g_1^p$. We find enhanced sensitivity to the asymptotic sign of $g_1^p$ from the linear combinations $a' \equiv (a+b)/2$ and $b' \equiv (a-b)/2$.  Then the dipole initial condition $G^{(0)} = a \, \eta + b \, s_{10} + c$ can be written as
\begin{align}
    G^{(0)} &= a' \, (\eta+s_{10}) + b' \, (\eta-s_{10}) + c \,.\label{e:param_transform}
\end{align}
%

    \begin{figure}[b!]
    \begin{centering}
    \includegraphics[width=0.75 \textwidth]{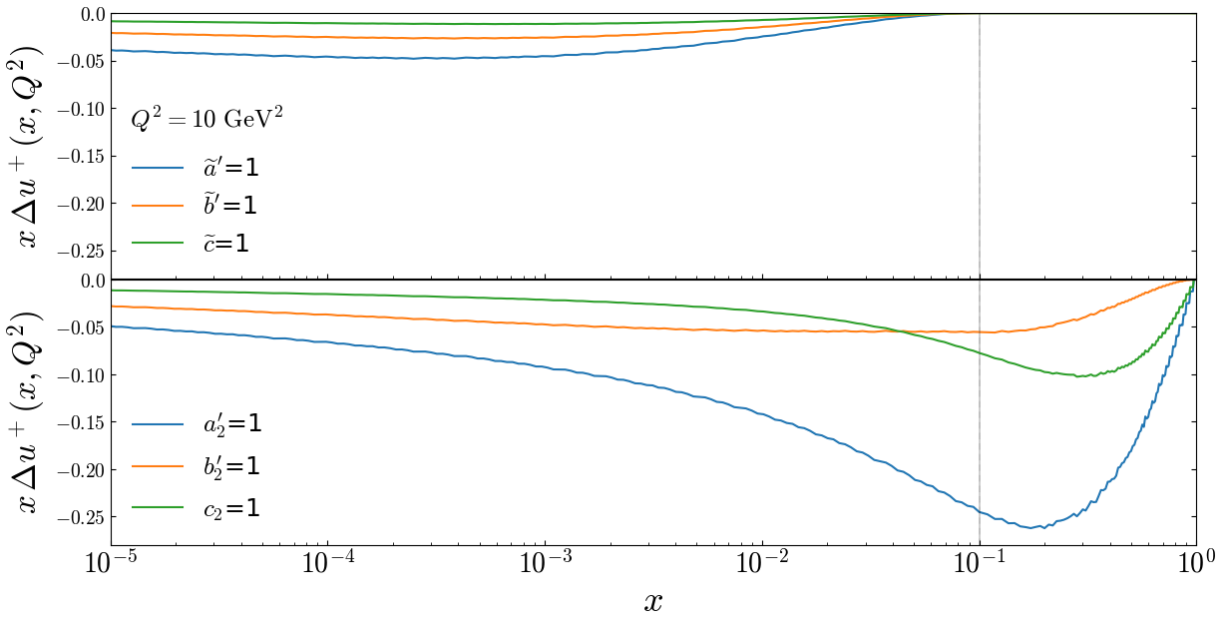}
    \caption{Basis functions analagous to those in \fig{xhpdfu_medx}, where instead of plotting the $\eta$, $s_{10}$, and $1$ contributions (displayed as the curves in \fig{xhpdfu_medx} labeled $a=1$, $b=1$, $c=1$, respectively), we instead show the contributions of $\eta+s_{10}$, $\eta-s_{10}$, and $1$ displayed as the curves labeled $a'=1$, $b'=1$, and $c'=c=1$. Here only the $\widetilde{G}$ and $G_2$ dipole amplitudes are shown.
        \label{Primed_basisfunctions}
    }
    \end{centering}
    \end{figure}

These new basis functions are displayed in \fig{Primed_basisfunctions}.  Compared to Fig.~\ref{xhpdfu_medx}, the alternative parameters $a', b'$ change the shape of the basis hPDFs.  In particular, we note that this greatly increases the separation between the $a_2^\prime = 1$ and $b_2^\prime = 1$ basis functions at large $x$ where the data provides constraints.  When we bin the replicas into asymptotically positive/negative $g_1^p$ at small $x$, we find that the parameter with the largest difference between the solutions is $\tilde{a}^\prime$.  
The asymptotically-positive solutions preferred a negative parameter $\tilde{a}^\prime = -1.56 \pm 2.32$, while the asymptotically-negative solutions preferred the positive $\tilde{a}^\prime = 1.42 \pm 2.34$.  No other systematic differences in parameters were observed.  

We can understand from the basis hPDFs shown in Fig.~\ref{Primed_basisfunctions} why asymptotically-positive/negative $g_1^p$ correlates, respectively, with negative/positive values of $\tilde{a}^\prime$, and why $\tilde{a}^\prime$ shows the greatest discrimination power.  First, we note that the basis hPDFs themselves are negative-definite functions of $x$ for positive values of the initial parameters $a',b',c$, which is simply a consequence of the explicit minus sign in Eq.~\eqref{q+}.  Second, we note that the hPDFs arising from both the $\widetilde{G}^{(0)}$ (with parameters $\tilde{a}', \tilde{b}', \tilde{c}$) and $G_2^{(0)}$ (with parameters $a_2^\prime, b_2^\prime, c_2$) initial conditions are comparably large at small $x$; the $a_2^\prime = 1$ basis function also being sizeable at large $x$, whereas the $\widetilde{a}'$ basis function only contributes meaningfully at small-$x$.  The large-$x$ behaviour means that the parameter $a_2^\prime$, while important for determining the small-$x$ asymptotics, is constrained by higher-$x$ experimental data, and it specifically prefers \textit{negative} values: $a_2^\prime = -0.98 \pm 1.00$.  The origin of the different asymptotic behaviors seen in Fig.~\ref{Plot_g1} therefore appears to be due to the dipole $\widetilde{G}$, which makes no contribution to the basis hPDFs at larger $x$, and, thus, the sign of $\tilde{a}^\prime$ evades experimental constraints. 

\begin{figure}[t!]
    \begin{centering}
    \includegraphics[width=475pt]{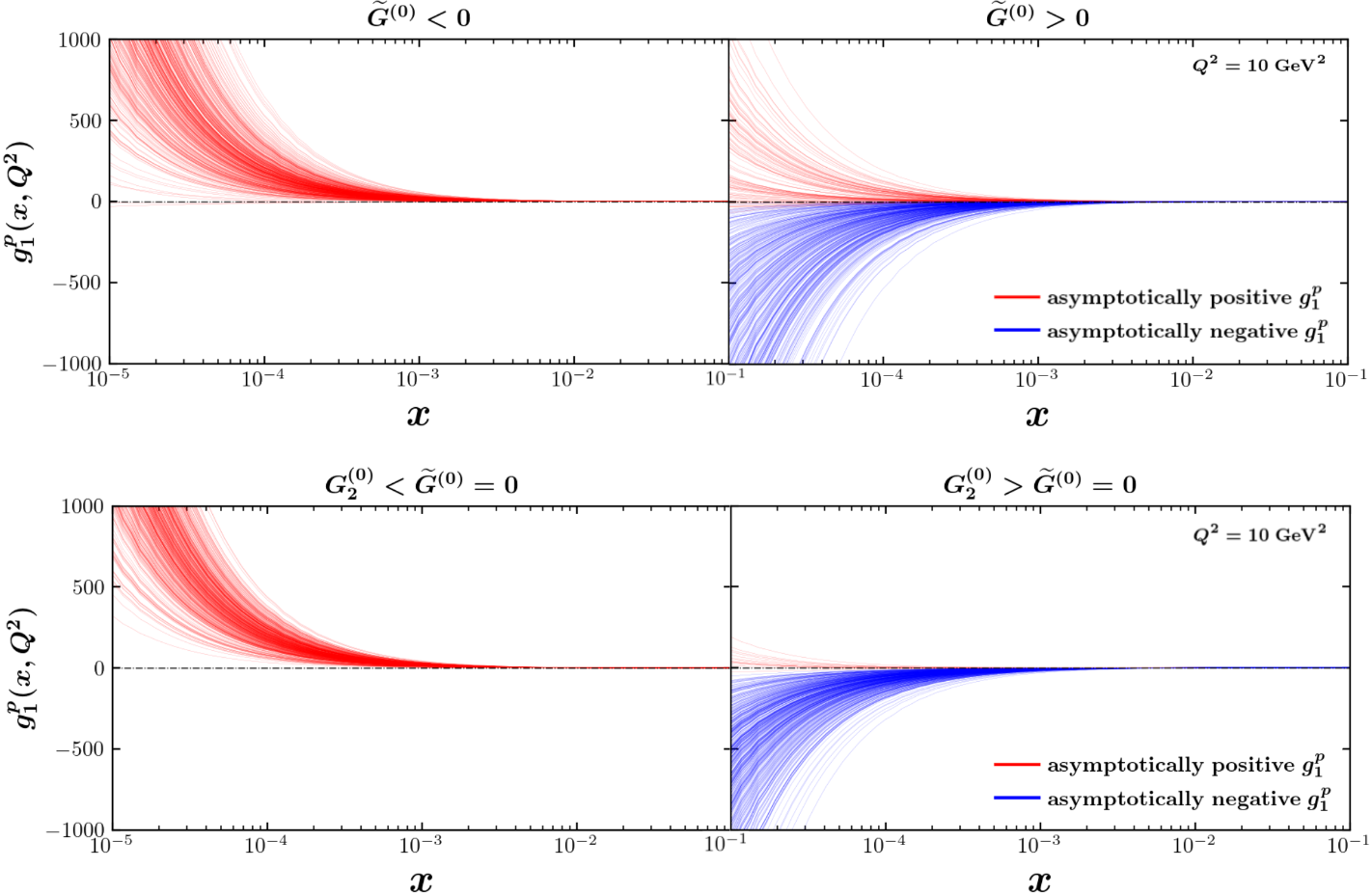}
    \caption{Comparing the effect $\widetilde{G}$ and $G_2$ has on the overall sign of $g_1^p(x)$ at small-$x$. Top row:~the priors are restricted so that (left) $\widetilde{G} \leq 0$  and (right) $\widetilde{G} \geq 0$. Bottom row:~the priors are restricted so that (left) $G_2 < \widetilde{G} = 0$, and (right) $G_2 > \widetilde{G} = 0$. All other parameters initially are randomly sampled just as they were in the fit shown in \fig{Plot_g1}. We see that controlling the sign of $\widetilde{G}$ strongly influences the sign of $g_1^p$, and that the sign of $G_2$ will also influence the sign of $g_1^p$.
        \label{GT_G2_Influence}
    }
    \end{centering}
    \end{figure}

To test this hypothesis, we ran fits where all of the $\widetilde{G}$ initial condition parameters ($\tilde{a}, \tilde{b}, \tilde{c}$) were restricted to either be negative-definite or positive-definite, with all other parameters unchanged. All $g_1^p$ replicas in the negative-definite $\widetilde{G}$ fit were  asymptotically positive. The positive-definite $\widetilde{G}$ fit was slightly less selective but still generated a 73\% majority preferring asymptotically negative $g_1^p$ replicas (recall the original fit in \fig{Plot_g1} had a 70\% \textit{positive} preference). The results, shown in the top row of \fig{GT_G2_Influence}, clearly demonstrate that the sign of the $\widetilde{G}$ dipole amplitude determines the small-$x$ asymptotics of $g_1^p$, as anticipated by the basis functions in Figs.~\ref{xhpdfu_medx} and \ref{Primed_basisfunctions}. 

The reason $\widetilde{G}$ leads to a $g_1^p$ that is poorly constrained at small $x$ can be seen directly from Eqs.~\eqref{g1_Dq}--\eqref{JM_DeltaG}, \eqref{eq_LargeNcNf} and Eqs.~\eqref{dq-}, \eqref{NSeq_LargeNcNf}, \eqref{g1h_final}:~$\widetilde{G}$ does not contribute directly to any hPDF.  Whereas all the other (non-neighbor) polarized dipole amplitudes directly enter a DIS/SIDIS observable, the effects of $\widetilde{G}$ are only felt indirectly through its impact on the evolution of the other amplitudes.  As a result, hPDFs mediated by $\widetilde{G}$  only become large at very small $x$ (see the top panel of Fig.~\ref{Primed_basisfunctions}), where there are no constraints from data.

While $\widetilde{G}$ is the driving factor in determining the small-$x$ asymptotics of $g_1^p$, $G_2$ also plays a role. In fact, if $\widetilde{G}$ was removed, $G_2$ would be the 
most important amplitude in controlling the small-$x$ asymptotics of $g_1^p$. We see this explicitly when setting the initial conditions for $\widetilde{G}$ all to zero ($\tilde{a} = \tilde{b} = \tilde{c} = 0$) and repeating the previous analysis of now restricting the $G_2$ initial condition parameters to be always positive or always negative. The result, shown in the bottom panel of \fig{GT_G2_Influence}, confirms that, although constrained by large-$x$ data, $G_2$ plays the second-most important role after $\widetilde{G}$ in determining the small-$x$ asymptotics of $g_1^p$. The negative-definite $G_2$ fit was 100\% selective of asymptotically positive $g_1^p$ replicas, while the positive-definite $G_2$ fit was 96\% selective of asymptotically negative $g_1^p$ replicas. 

\fig{GT_G2_Influence} then compactly summarizes the origin of the asymptotic behavior seen in \fig{Plot_g1}.  The origin of the huge uncertainty band at small $x$ is due to the inability to constrain the sign of $\widetilde{G}$ from large-$x$ data, and  the overall preference of the central curve in \fig{Plot_g1} favoring positive solutions is due to the fact that there \textit{is} an experimental constraint which prefers $G_2 < 0$, leading to $g_1^p > 0$.

\begin{figure}[t!]
    \begin{centering}
    \includegraphics[width=525pt]{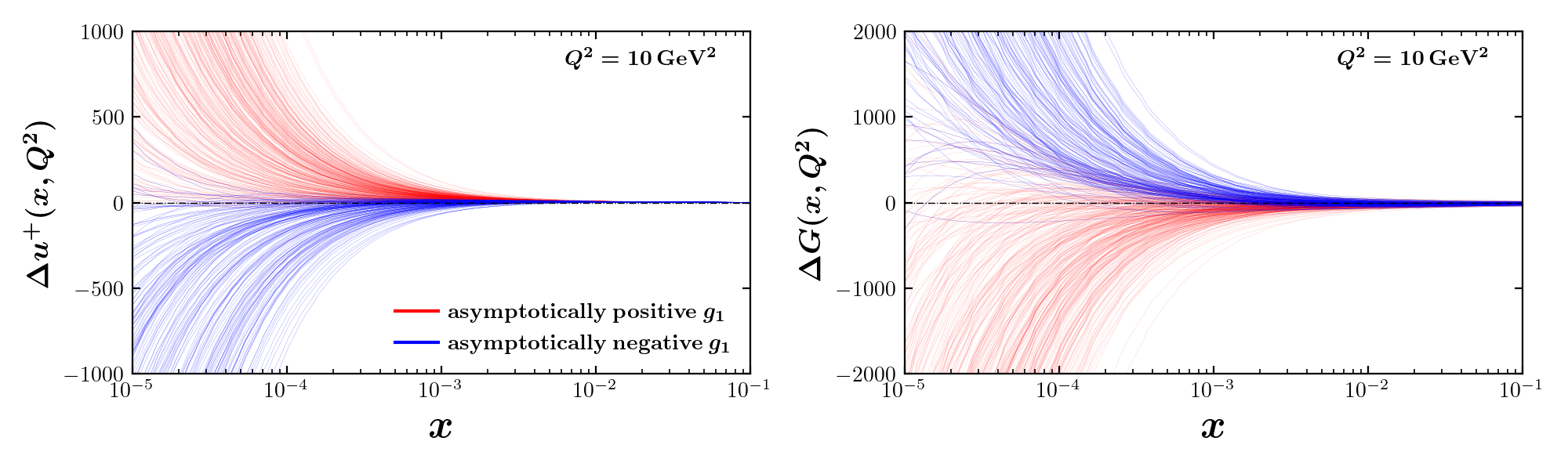}
    \caption{Color coding the hPDF replicas according to the asympotic sign of $g_1^p$ shows that there is a novel correlation:~at small $x$, quark hPDFs (left) have the same sign as $g_1^p$  (only $\Delta u^+$ is shown) while the gluon hPDF (right) has the opposite sign as $g_1^p$.
        \label{hPDF_Correlation}
    }
    \end{centering}
    \end{figure}

Knowing now that the dipole amplitude $\widetilde{G}$ controls the small-$x$ asymptotics of $g_1^p$ gives us powerful insight into the hPDF correlations which characterize the fits.  Comparing Eqs.~\eqref{g1_Dq},~\eqref{q+}, and~\eqref{JM_DeltaG} we can draw the conclusion that at asymptotically small $x$ these quantities are simply related by
\begin{align}\label{e:proportionality}
    g_1^p(x)\propto\Delta q^+(x)&\sim-(Q_q+2G_2) \to -\widetilde{G}, \\
    \Delta G(x)&\sim G_2\to\widetilde{G}\,,\notag
\end{align}
where the last step in each line represents the fact that the evolution of $Q_q$ and $G_2$ is driven by $\widetilde{G}$ (see Eqs.~(\ref{eq_LargeNcNf})). At small $x$, the two hPDFs $\Delta q^+$ and $\Delta G$ are both driven by the same polarized dipole amplitude $\widetilde{G}$, but have opposite signs.  Since $g_1^p$ is proportional to $\Delta q^+$ (weighted by quark electric charge squared and summed over flavors), it follows that if the quark hPDFs for all flavors have the same sign, then, at small $x$, $g_1^p$ will have the same sign as the quark hPDFs and opposite sign as the gluon hPDF. These anticipated (anti)correlations among the hPDFs are shown in \fig{hPDF_Correlation}, where we plot only $\Delta u^+$ and $\Delta G$ for brevity. Note that the color coding used for the replicas in \fig{hPDF_Correlation} indicates the ultimate asymptotic sign of $g_1^p$, not the hPDF itself.  That is, an hPDF replica is colored red (blue) if the corresponding $g_1^p$ replica is asymptotically positive (negative). The fact that the asymptotic signs of $\Delta q^+$ and $\Delta G$ are, respectively, correlated and anti-correlated to the sign of $g_1^p$ at small $x$ is a robust, novel prediction of the small-$x$ helicity evolution framework.
\footnote{We note that no such relationship is exhibited by the nonsinglet hPDFs.  When attempting the same strategy of color coding the nonsinglet hPDFs (not shown) according to the asymptotic sign of the proton SIDIS structure function $g_1^{p\to h}$, no correlations could be identified.}$^,$
\footnote{We note that in Ref.~\cite{Borsa:2020lsz} a connection was found at small $x$ between $\Delta G(x,Q^2)$ and the $\log Q^2$ derivative of $g_1 (x,Q^2)$: $\Delta G (x, Q^2) \approx - \partial g_1 (x,Q^2) /\partial \ln Q^2$. Our result, however, demonstrates anti-correlation of the signs of $\Delta G(x,Q^2)$ and $g_1^p(x,Q^2)$ (and not of the $\log Q^2$ derivative of $g_1^p(x,Q^2)$). In addition, we note that the calculation in  Ref.~\cite{Borsa:2020lsz} was in a DGLAP-based NLO perturbative QCD framework, while our calculation involves the all-order DLA-resummed coefficient functions (see the discussion around Eq.~\eqref{g1_coef_ftns}).} 
Thus, in order to better predict the asymptotic sign of $g_1^p$, $\Delta q^+$ and $\Delta G$, we need to better constrain the polarized dipole amplitude $\widetilde{G}$.  One option is data from the future EIC, discussed in the Sec.~\ref{sec:results-EICimpact}.  We also outline several additional ways in Sec.~\ref{sec:additional_constraints}.

\subsection{Extracted helicity PDFs and calculation of net parton spin and axial-vector charges at small \textbf{{\it x}}} \label{s:spin}

\begin{figure}[t!]
        \begin{centering}
        \includegraphics[width=500pt]{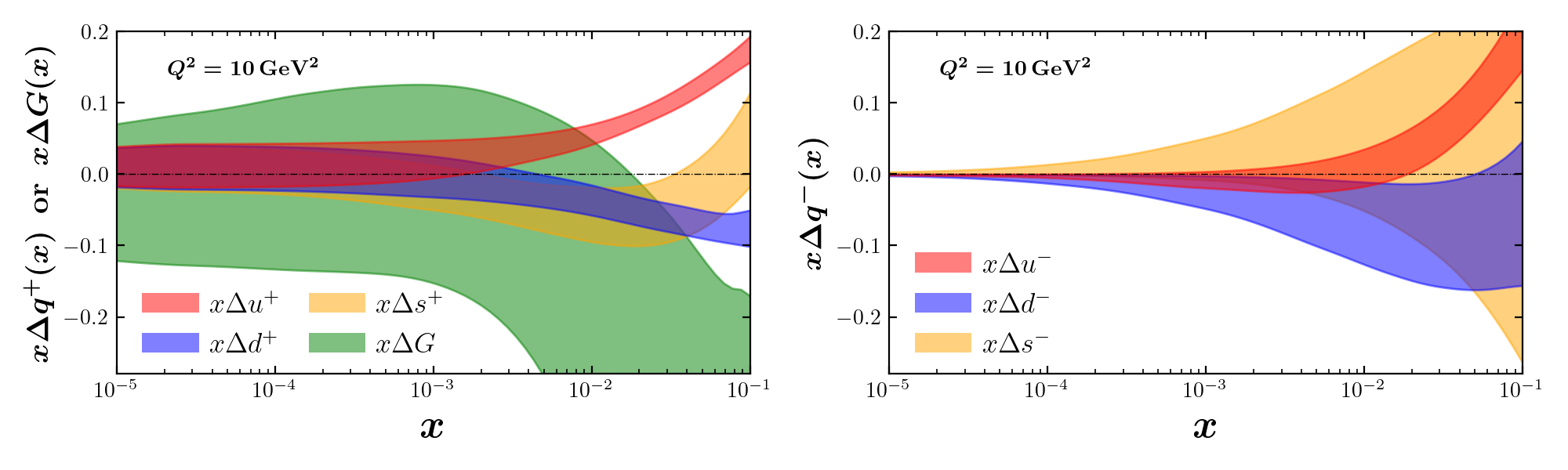}
        \caption{(Left) $C$-even hPDFs $x\Delta u^+$ (red), $x\Delta d^+$ (blue), $x\Delta s^+$ (orange) and $x\Delta G$ (green) extracted from existing low-$x$ experimental data.
        (Right) Same as left panel but for the flavor nonsinglet $C$-odd hPDFs $x\Delta u^-$ (red), $x\Delta d^-$ (blue), $x\Delta s^-$~(orange).
            \label{hPDF_bands}
        }
        \end{centering}
        \end{figure}

Our results for the hPDFs are shown in \fig{hPDF_bands}. Since our small-$x$ analysis is only valid for $x < x_0 = 0.1$, we restrict the plots to that region. As with the $g_1^p$ structure function shown in \fig{Plot_g1}, the hPDFs themselves also exhibit broad uncertainty bands at small $x$.
\footnote{Note that in \fig{hPDF_bands} we plot $x$ mutliplied by hPDF on the vertical axis:~this explains why the error bands in \fig{hPDF_bands} appear to be smaller than those in \fig{Plot_g1_EIC_impact}, with the latter showing $g_1^p$ not multiplied by $x$.}
The uncertainty bands for all four hPDFs span zero below $x \lesssim 10^{-3}$, indicating that the hPDFs in that region may be positive, negative, or consistent with zero.  By far the largest uncertainty is seen in $\Delta G$, which, unlike $\Delta q^+$, is not directly sensitive to inclusive DIS constraints on $g_1^p$ (\eq{g1_Dq}).  As shown in Figs.~\ref{GT_G2_Influence} and \ref{hPDF_Correlation}, the large uncertainty in $\Delta G$ is due to the lack of sufficient constraints on the dipole amplitudes $\widetilde{G}$ and $G_2$ that dominate both $\Delta q^+$ and $\Delta G$ at small $x$. This conclusion is further supported by the left panel of \fig{hPDF_bands}, where $\Delta u^+$, $\Delta d^+$ and $\Delta s^+$ exhibit approximately the same error band  below $x \approx 10^{-4}$. At larger $x$, where the hPDF behavior is driven more by the $Q_q$ dipole amplitudes, we can observe flavor separation between the three quarks. The uncertainty of the $\Delta s^+$ distribution then becomes much larger than that for $\Delta u^+$ and $\Delta d^+$, most likely due to the limited SIDIS kaon data. The similar error bands at small $x$ for $\Delta u^+$, $\Delta d^+$ and $\Delta s^+$ are in contrast to markedly distinct error bands for $\Delta u^-$, $\Delta d^-$ and $\Delta s^-$, shown in the right panel of \fig{hPDF_bands}, which exhibit significant flavor separation even down to small $x$. Recall that the flavor nonsinglet hPDFs are driven by a different polarized dipole amplitude, $G^{\rm NS}$ (see Eq.~(\ref{dq-})), which is sensitive to flavor separation through the SIDIS data.  As a result of the different evolution, the $x \Delta q^-$ distributions converge quickly to zero at small $x$, unlike the $x \Delta q^+$ distributions, due to the smaller intercept at small $x$ (see also Appendix~\ref{NS_Crosscheck_apdx}).  The similarity of the error bands for $\Delta u^+$, $\Delta d^+$ and $\Delta s^+$ appears to be driven by the error band of the polarized dipole amplitude $G_2$, which affects all quark flavors in the same way, per \eq{q+2}.  Consequently, additional input which can better constrain $\widetilde{G}$ and/or $G_2$ may well reduce this uncertainty by forcing the hPDFs to choose a definite sign at small $x$.  We discuss possible strategies to achieve this in Sec.~\ref{sec:additional_constraints}.

    \begin{figure}[t!]
    \begin{centering}
    \includegraphics[width=0.55\textwidth]
    {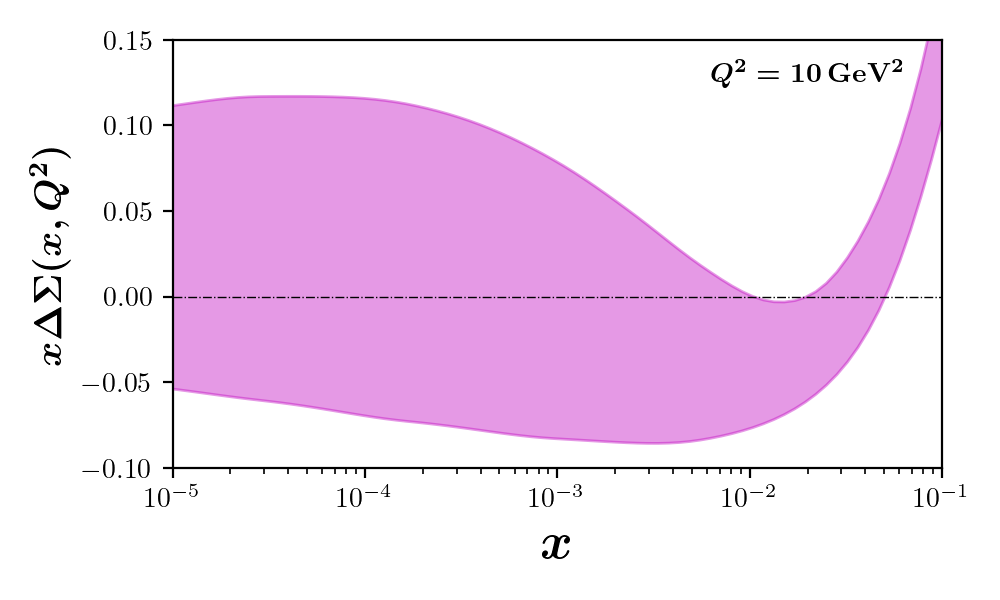}
    \vspace{-0.3cm}
    \caption{Quark flavor singlet helicity distribution $x\Delta \Sigma(x,Q^2)$ calculated from hPDFs extracted from existing low-$x$ experimental data. 
        \label{xDelta_Sigma_plot}
    }
    \end{centering}
    \end{figure}

One feature of note in our hPDFs from \fig{hPDF_bands} is that $\Delta s^+$ and $\Delta G$ are much larger in magnitude than the same hPDFs obtained in the JAM framework using the DGLAP-based approach~\cite{Ethier:2017zbq, Zhou:2022wzm, Cocuzza:2022jye}.  In particular, our extracted $\Delta s^+$ distribution is below zero at about the $1\sigma$ level at $x \approx 10^{-2}$.  This is to be compared with Fig.~6 of Ref.~\cite{Zhou:2022wzm}, which exhibits a $\Delta s^+$ consistent with zero across the entire considered range $5 \times 10^{-3} \leq x \leq 0.9$. Note that the global analyses conducted in Refs.~\cite{Ethier:2017zbq, Zhou:2022wzm, Cocuzza:2022jye} are quite different than the one we present here, {\it e.g.}, they use DGLAP evolution within collinear factorization, include data across the full range of $x$, and in some cases impose SU(2) and SU(3) flavor symmetries. Nevertheless, it is a valuable cross-check to see whether zero strangeness polarization is consistent with our results as well.
To that end, we have separately re-fit the data, setting the strangeness polarization identically to zero: $\Delta s^+ (x, Q^2) = \Delta s^- (x, Q^2) =0$.  The overall quality $\chi^2/N_{\rm pts} = 1.04$ of the zero-strangeness fit is slightly worse than the quality $\chi^2/N_{\rm pts} = 1.03$ of the default fit, with the asymmetries $A_1^h$ from tagged kaon SIDIS being the most affected by the change.  For that subset of the data, the quality-of-fit degraded from $\chi^2 / N_\mathrm{pts} = 0.81$ in the default fit to $\chi^2 / N_\mathrm{pts} = 1.05$ in the zero-strangeness fit.  This marginal degradation of the fit quality is consistent with the $1\sigma$ departure of $\Delta s^+$ from zero preferred by the default fit in \fig{hPDF_bands}, with the tagged kaon data only accounting for 26/226 data points in total. Therefore, we conclude that small $\Delta s^+$ is indeed consistent with our formalism, and that there is a real (but weak) preference from the data for nonzero $\Delta s^+$ at $x \sim 0.01$ within our small-$x$ framework. 

Next, we address the contribution to the proton spin and axial-vector charges from small $x$. 
The flavor singlet quark helicity distribution is given by
\begin{align}
    \Delta \Sigma (x, Q^2) \equiv \Delta u^+ (x, Q^2) + \Delta d^+ (x, Q^2) + \Delta s^+ (x, Q^2)\,,
\end{align}
for the light flavors considered in this work.  Using  the hPDFs in \fig{hPDF_bands}, we show $x\Delta \Sigma (x, Q^2)$ in \fig{xDelta_Sigma_plot}. Again, the uncertainty band at small $x$ based on current experimental data is rather wide, spanning zero so that the sign of $\Delta \Sigma$ is uncertain.

    \begin{figure}[t!]
    \begin{centering}
    \includegraphics[width=400pt]{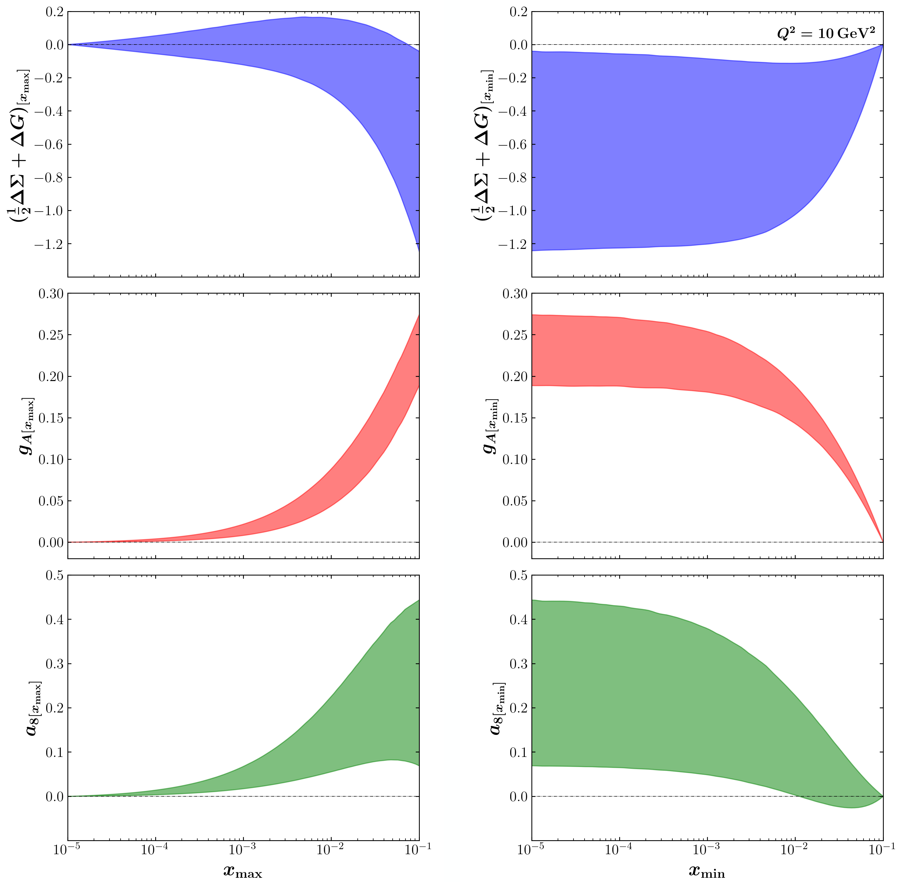}
    \caption{
    Truncated moments of $(\tfrac{1}{2}\Delta\Sigma+\Delta G)(x,Q^2)$, $g_A(x,Q^2)$ and $a_8(x,Q^2)$, defined in Eqs.~(\ref{SG_xmax}), versus $x_{\rm max}$ (left) and $x_{\rm min}$ (right) at $Q^2=10~\mathrm{GeV}^2$. 
    \label{integrated_plots}
    }
    \end{centering}
    \end{figure}

From $\Delta\Sigma(x,Q^2)$ and $\Delta G(x,Q^2)$ we can determine how much net parton spin (see Eq.~\eqref{eqn:SqSG}) resides at small $x$ by computing truncated moments of the distributions.  We can similarly determine the small-$x$ contributions to the triplet $g_A$ and octet $a_8$ axial-vector charges from truncated moments of the appropriate linear combinations of quark hPDFs.  Focusing on the $x$ region $10^{-5} \leq x \leq 10^{-1}$ of our analysis, we consider the following truncated moments:
\begin{subequations}\label{SG_xmax}
\begin{align}
    \left(\frac{1}{2}\Delta \Sigma+\Delta G\right)_{\![x_{\rm max(min)}]} \! & \equiv \int\limits_{x_1}^{x_2} \dd x \, \left(\frac{1}{2}\Delta \Sigma+\Delta G\right) \!(x, Q^2)\,,\\
    g_{A[x_{\rm max(min)}]}\! \equiv \int\limits_{x_1}^{x_2} \dd x \,  g_A(x,Q^2)&\equiv\int\limits_{x_1}^{x_2} \dd x \, \left[ \Delta u^+ (x, Q^2) - \Delta d^+ (x, Q^2) \right], \\
    a_{8[x_{\rm max(min)}]}\! \equiv \int\limits_{x_1}^{x_2} \dd x \, a_8(x,Q^2)\equiv\int\limits_{x_1}^{x_2} \dd x \,& \left[ \Delta u^+ (x, Q^2) + \Delta d^+ (x, Q^2)  - 2 \, \Delta s^+ (x, Q^2) \right]    \: .
\end{align}
\end{subequations}
Here we consider two representations of the truncated moments: either as a function of the upper limit $x_{\rm max}$ with fixed lower limit $10^{-5}$, or as a function of the lower limit $x_{\rm min}$ with fixed upper limit $0.1$.  That is, in the notation of \eq{SG_xmax}, we have $(x_1,x_2)=(10^{-5},x_{\rm max})$ for $[x_{\rm max}]$ and $(x_1,x_2)=(x_{\rm min},0.1)$ for $[x_{\rm min}]$.  We have also dropped the $Q^2$ dependence of the truncated moments on the left-hand side of \eq{SG_xmax} for brevity.

Both $[x_{\rm max}]$ and $[x_{\rm min}]$ representations of the truncated moments are plotted in \fig{integrated_plots}.  From the truncated moment of the total parton helicity $\left( \frac{1}{2}\Delta \Sigma+\Delta G\right)_{[x_{\rm max(min)}]}$, we conclude that, despite the sizable uncertainties, the amount of the proton spin coming from the net spin of small-$x$ partons could be quite large.  The outer bounds of these truncated moments also allow for the possibility that the net quark and gluon spin contained within the small-$x$ region may be even more significant than what has been computed at large $x$. We observe that, despite the wide error bands in $\Delta G(x,Q^2)$ and $\Delta \Sigma(x,Q^2)$ separately, the error in the truncated moment $\left( \frac{1}{2}\Delta \Sigma+\Delta G\right)$ is narrower than if the two were uncorrelated. Because of the replica-by-replica anticorrelation between $\Delta q^+(x,Q^2)$ and $\Delta G(x,Q^2)$ seen in \fig{hPDF_Correlation}, there is a systematic cancellation between them, resulting in a truncated moment $\left( \frac{1}{2}\Delta \Sigma+\Delta G\right)$ which skews net negative and is more tightly constrained than either $\Delta \Sigma(x,Q^2)$ or $\Delta G(x,Q^2)$ alone. In addition, the nonzero slope of $\left( \frac{1}{2}\Delta \Sigma+\Delta G\right)_{[x_{\rm max}]}$ as one approaches $x_{\rm max}=10^{-5}$ indicates that this truncated moment has not fully saturated at that point in $x$. In contrast, the small-$x$ contribution to $g_A$ and $a_8$ appears to saturate around $x=10^{-4}$, giving a finite, non-negligible contribution from small-$x$ partons.  

Taken at face value, our formalism strikingly predicts a \textit{negative} contribution to the proton spin from the net spin of small-$x$ partons even when accounting for the $1 \sigma$ error band. In this scenario favored by our default fit, a significant positive contribution from orbital angular momentum would be needed to satisfy the Jaffe-Manohar sum rule \eqref{eqn:JM}.  Interestingly, similar observations have been made in using AdS/CFT to analyze $g_1^p$~\cite{Hatta:2009ra,Kovensky:2018xxa,Jorrin:2022lua,Borsa:2023tqr}.   We also predict that approximately $15$-$21\%$ of the known value of $g_A$ and $12$-$77\%$ of the known value of $a_8$ are generated from partons with $10^{-5} \leq x \leq 10^{-1}$, where the values of the moments over the full range $x\in[0,1]$ are known from neutron and hyperon
$\beta$-decays~\cite{Jimenez-Delgado:2013boa}:~$g_A=1.269(3)$ and $a_8=0.586(31)$.   

However, we caution the reader that our small-$x$ analysis is strongly dependent on the large-$x$ initial conditions to our evolution, and that the error bands shown throughout this work are \textit{strictly statistical} in nature.  These are an accurate representation of the uncertainty coming from the experimental data and from the Monte Carlo sampling procedure, but in particular they do not reflect the systematic bias that comes from omitting large-$x$ data that cannot be captured in this formalism.  Combining our small-$x$ evolution equations with external input from large $x$ can therefore possibly result in large, systematic changes to the extracted hPDFs beyond the $1\sigma$ statistical error bands.  This suggests that an appropriate matching procedure onto hPDFs extracted from a large-$x$, DGLAP-based analysis like JAM~\cite{Ethier:2017zbq, Zhou:2022wzm, Cocuzza:2022jye} will be crucial to determining the proton spin budget.  Moreover, since JAM found both viable positive $\Delta G(x,Q^2)$ and negative $\Delta G(x,Q^2)$ solutions~\cite{Zhou:2022wzm, Cocuzza:2022jye}, the predictions for the small-$x$ truncated moments may even depend on \textit{which} large-$x$ solution is chosen for the matching. Indeed, as we show in Fig.~\ref{Plot: Delta G matching} below, matching to the positive gluon hPDF solution could lead to a substantially different outcome for $\Delta G(x,Q^2)$, deviating beyond the $1\sigma$ error band over a significant range of $x$. Clearly a rigorous implementation of such a matching will be an important aspect of future analyses; a first attempt is detailed in Section \ref{sec:additional_constraints} below.
Having emphasized this vital caveat, we summarize our results for the small-$x$ truncated moments of $(\tfrac{1}{2}\Delta\Sigma+\Delta G)(x,Q^2)$, $g_A(x,Q^2)$ and $a_8(x,Q^2)$ over the small-$x$ window $x\in [10^{-5}, 0.1]$ for $Q^2=10\,{\rm GeV^2}$:
\begin{align}
\label{e:moment_sum}
\int\limits_{10^{-5}}^{0.1} \dd x \, \left(\frac{1}{2}\Delta \Sigma+\Delta G\right) \!(x) = -0.64 \pm 0.60\,,\quad\quad
\int\limits_{10^{-5}}^{0.1} \dd x \,g_A(x) = 0.23 \pm 0.04\,, \quad\quad
\int\limits_{10^{-5}}^{0.1} \dd x \,a_8(x) = 0.26 \pm 0.19\,. 
\end{align}
%

\subsection{Impact of EIC data on $g_1^p$}
\label{sec:results-EICimpact}

In order to study the impact of lower $x$ measurements on our ability to predict the behavior of $g_1^p$ and the hPDFs at even smaller $x$, we utilized EIC pseudodata for the kinematic region of  $10^{-4}<x<10^{-1}$ and $1.69 ~\text{GeV}^2 < Q^2 <50 ~\text{GeV}^2$. The EIC will be capable of going lower in $x$ by reaching higher $Q^2$, but we do not expect our formalism to be applicable for arbitrarily large $Q^2$ (DGLAP resummation is needed to fully describe the $Q^2$ dependence). For DIS  on the proton, the pseudodata was at center-of-mass energies $\sqrt{s} = \lbrace 29, 45, 63, 141\rbrace$~GeV with integrated luminosity of $100 \,\text{fb}^{-1}$, while for the deuteron and $^3$He beams the pseudodata spanned $\sqrt{s} = \lbrace 29, 66, 89\rbrace$~GeV with $10 \,\text{fb}^{-1}$ integrated luminosity. These are consistent with the EIC detector design of the Yellow Report, including 2$\%$ point to-point uncorrelated systematic uncertainties~\cite{AbdulKhalek:2021gbh}. For SIDIS on a proton, the pseudodata was at $\sqrt{s} = {141}$~GeV, also with a $2\%$ systematic uncertainty \cite{VANHULSE2023168563}. In our earlier work \cite{Adamiak:2021ppq}, we had relied on parity violating DIS pseudodata in order to disentangle the three light quark $C$-even hPDFs $\Delta q^+$. With the inclusion of SIDIS data, that is no longer necessary. The EIC could provide such data \cite{AbdulKhalek:2021gbh}, and it would serve as an additional constraint in the future, but we do not consider its impact in the present analysis. 

    \begin{figure}[b!]
            \begin{centering}
            \includegraphics[width=0.65\textwidth]
            {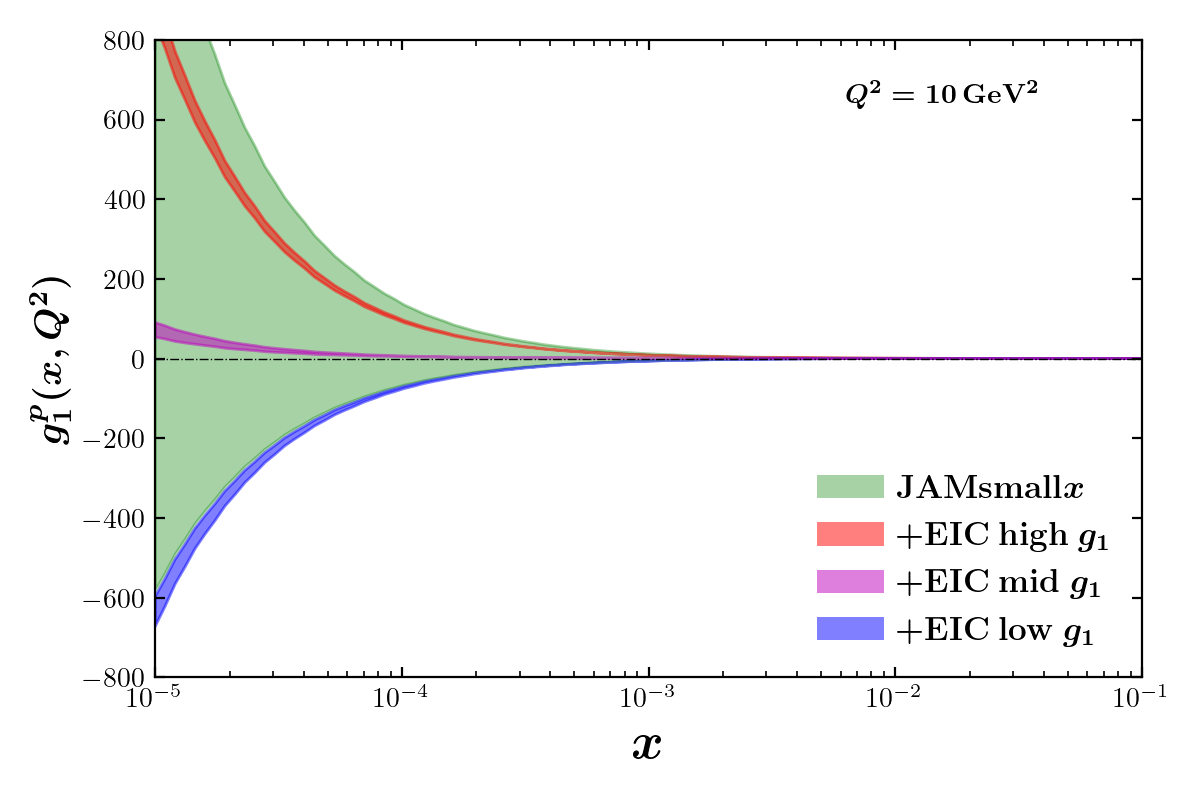}
            \vspace{-0.3cm}
            \caption{Extraction of $g_1^p$ from the current low-$x$ experimental data (green, same as Fig.~\ref{Plot_g1}) and with EIC pseudodata generated from the mean of the asymptotically positive $g_1^p$ replicas (red), mean of the asymptotically negative $g_1^p$ replicas (blue), and mean of replicas restricted such that $|g_1^p|<100$ at $x=10^{-4}$ (magenta).
                \label{Plot_g1_EIC_impact}
            }
            \end{centering}
    \end{figure}

    \begin{figure}[b!]
            \begin{centering}
            \includegraphics[width=0.65\textwidth]
            {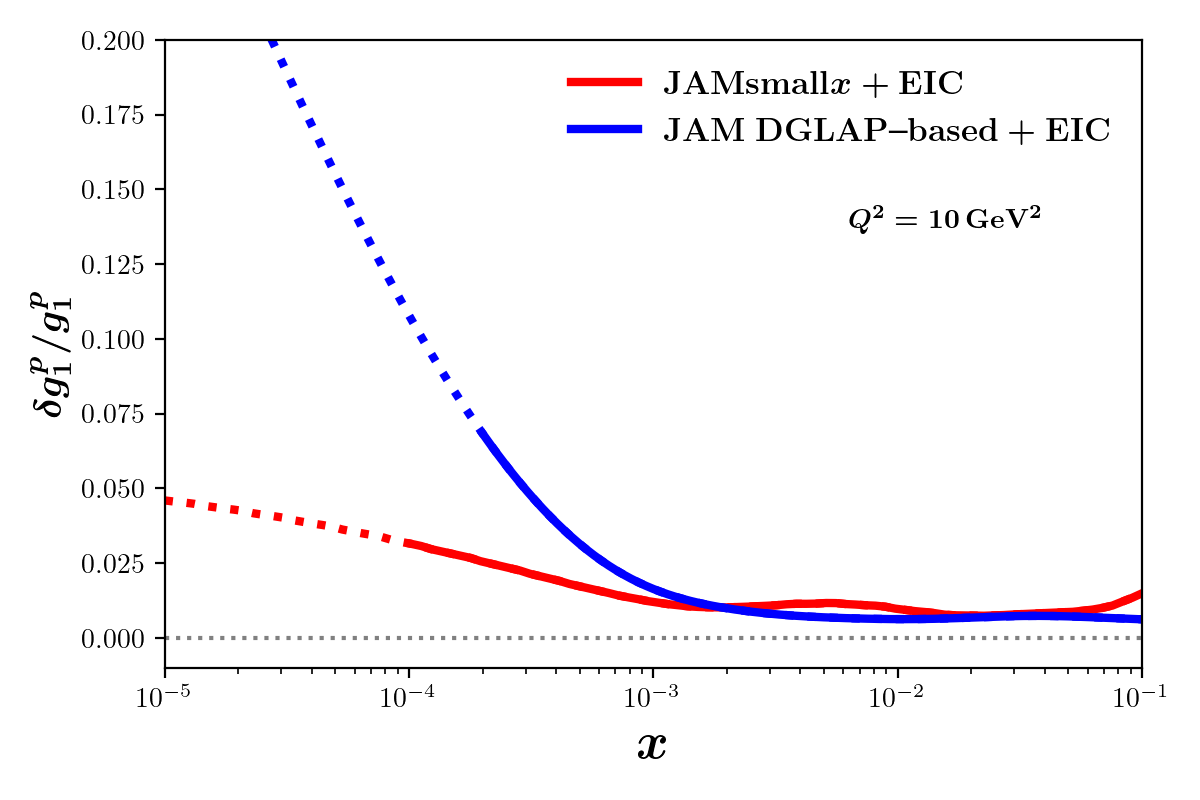}
            \vspace{-0.3cm}
            \caption{Relative uncertainty for both this work (red) and a JAM DGLAP-based extraction~\cite{Zhou:2021llj} (blue) for EIC impact studies using the high $g_1^p$ scenario. Dotted lines denote extrapolating beyond lowest $x$ for which pseudodata was generated. For this work, pseudodata was generated down to $x=10^{-4}$. For the JAM DGLAP-based fit, pseudodata was generated down to $x=2\times 10^{-4}$~\cite{Zhou:2021llj}. 
                \label{Plot_g1_uncertainty}
            }
            \end{centering}
    \end{figure}

Our current extrapolation of $g_1^p$ covers a wide range of possibilities at small $x$, so we generate the pseudodata based on three scenarios for $g_1^p$ that are consistent with present data: (1) the mean of the asymptotically positive replicas (``high $g_1$''), (2) the mean of the asymptotically negative replicas (``low $g_1$'') and (3) the mean of a fit where $g_1^p$ was constrained to have $|g_1^p|<100$ at $x=10^{-4}$ (``mid $g_1$''). These three options have qualitatively distinct behaviors and comparing them should inform us if the impact of the EIC is dependent on the precise small-$x$ asymptotics of $g_1^p$. The results are shown in \fig{Plot_g1_EIC_impact}. We find a dramatic decrease in uncertainties for all three scenarios, even in the extrapolated region of $x<10^{-4}$. In \fig{Plot_g1_uncertainty} we plot the relative uncertainty of $g_1^p$ compared to that of a JAM DGLAP-based extraction in Ref.~\cite{Zhou:2021llj} using EIC pseudodata. The results confirm the observation above that, when using the genuine predictability of the small-$x$ helicity evolution, control over uncertainties is maintained as we extrapolate to smaller $x$. In contrast, since the DGLAP-based fit must use an ad-hoc parameterization of the $x$ dependence, it cannot maintain control over the uncertainties into the extrapolation region. 

\subsection{Imposing additional constraints}
\label{sec:additional_constraints}

While future data from the EIC is a promising way to resolve the issue of sizeable uncertainties in our extracted hPDFs at small $x$, it is worth considering other options that might be more immediately accessible. Ideally these constraints would enter in the form of existing data or as theoretical constraints on the initial conditions.

The hPDF with largest uncertainty that we have extracted is $\Delta G(x,Q^2)$, as demonstrated in \fig{hPDF_bands}, so we explored a few options to constrain it. The first such constraint is positivity, which is the statement that the number densities for  positive and negative helicity partons are positive. In particular, for gluons this leads to
\begin{align}\label{eq: positivity}
    |\Delta G(x,Q^2)| < G(x,Q^2) \,,
\end{align}
where $G(x,Q^2)$ is the unpolarized gluon PDF.  (We will set aside issues as to whether Eq.~\eqref{eq: positivity} is strictly satisfied under ($\overline{\rm MS}$) renormalization~\cite{Candido:2020yat,Collins:2021vke,Candido:2023ujx}.) We impose this constraint by checking the value of $\Delta G(x,Q^2)$ in the region $x<x_0=0.1$, and punishing the $\chi^2$ of the fit if the positivity constraint is violated. Unfortunately, by the time our evolution begins, our baseline fit for $\Delta G(x,Q^2)$ and the JAM DGLAP-based $G(x,Q^2)$~\cite{Cocuzza:2022jye,Zhou:2022wzm} are of comparable size. The latter grows much faster at small $x$ than our extraction for $\Delta G(x,Q^2)$, causing the positivity constraint to have a negligible effect. This is perhaps not surprising, given that at small $x$ the unpolarized gluon distribution $G(x,Q^2)$ is eikonal, while $\Delta G(x,Q^2)$ is sub-eikonal, and, hence, suppressed by a power of $x$.

    \begin{figure}[b!]
    \begin{centering}
    \includegraphics[width=0.6\textwidth]
    {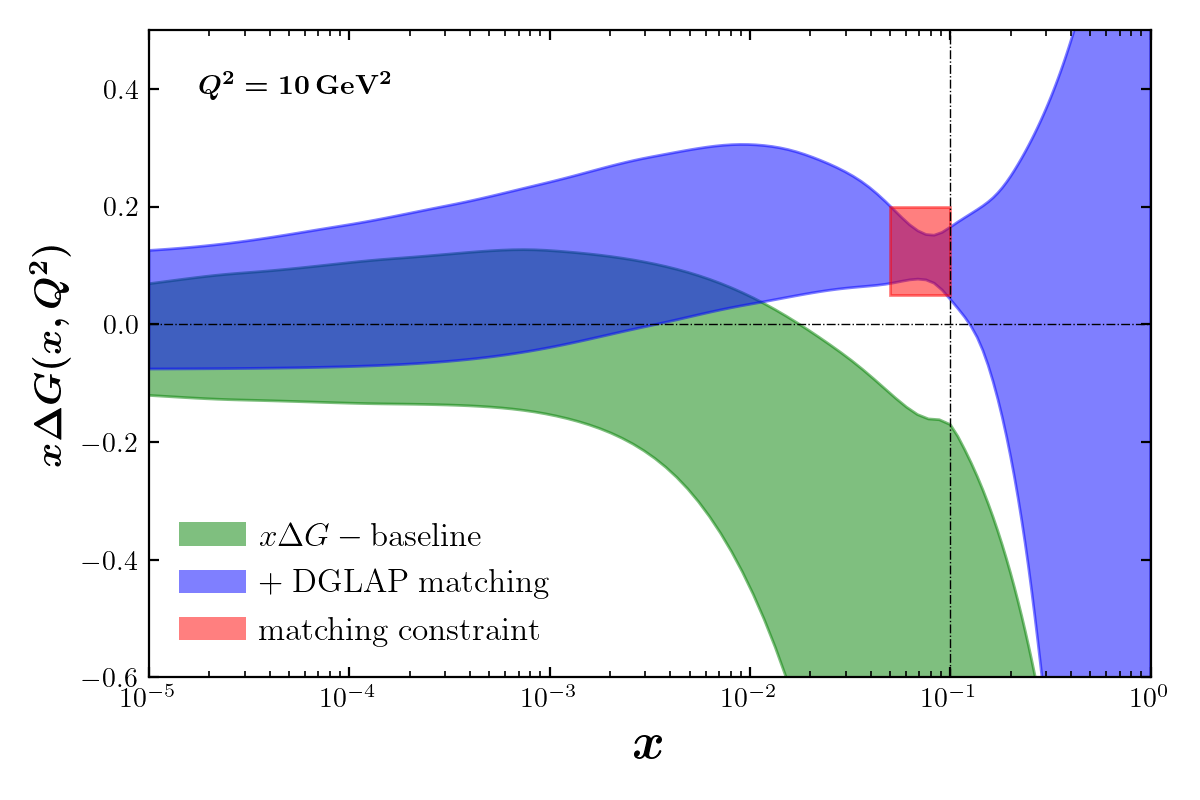}
    \caption{The result of matching onto the $\Delta G(x)$ extraction from DGLAP \cite{Zhou:2022wzm,Cocuzza:2022jye} at intermediate $x$. The green band is our baseline fit. The blue band is the result of matching. The light red square is the region where we enforce matching. 
    \label{Plot: Delta G matching}
    }
    \end{centering}
    \end{figure}

Another constraint on $\Delta G(x,Q^2)$ that we explored was a preliminary matching onto the (large-$x$) JAM DGLAP-based extraction of $\Delta G(x,Q^2)$ in Refs.~\cite{Zhou:2022wzm,Cocuzza:2022jye}, in particular the SU(3)+positivity scenario. The result is shown in \fig{Plot: Delta G matching}; the red box is bounded by $10^{-1.3} < x < 10^{-1}$ and $0.05 < \Delta G(x,Q^2) < 0.2$. The motivation is that any complete description of $\Delta G(x,Q^2)$ should agree with  DGLAP extractions in this region. The matching is performed in a simple way, by choosing an intermediate region in $x$ and forcing our fit of $\Delta G(x,Q^2)$ to qualitatively agree with the JAM DGLAP-based extraction. This is done in a similar way to the positivity constraint described above, whereby we punish the $\chi^2$ whenever $\Delta G(x,Q^2)$ strays outside of the matching region (red rectangle in \fig{Plot: Delta G matching}). This constraint causes our extracted $\Delta G(x,Q^2)$ to take on mostly positive values at small $x$, seemingly changing sign from our original extraction.  However,  note that while the baseline extraction \textit{uncertainty band} grew negative for large $x$, there were still a significant number of replicas (with good $\chi^2$) that grew positive at large $x$ and overlapped with the red region.  Forcing $\Delta G(x,Q^2)$ to pass through that area then preferentially selects those replicas.  Consequently, the whole uncertainty band for $\Delta G(x,Q^2)$  remains shifted upward even in the  small-$x$ region.  Given that $g_1^p(x,Q^2)\propto -\Delta G(x,Q^2)$ (see Eq.~\eqref{e:proportionality}), the matching constraint leads to a quantitative change to the distribution of $g_1^p$ replicas:~they are now 40\% positive and 60\% negative. As we emphasized previously, input on hPDFs from large $x$ can have a significant effect on predictions made at small $x$, motivating future work into a more rigorous matching to DGLAP-based hPDF fits.

Furthermore, the issue with constraining $\widetilde{G}$ could be alleviated by a more rigorous way of handling the starting point of evolution $x_0$. In this work, we chose $x_0=0.1$ and then used experimental data to fit initial conditions for the polarized dipole amplitudes in order to obtain the correct starting values for all of the extracted hPDFs. Only after these starting values have been determined do we then evolve the distributions in a region dominated by our double logarithmic resummation. In reality, evolution in $x$ begins at $x=1$, but is sub-leading, with the dominant contribution at large $x$ given by DGLAP-driven large-$x$ dynamics. The method of matched asymptotic expansions~\cite{MuellerDingle1962, OMalley2014} suggests that we start the evolution at $x_0 =1$, include the DGLAP contributions, but subtract the double counting of logarithms that are present in both resummations. By starting evolution earlier, $\widetilde{G}$ might become more sensitive to the data. As discussed at the end of Sec.~\ref{sec:global}, the challenge in constraining $\widetilde{G}$ stems from the fact that it has a small magnitude in the region where there are measurements (see \fig{xhpdfu_medx}). The magnitude of the $\widetilde{G}$ contribution to $\Delta u^+$ is so small at larger $x$ partly  because $\widetilde{G}$ enters only through evolution, and evolution is delayed until $x_0 =0.1$. If $x_0=1$, $\widetilde{G}$ will start growing sooner, and it might then have a large enough contribution to be sensitive to the experimental data.

Moreover, perhaps the most direct way to constrain $\Delta G$ is to include in the analysis an observable directly sensitive to it. (Recall that in the polarized DIS and SIDIS processes considered here the contribution from the gluon hPDF is suppressed by a factor of $\alpha_s$.) Two possibilities, which have been used in DGLAP-based extractions~\cite{deFlorian:2014yva,Leader:2010rb,Nocera:2014gqa,Zhou:2022wzm,Cocuzza:2022jye}, are jet and hadron production in polarized proton-proton collisions. The numerator of the double-longitudinal spin asymmetry $A_{LL}$ in ${\vec p} + {\vec p}$ collisions takes the following form
\begin{align}\label{pp_coll}
    \sigma^{\downarrow \Uparrow}-\sigma^{\uparrow \Uparrow} = \sum\limits_{a,b}\Delta f_{a/A}\otimes f_{b/B}\otimes \sigma_{ab}\,,
\end{align}
where $\Delta f$ is the parton hPDF for either the quarks or gluon, $a (b)$ is the parton coming from proton $A (B)$, and $\sigma_{ab}$ is the partonic cross section of parton $a$ interacting with parton $b$. For hadron production, \eq{pp_coll} needs also to be convoluted with the $D_1$ FF. More work is needed to derive an analogue of \eq{pp_coll} in the KPS-CTT small-$x$ evolution framework, and initial developments can be found in Ref.~\cite{Li:2023tlw}. 

Lastly, in the future, it will also be interesting to attempt to constrain the large-$x$ behavior of the hPDFs by direct matching onto nonperturbative calculations from lattice QCD. Such matching in the vicinity of $x \sim 0.1$ is actually feasible for the double-logarithmic helicity evolution, unlike for the case of single-logarithmic unpolarized small-$x$ evolution which would require reliable lattice data down to much smaller $x$.
In addition, recently a new approach to determining the initial conditions for small-$x$ evolution by starting at the level of the proton wave function has been developed in Ref.~\cite{Dumitru:2020gla}. While that work was done in the context of unpolarized small-$x$ evolution, it is possible that it could be extended to the polarized case, helping us constrain the initial conditions for helicity evolution at hand.

%
\section{Conclusions}
%

\label{sec:conclusions}

In this paper we have presented the first phenomenological implementation of the KPS-CTT theoretical framework~\cite{Kovchegov:2015pbl, Kovchegov:2018znm,Cougoulic:2022gbk} for the evolution of hPDFs. This work represents a significant improvement over our previous study~\cite{Adamiak:2021ppq} by utilizing the revised evolution equations instead of the original KPS equations. On top of that, we have adopted the large-$N_c \& N_f$ limit, which enables a more realistic description of the physics, now including quarks in addition to gluons. 
Another key advancement of this research is an expansion of our analysis beyond just polarized DIS data by also incorporating polarized semi-inclusive DIS measurements. This allowed us to extract both the $C$-even and $C$-odd quark hPDFs $\Delta q^+$ and $\Delta q^-$, along with the gluon hPDF $\Delta G$. To extract $\Delta q^-$ we had to, for the first time, implement the numeric solution for the KPS evolution of the nonsinglet hPDFs. Moreover, we have included running coupling corrections in the evolution of $\Delta q^+$, $\Delta q^-$, and $\Delta G$, which is another feature of the analysis that makes our approach more rigorous. 

Through the application of the JAM Bayesian Monte Carlo framework, we have successfully described all available polarized DIS and SIDIS data below the threshold $x_0=0.1$, achieving a very good fit with $\chi^2/N_{\rm pts} = 1.03$. However, when attempting to extend our predictions to lower values of $x$, the uncertainty associated with our results was found to be substantial. This large uncertainty arises from the inherent insensitivity of the data to the polarized dipole amplitudes $G_2$ and $\widetilde G$. To address this challenge, we discussed several potential future improvements, among which investigating jet or hadron production in longitudinally polarized proton-proton collisions emerges as a promising medium-term solution. However, more theoretical developments are desirable in the short term, where one must identify the observables which can be expressed in terms of the polarized dipole amplitudes $G_2$ and $\widetilde G$.

Another issue which needs to be clarified in the medium term is the impact of the axial anomaly on the $g_1$ structure function and hPDFs at small $x$. The role of the axial anomaly in the polarized structure functions, originally pointed out in Refs.~\cite{Altarelli:1988nr, Jaffe:1989jz, Shore:1991dv}, has been recently revisited in Refs.~\cite{Tarasov:2020cwl, Tarasov:2021yll, Bhattacharya:2022xxw, Bhattacharya:2023wvy}. The effect appears to be distinct from the DLA of BER and KPS-CTT evolution. Developing the corresponding phenomenology is left for future work.

Based on current experimental data, we find that there could be significant {\it negative} net spin, as well as non-negligible contributions to the triplet and octet axial-vector charges, coming from small-$x$ partons.  However, there are large uncertainties in our estimates, including unaccounted-for systematics in matching onto large-$x$ DGLAP-based fits, which will be important to implement in future work.  Nevertheless, in such a scenario (negative net parton spin), significant OAM would be needed to satisfy the (Jaffe-Manohar) spin sum rule. The inclusion of EIC data in the long term would greatly enhance our understanding of hPDFs, as our impact study showed, and enable more precise statements about the distribution of (spin and orbital) angular momentum within the proton.  

\section*{Acknowledgments}
\label{sec:acknowledgement}

We thank Yiyu Zhou and Chris Cocuzza for providing us replicas and calculations for the JAM DGLAP-based analyses. This work has been supported by the U.S. Department of Energy, Office of Science, Office of Nuclear Physics under Award Number DE-SC0004286 (DA and YK), No. DE-SC0020081 (AT), No.~DE-AC05-06OR23177 (DA, WM and NS) under which Jefferson Science Associates, LLC, manages and operates Jefferson Lab, and the National Science Foundation under Grants No.~PHY-2011763 and No.~PHY-2308567 (DP). The work of NS was supported by the DOE, Office of Science, Office of Nuclear Physics in the Early Career Program, and the work of MS was supported by the DOE, Office of Science, Office of Nuclear Physics under the Funding for Accelerated, Inclusive Research Program.  MS would also like to thank the BNL EIC Theory Institute for hospitality and support during the course of this work.

This work is also supported by the U.S. Department of Energy, Office of Science, Office of Nuclear Physics, within the framework of the Saturated Glue (SURGE) Topical Theory Collaboration. The work of YT has been supported by the Academy of Finland, by the Centre of Excellence in Quark Matter and project no. 346567, and under the European Union’s Horizon 2020 research and innovation programme by the STRONG-2020 project (grant agreement No 824093) and by the European Research Council, grant agreement ERC-2015-CoG-681707. The content of this article does not reflect the official opinion of the European Union and responsibility for the information and views expressed therein lies entirely with the authors.


\appendix

\section{Discretization of the flavor singlet and nonsinglet evolution equations}
\label{Disc_apdx}

In this Appendix we present the process of discretizing Eqs.~(\refeq{eq_LargeNcNf}) and \eqref{NSeq_LargeNcNf} in order to perform the computation to obtain their numerical solutions. In addition, we implement the constraints corresponding to the fact that the starting point of our evolution is at $x = x_0 < 1$. See the discussion above Eqs.~\eqref{DiscreteEvol} for more detail.

We start with the flavor singlet case.  Imposing the $\eta - s_{10} > y_0$, $\eta' - s_{21} > y_0$, $\eta'' - s_{32} > y_0$ constraints with $y_0 = \sqrt{\frac{N_c}{2\pi}}\,\ln\frac{1}{x_0}$, we re-write Eqs.~(\refeq{eq_LargeNcNf}) in terms of the variables \eqref{varchange} as 
\begin{subequations}\label{setaint}
\begin{align}
    Q_q(s_{10},\eta) & = Q_q^{(0)}(s_{10},\eta) + \int_{s_{10}+y_0}^{\eta}\dd\eta'\int_{s_{10}}^{\eta'-y_0}\dd s_{21}\,\alpha_s(s_{21})\Bigl[Q_q(s_{21},\eta')+2\widetilde{G}(s_{21},\eta')+2\widetilde{\Gamma}(s_{10},s_{21},\eta') \label{setaint_Q} \\
    &\;\;\;\;\;\;\;\;\;\;-\overline{\Gamma}_q(s_{10},s_{21},\eta')+2G_2(s_{21},\eta')+2\Gamma_2(s_{10},s_{21},\eta')\Bigr] \notag \\
    &\;\;\;\;\;+ \frac{1}{2}\int_{y_0}^{\eta}\dd\eta'\int_{\max[0, s_{10}+\eta'-\eta]}^{\eta'-y_0}\dd s_{21}\,\alpha_s(s_{21})\Bigl[Q_q(s_{21},\eta') + 2G_2(s_{21},\eta')\Bigr], \notag \\[0.5cm]
    \overline{\Gamma}_q(s_{10},s_{21},\eta') & = Q_q^{(0)}(s_{10},\eta') + \int_{s_{10}+y_0}^{\eta'}\dd\eta''\int_{\max[s_{10},s_{21}-\eta'+\eta'']}^{\eta''-y_0}\dd s_{32}\,\alpha_s(s_{32})\Bigl[Q_q(s_{32},\eta'')+2\widetilde{G}(s_{32},\eta'') \label{setaint_Gmb} \\
    &\;\;\;\;\;\;\;\;\;\;+2\, \widetilde{\Gamma}(s_{10},s_{32},\eta'') -\overline{\Gamma}_q(s_{10},s_{32},\eta'')+2G_2(s_{32},\eta'')+2\Gamma_2(s_{10},s_{32},\eta'')\Bigr] \notag \\
    &\;\;\;\;\;+ \frac{1}{2}\int_{y_0}^{\eta'}\dd\eta''\int_{\max[0, s_{21}+\eta''-\eta']}^{\eta''-y_0}\dd s_{32}\,\alpha_s(s_{32})\Bigl[Q_q(s_{32},\eta'') + 2G_2(s_{32},\eta'')\Bigr], \notag \\[0.5cm]
    \widetilde{G}(s_{10},\eta) & = \widetilde{G}^{(0)}(s_{10},\eta) + \int_{s_{10}+y_0}^{\eta}\dd\eta'\int_{s_{10}}^{\eta'-y_0}\dd s_{21}\,\alpha_s(s_{21})\Bigl[3\widetilde{G}(s_{21},\eta')+\widetilde{\Gamma}(s_{10},s_{21},\eta') \label{setaint_Gt} \\
    &\;\;\;\;\;\;\;\;\;\;+2G_2(s_{21},\eta')+\bigl(2-\frac{N_f}{2N_c}\bigr)\Gamma_2(s_{10},s_{21},\eta')-\frac{1}{4N_c}\sum_q\overline{\Gamma}_q(s_{10},s_{21},\eta')\Bigr] \notag \\
    &\;\;\;\;\;- \frac{1}{4N_c}\int_{y_0}^{\eta}\dd\eta'\int_{\max[0, s_{10}+\eta'-\eta]}^{\min[s_{10},\eta'-y_0]}\dd s_{21}\,\alpha_s(s_{21})\Bigl[\sum_qQ_q(s_{21},\eta') + 2N_fG_2(s_{21},\eta')\Bigr],\notag  \\[0.5cm]
    \widetilde{\Gamma}(s_{10},s_{21},\eta') & = \widetilde{G}^{(0)}(s_{10},\eta') + \int_{s_{10}+y_0}^{\eta'}\dd\eta''\int_{\max[s_{10},s_{21}-\eta'+\eta'']}^{\eta''-y_0}\dd s_{32}\,\alpha_s(s_{32})\Bigl[3\widetilde{G}(s_{32},\eta'')+\widetilde{\Gamma}(s_{10},s_{32},\eta'') \label{setaint_Gmt} \\
    &\;\;\;\;\;\;\;\;\;\;+2G_2(s_{32},\eta'')+\bigl(2-\frac{N_f}{2N_c}\bigr)\Gamma_2(s_{10},s_{32},\eta'') - \frac{1}{4N_c}\sum_q\overline{\Gamma}_q(s_{10},s_{32},\eta'')\Bigr] \notag \\
    &\;\;\;\;\;- \frac{1}{4N_c}\int_{y_0}^{\eta'+s_{10}-s_{21}}\dd\eta''\int_{\max[0, s_{21}+\eta''-\eta']}^{\min[s_{10},\eta''-y_0]}\dd s_{32}\,\alpha_s(s_{32})\Bigl[\sum_qQ_q(s_{32},\eta'') + 2N_fG_2(s_{32},\eta'')\Bigr], \notag \\[0.5cm]
    G_2(s_{10},\eta) & = G_2^{(0)}(s_{10},\eta)+2\int_{y_0}^{\eta}\dd\eta'\int_{\max[0,s_{10}+\eta'-\eta]}^{\min[s_{10},\eta'-y_0]}\dd s_{21}\,\alpha_s(s_{21})\Bigl[\widetilde{G}(s_{21},\eta')+2G_2(s_{21},\eta')\Bigr], \label{setaint_G2} \\[0.5cm]
    \Gamma_2(s_{10},s_{21},\eta') & = G_2^{(0)}(s_{10},\eta')+2\int_{y_0}^{\eta'+s_{10}-s_{21}}\dd\eta''\int_{\max[0,s_{21}+\eta''-\eta']}^{\min[s_{10},\eta''-y_0]}\dd s_{32}\,\alpha_s(s_{32})\Bigl[\widetilde{G}(s_{32},\eta'')+2G_2(s_{32},\eta'')\Bigr]. \label{setaint_Gm2}
\end{align}
\end{subequations}
Following Refs.~\cite{Kovchegov:2020hgb,Cougoulic:2022gbk,Adamiak:2023okq}, the evolution equations \eqref{setaint} can be iterated more optimally by considering the recursive form of their Riemann sums. To do so, we begin by writing Eqs. \eqref{setaint_Q}, \eqref{setaint_Gt} and \eqref{setaint_G2} as the first-order Taylor expansions in $\eta$, {\it e.g.},
\begin{align}\label{TaylorEta}
    &Q_q(s_{10},\eta+\Delta) = Q_q(s_{10},\eta) + \Delta\,\frac{\partial}{\partial\eta}Q_q(s_{10},\eta) + {\cal O} (\Delta^2) \, ,
\end{align}
and Eqs. \eqref{setaint_Gmb}, \eqref{setaint_Gmt} and \eqref{setaint_Gm2} as the first-order Taylor expansions in $\eta'$ and $s_{21}$, {\it e.g.}, 
\begin{align}
    &\overline{\Gamma}_q(s_{10},s_{21}+\Delta,\eta'+\Delta) = \overline{\Gamma}_q(s_{10},s_{21},\eta') + \Delta\,\frac{\partial}{\partial\eta'}\overline{\Gamma}_q(s_{10},s_{21},\eta') + \Delta\,\frac{\partial}{\partial s_{21}}\overline{\Gamma}_q(s_{10},s_{21},\eta') + {\cal O} (\Delta^2) \, .
\end{align}
The expansions for other (neighbor) dipole amplitudes are similar. Note that the transverse sizes in neighbor dipoles are always ordered such that $x_{32}<x_{21}<x_{10}$, which implies that $s_{32}>s_{21}>s_{10}$. Neglecting order-$\Delta^2$ terms for small step sizes $\Delta \ll 1$, Eqs.~\eqref{setaint} can be written as
\begin{subequations}\label{Taylor}
\begin{align}
    Q_q(s_{10},\eta+\Delta) & = Q_q(s_{10},\eta) + Q_q^{(0)}(s_{10},\eta+\Delta) - Q_q^{(0)}(s_{10},\eta)  \label{Taylor_Q} \\
    &\;\;\;\;\;+ \Delta \int_{s_{10}}^{\eta-y_0}\dd s_{21}\,\alpha_s(s_{21})\Bigl[\frac{3}{2}Q_q(s_{21},\eta)+2\widetilde{G}(s_{21},\eta)+2\widetilde{\Gamma}(s_{10},s_{21},\eta) \notag \\
    &\;\;\;\;\;\;\;\;\;\;-\overline{\Gamma}_q(s_{10},s_{21},\eta)+3G_2(s_{21},\eta)+2\Gamma_2(s_{10},s_{21},\eta)\Bigr] \notag \\
    &\;\;\;\;\;+ \frac{1}{2}\Delta\int_{\eta-s_{10}}^{\eta}\dd\eta' \,\alpha_s(s_{10}+\eta'-\eta)\Bigl[Q_q(s_{10}+\eta'-\eta,\eta') + 2G_2(s_{10}+\eta'-\eta,\eta')\Bigr] \, , \notag \\[0.5cm]
  \overline{\Gamma}_q(s_{10},s_{21}+\Delta,\eta'+\Delta) & = Q_q(s_{10},\eta) + Q_q^{(0)}(s_{10},\eta+\Delta) - Q_q^{(0)}(s_{10},\eta) \label{Taylor_Gmb} \\ 
    &\;\;\;\;\;+ \Delta \int_{s_{21}}^{\eta'-y_0}\dd s_{32}\,\alpha_s(s_{32})\Bigl[\frac{3}{2}Q_q(s_{32},\eta')+2\widetilde{G}(s_{32},\eta') \notag \\
    &\;\;\;\;\;\;\;\;\;\;+2\, \widetilde{\Gamma}(s_{10},s_{32},\eta') -\overline{\Gamma}_q(s_{10},s_{32},\eta')+3G_2(s_{32},\eta')+2\Gamma_2(s_{10},s_{32},\eta')\Bigr] \, , \notag \\[0.5cm] 
    \widetilde{G}(s_{10},\eta+\Delta) & = \widetilde{G}(s_{10},\eta) + \widetilde{G}^{(0)}(s_{10},\eta+\Delta) - \widetilde{G}^{(0)}(s_{10},\eta) \label{Taylor_Gt} \\ 
        &\;\;\;\;\;+\Delta\int_{s_{10}}^{\eta-y_0}\dd s_{21}\,\alpha_s(s_{21})\Bigl[3\widetilde{G}[i',j-1]+\widetilde{\Gamma}(s_{10},s_{21},\eta) \notag\\
        &\;\;\;\;\;\;\;\;\;\;+2G_2(s_{21},\eta)+\bigl(2-\frac{N_f}{2N_c}\bigr)\Gamma_2(s_{10},s_{21},\eta)-\frac{1}{4N_c}\sum_q\overline{\Gamma}_q(s_{10},s_{21},\eta)\Bigr] \notag \\
        &\;\;\;\;\;- \Delta\frac{1}{4N_c}\int_{\eta-s_{10}}^{\eta}\dd\eta'\,\alpha_s(s_{10}+\eta'-\eta)\Bigl[\sum_qQ_q(s_{10}+\eta'-\eta,\eta') + 2N_fG_2(s_{10}+\eta'-\eta,\eta')\Bigr]  \, ,\notag  \\[0.5cm]
    \widetilde{\Gamma}(s_{10},s_{21}+\Delta,\eta'+\Delta) &= \widetilde{\Gamma}(s_{10},s_{21},\eta')+ \widetilde{G}^{(0)}(s_{10},\eta'+\Delta) - \widetilde{G}^{(0)}(s_{10},\eta') \label{Taylor_Gmt} \\
        &\;\;\;\;\;+ \Delta\int_{s_{21}}^{\eta'-y_0}\dd s_{32}\,\alpha_s(s_{32})\Bigl[3\widetilde{G}(s_{32},\eta')+\widetilde{\Gamma}(s_{10},s_{32},\eta') \notag \\
        &\;\;\;\;\;\;\;\;\;\;+2G_2(s_{32},\eta')+\bigl(2-\frac{N_f}{2N_c}\bigr)\Gamma_2(s_{10},s_{32},\eta') - \frac{1}{4N_c}\sum_q\overline{\Gamma}_q(s_{10},s_{32},\eta')\Bigr] \, , \notag \\[0.5cm]
    G_2(s_{10},\eta+\Delta) &= G_2(s_{10},\eta) + G_2^{(0)}(s_{10},\eta+\Delta)-G_2^{(0)}(s_{10},\eta) \label{Taylor_G2} \\
        &\;\;\;\;\;+2\Delta\int_{\eta-s_{10}}^{\eta}\dd\eta'\,\alpha_s(s_{10}+\eta'-\eta)\Bigl[\widetilde{G}(s_{10}+\eta'-\eta,\eta')+2G_2(s_{10}+\eta'-\eta,\eta')\Bigr] \,  , \notag \\[0.5cm]
   \Gamma_2(s_{10},s_{21}+\Delta,\eta'+\Delta) &= \Gamma_2(s_{10},s_{21},\eta') + G_2^{(0)}(s_{10},\eta'+\Delta) - G_2^{(0)}(s_{10},\eta') \,  . \label{Taylor_Gm2}
\end{align}
\end{subequations}
Next we discretize the remaining integrals via a left-hand Riemann sum in order to be able to iteratively compute the amplitudes at higher rapidities $\eta$, which are required for the calculation of hPDFs and the $g_1$ structure function at small $x$. This step is most conveniently performed once we make the change of variables, $\{\eta,s_{10},s_{21}\}\to \{j,i,k\}\cdot\Delta$. At the end, Eqs.~\eqref{Taylor} reduce to the discretized Eqs.~\eqref{DiscreteEvol} in the main text. 

The numerical implementation of the flavor nonsinglet evolution equation \eqref{NSeq_LargeNcNf} parallels that of the flavor singlet evolution considered above. We use the variable change from Eqs.~\eqref{varchange} and also require that the flavor nonsinglet evolution starts at $x=x_0$, such that $\eta-s_{10} \approx \sqrt{\frac{N_c}{2 \pi}}\ln{\frac{1}{x}}>\sqrt{\frac{N_c}{2\pi}}\ln{\frac{1}{x_0}} \equiv y_0$. Implementing these modifications in \eq{NSeq_LargeNcNf} gives us the following evolution for $G^\textrm{NS}$, 
\begin{equation}\label{NSeq_LargeNcNf2}
    G^\textrm{NS}(s_{10},\eta) = G^{\textrm{NS} \,(0)}(s_{10},\eta) + \frac{1}{2}\int\limits_{y_0}^{\eta}\dd\eta'\int\limits_{\mathrm{max}[0,s_{10}-\eta+\eta']}^{\eta'-y_0}\dd s_{21} \, \alpha_s (s_{21}) \, G^\textrm{NS}(s_{21},\eta')\, .
\end{equation}
Discretizing the nonsinglet evolution is mostly similar to that of the singlet evolution.  First, we produce a recursion relation using the first order Taylor expansion, simplify it, and discretize it using the left-handed Riemann sum. Differentiating \eq{NSeq_LargeNcNf2} yields
\begin{align}
    \frac{\partial}{\partial\eta}G^\textrm{NS}(s_{10},\eta) = & \, \frac{\partial}{\partial\eta}G^{\textrm{NS} \,(0)}(s_{10},\eta)+\frac{1}{2}\int\limits_{s_{10}}^{\eta-y_0}\dd s_{21}~\alpha_s(s_{21}) \, G^\textrm{NS}(s_{21},\eta) \\ 
    & + \frac{1}{2}\int\limits_{\eta-s_{10}}^{\eta}\dd\eta'~\alpha_s(s_{10}-\eta +\eta ')G^\textrm{NS}(s_{10}-\eta+\eta', \eta')\,, \notag
\end{align}
where we have also employed the $s_{10}>0,\, \eta-s_{10}>y_0$ conditions. Using the Taylor expansion in $\eta$, c.f. \eq{TaylorEta}, we obtain a recursive form of our flavor nonsinglet evolution
\begin{align} \label{e:NSapp_steps}
        G^\textrm{NS}(s_{10},\eta) = &\;G^\textrm{NS}(s_{10},\eta-\Delta\eta)+G^{\textrm{NS} \,(0)}(s_{10},\eta) - G^{\textrm{NS} \,(0)}(s_{10},\eta-\Delta\eta) \\
        & +\frac{1}{2}\Delta\eta\int\limits_{s_{10}}^{\eta-y_0}\dd s_{21}~\alpha_s(s_{21}) \, G^\textrm{NS}(s_{21},\eta)+\frac{1}{2}\Delta\eta\int\limits_{\eta-s_{10}}^{\eta}\dd\eta'~\alpha_s(s_{10}-\eta +\eta ') \, G^\textrm{NS}(s_{10}-\eta+\eta',\eta')\,. \notag
\end{align}
In order to have a numerical solution consistent with the flavor singlet numerical evolution, we again define $\Delta\eta = \Delta s \equiv \Delta$. We also index our numerics in the same way as in the flavor singlet case, $\{\eta,\,\eta',\, s_{10},\,s_{21}\}\to\{j,\,j',\,i,\,i'\}\cdot\Delta$.  Ultimately, Eq.~\eqref{e:NSapp_steps} yields the discretized Eq.~\eqref{eq_nonsinglet_numerical_solution} in the main text.

\section{Analytic cross-check of the numerical solution for the flavor nonsinglet evolution}\label{NS_Crosscheck_apdx}

Finding an analytic solution for the large-$N_c$ flavor nonsinglet evolution equation that enforces all of our physical assumptions \textit{and} includes running coupling is, unfortunately, outside the scope of this paper. However, an analytic solution does exist  for the large-$N_c$ evolution equations with \textit{fixed} coupling \cite{Kovchegov:2016zex}, which ignores the $1/\Lambda$ IR cutoff on the transverse size of the dipoles. We can perform a limited cross-check by modifying our numerical solution to use a fixed coupling $\alpha_s=0.3$, and expand our domain of $s_{10}$ by removing the IR dipole size cutoff, $x_{21} < 1/\Lambda$, employed in Eq.~\eqref{NSeq_LargeNcNf}. Since the dipole size constraint is enforced by the relation $s_{10} > 0$, we refer to the analytic cross-check regime as the all-$s_{10}$ ($\pm s$) regime. 

The revised evolution equation becomes (c.f. Eq.~\eqref{NSeq_LargeNcNf2})
\begin{equation}
    G_{\pm s}^\textrm{NS}(s_{10},\eta) = G_{\pm s}^{\rm NS\,(0)}(s_{10},\eta) + \frac{\alpha_s}{2}\int\limits_0^{\eta}\dd\eta'\int\limits_{s_{10}-\eta+\eta'}^{\eta'-y_0}\dd s_{21} \, G_{\pm s}^\textrm{NS}(s_{21},\eta')\,,
\end{equation}
where relaxing the $s_{ij} > 0$ constraint has extended the lower limits of the $\eta'$ and $s_{21}$ integrals. As expected, changing the phase space of the evolution equation had an effect on our numerical solution, with the discretized flavor nonsinglet equation now being
\begin{align}\label{NS_Num_CC}
        G_{\pm s}^\textrm{NS}[i,j] & = G_{\pm s}^\textrm{NS}[i,j-1]+G_{\pm s}^{\rm NS\,(0)}[i,j] - G_{\pm s}^{\rm NS\,(0)}[i,j-1]  \\ 
        &+\frac{\alpha_s}{2}\Delta^2\Biggl[\sum\limits_{i'=i}^{j-2-y_0}G_{\pm s}^\textrm{NS}[i',j-1]+\sum\limits_{j'=0}^{j-2}G_{\pm s}^\textrm{NS}[i-j+1+j',j']\Biggr], \notag
\end{align}
where the notable modifications compared to Eq.~\eqref{eq_nonsinglet_numerical_solution} are the factoring out of the fixed coupling $\alpha_s$ in front of the sum and the different starting point $j'=0$ of the summation.

We can solve the all-$s_{10}$ evolution equation analytically using Laplace-Mellin transforms (c.f. Ref.~\cite{Kovchegov:2016zex}). To enforce the small-$x$ assumption on our conjugate variables, we define the forward and inverse transforms
\begin{subequations}\begin{align}
    G_{\pm s}^\textrm{NS}(s_{10},\eta) & = \int\frac{\dd\omega}{2\pi i}e^{\omega\eta}\int\frac{\dd\lambda}{2\pi i}e^{(\eta - s_{10}-y_0)\lambda}G_{\pm s}^\textrm{NS}(\omega,\lambda)\,, \\
    G_{\pm s}^\textrm{NS}(\omega,\lambda)\,, & = \int_0^{\infty}\dd(\eta-s_{10}-y_0)e^{-\lambda(\eta-s_{10}-y_0)}\int_0^{\infty}\dd\eta\, e^{-\omega\eta}\, G_{\pm s}^\textrm{NS}(s_{10},\eta)\,.
\end{align}\end{subequations}
In Mellin space, the solution presents itself just as it did in Ref.~\cite{Kovchegov:2016zex},
\begin{equation}
    G_{\pm s}^\textrm{NS}(\omega,\lambda) = \frac{\omega\lambda}{\lambda-\frac{\alpha_s}{2\omega}}\frac{1}{\omega}G_{\pm s}^{\rm NS\,(0)}(\omega,\lambda)\,.
\end{equation}
This is convenient since we only have three distinct initial conditions, $G_{\pm s}^{\rm NS\,(0)} = \eta,\,s_{10},\,1$. First, we will evaluate the nonsinglet evolution beginning with the constant contribution, $G_{\pm s}^{\rm NS\,(0)}(\eta,s_{10}) = 1$.
\begin{align}
    G_{\pm s}^{\rm NS\,(0)}(\omega,\lambda) & = \int_0^{\infty}d(\eta-s_{10}-y_0)\,e^{-\lambda(\eta-s_{10}-y_0)}\int_0^{\infty}\dd\eta\, e^{-\omega\eta} 
     = \frac{1}{\omega\lambda} \,.
\end{align}
Plugging this into the evolution equation leads to another contour integral with a pole at $\lambda = \alpha/(2\omega)$, which is evaluated via the residue theorem to give
\begin{align}
    G_{\pm s}^\textrm{NS}(\eta,s_{10}) & = \int\frac{\dd\omega}{2\pi i}\,e^{\omega\eta}\int\frac{\dd\lambda}{2\pi i}\,e^{(\eta-s_{10}-y_0)\lambda}\frac{1}{\lambda-\frac{\alpha_s}{2\omega}}\frac{1}{\omega}  
     = \int\frac{\dd\omega}{2\pi i}e^{\omega\eta+\frac{\alpha_s}{2\omega}(\eta-s_{10}-y_0)}\frac{1}{\omega}\,. 
\end{align}
Now we can Taylor-expand the singular ($\sim 1/\omega$) part of the exponential, use the residue theorem, and simplify the result, obtaining
\begin{align}
    G_{\pm s}^\textrm{NS}(\eta,s_{10}) & = \int\frac{\dd\omega}{2\pi i}e^{\omega\eta}\sum\limits_{n=0}^{\infty}\frac{1}{n!}\left(\frac{\alpha_s (\eta-s_{10}-y_0)}{2\omega}\right)^n\frac{1}{\omega}  
    = \sum\limits_{n=0}^{\infty}\frac{1}{(n!)^2}\left(\frac{\alpha_s}{2}\eta(\eta-s_{10}-y_0)\right)^n \,.
\end{align}
This infinite sum is equivalent to the modified Bessel function of the first kind, $I_m (z)$ at $m=0$,
\begin{equation}
    G_{\pm s}^{NS,\,1}(\eta,s_{10}) = I_0\bigl(\sqrt{\alpha_s}\sqrt{2\eta(\eta-s_{10}-y_0)}\bigr).
\end{equation}
%

    \begin{figure}[b]
    \begin{centering}
    \includegraphics[width=500pt]{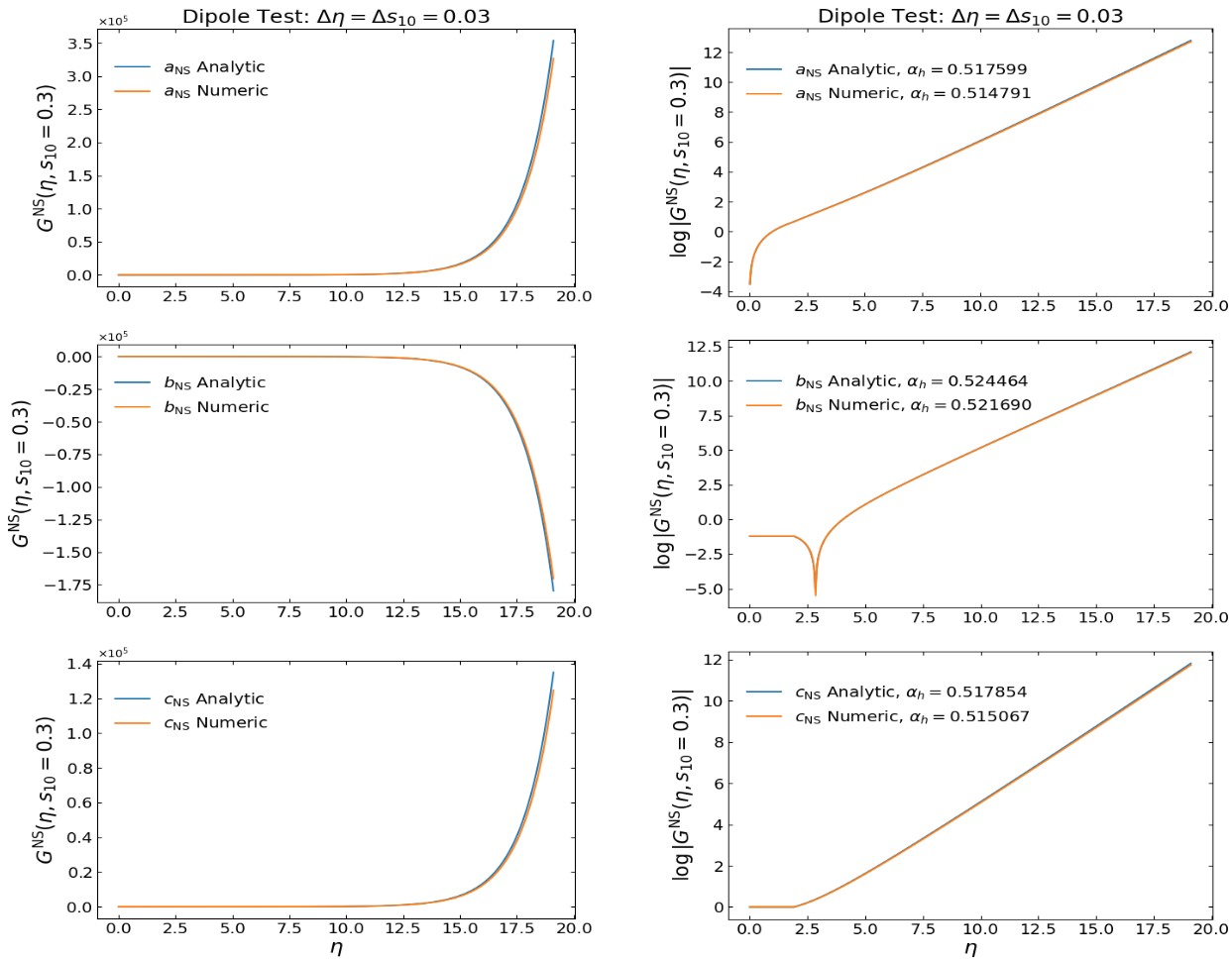}
    \caption{$G^{\textrm{NS}}\,(s_{10}, \eta)$ (left) and $\log{\abs{G^\textrm{NS}\,(s_{10}, \eta)}}$ (right) plotted as functions of $\eta$ for a fixed value of $s_{10} = 0.3$. The large-$\eta$ behavior corresponds to the small-$x$ behavior, and this allows us to see how and when our numerical solution deviates from the analytic. The absolute value of the logarithm allows us to investigate the sign change (the cusp), and the slope of the logarithmic plot will give us a dipole amplitude-level intercept.
        \label{NS_Crosscheck}
    }
    \end{centering}
    \end{figure}

This process is repeated for the $\eta$ initial condition, $G_{\pm s}^{\rm NS\,(0)}(\eta,s_{10}) = \eta$, giving
\begin{align}
    G_{\pm s}^{\rm NS\,(0)}(\omega,\lambda) & = \int_0^{\infty}\dd(\eta-s_{10}-y_0)\,e^{-\lambda(\eta-s_{10}-y_0)}\int_0^{\infty}\dd\eta\, e^{-\omega\eta}\eta  = \frac{1}{\omega^2\lambda} \,.
\end{align}
This contour integral has the same pole at $\lambda=\alpha_s/(2\omega)$, resulting in a similar integral,
\begin{equation}
    G_{\pm s}^\textrm{NS}(\eta,s_{10})  = \int\frac{\dd\omega}{2\pi i}e^{\omega\eta+\frac{\alpha}{2\omega}(\eta-s_{10}-y_0)}\frac{1}{\omega^2}\,.
\end{equation}
We use the same Taylor expansion and the above expression for the $\omega$ contour integral to obtain
\begin{align}
    G_{\pm s}^\textrm{NS}(\eta,s_{10}) & = \int\frac{\dd\omega}{2\pi i}e^{\omega\eta}\sum\limits_{n=0}^{\infty}\frac{1}{n!}\left(\frac{\alpha_s (\eta-s_{10}-y_0)}{2}\right)^n\frac{1}{\omega^{n+2}}  = \eta\,\sum\limits_{n=0}^{\infty}\frac{1}{(n!)(n+1)!}\left(\frac{\alpha_s}{2}\eta(\eta-s_{10}-y_0)\right)^n. 
\end{align}
This, too, is proportional to a modified Bessel function of the first kind, now for $m=1$. The solution is then rewritten as
\begin{equation}
    G_{\pm s}^{NS,\,\eta}(\eta,s_{10}) = \frac{1}{\sqrt{\alpha_s}}\sqrt{\frac{2 \, \eta}{\eta-s_{10}-y_0}} \, I_1\bigl(\sqrt{\alpha_s}\sqrt{2\eta(\eta-s_{10}-y_0)}\bigr).
\end{equation}
Lastly, we must solve for the initial condition term $G_{\pm s}^\textrm{NS}(\eta,s_{10}) = s_{10}$. Noting that $s_{10} = \eta-(\eta-s_{10}-y_0)-y_0$,
\begin{align}
    G_{\pm s}^{\rm NS\,(0)}(\omega,\lambda) & = \int_0^{\infty}\dd(\eta-s_{10}-y_0)\,e^{-\lambda(\eta-s_{10}-y_0)}\int_0^{\infty}\dd\eta\, e^{-\omega\eta}(\eta-(\eta-s_{10}-y_0)-y_0)  = \frac{\lambda-\omega-y_0\omega\lambda}{(\omega\lambda)^2}\,.
\end{align}
In this case we have two poles: $\lambda = 0,\,\alpha_s/(2\omega)$. Conveniently, there are no poles in $\omega$ at $\lambda = 0$, so that particular integral vanishes. Moving forward with the other $\lambda$ pole we write
\begin{align}
    G_{\pm s}^\textrm{NS}(\eta,s_{10}) & = \int\frac{\dd\omega}{2\pi i}\,e^{\omega\eta}\int\frac{\dd\lambda}{2\pi i}\,e^{(\eta-s_{10}-y_0)\lambda}\frac{\omega\lambda}{\lambda-\frac{\alpha_s}{2\omega}}\frac{1}{\omega}\frac{\lambda-\omega-y_0\omega\lambda}{(\omega\lambda)^2} \notag \\ 
    & = \int\frac{\dd\omega}{2\pi i} \, e^{\omega\eta+\frac{\alpha_s}{2\omega}(\eta-s_{10}-y_0)}\left(\frac{1}{\omega^2} - \frac{2}{\alpha_s} - \frac{y_0}{\omega}\right). 
\end{align}
This equation is a linear combination of the two other contributions we derived, plus a new term. This new term can be evaluated in the same way as the previous two. We obtain the following result for the $s_{10}$ contribution:
\begin{align}
    G_{\pm s}^{\textrm{NS},\,s_{10}}(\eta,s_{10}) & = \frac{1}{\sqrt{\alpha_s}}\sqrt{\frac{2 \, \eta}{\eta-s_{10}-y_0}}I_1\bigl(\sqrt{\alpha_s}\sqrt{2\eta(\eta-s_{10}-y_0)}\bigr) \\
     &\;\;\; - \frac{1}{\sqrt{\alpha_s}}\sqrt{\frac{2(\eta-s_{10}-y_0)}{\eta}}I_1\bigl(\sqrt{\alpha_s}\sqrt{2\eta(\eta-s_{10}-y_0)}\bigr) \notag \\
     & \;\;\; -y_0\,I_0\bigl(\sqrt{\alpha_s}\sqrt{2\eta(\eta-s_{10}-y_0)}\bigr). \notag
\end{align}
In the end, we arrive at an analytic solution for the flavor nonsinglet evolution equation in the all-$s_{10}$ regime,
\begin{align}\label{e:NS_analytic}
    G_{\pm s}^\textrm{NS}(\eta,s_{10}) & = a^{\mathrm{NS}}\,G_{\pm s}^{NS,\,\eta} + b^{\mathrm{NS}}\,G_{\pm s}^{NS,\,s_{10}} + c^{\mathrm{NS}}\,G_{\pm s}^{NS,\,1} \\
    &=\; a^{\mathrm{NS}} \,  \frac{1}{\sqrt{\alpha_s}}\sqrt{\frac{2 \, \eta}{\eta-s_{10}-y_0}} \, I_1\bigl(\sqrt{\alpha_s}\sqrt{2\eta(\eta-s_{10}-y_0)}\bigr) + \frac{b^{\mathrm{NS}}}{\sqrt{\alpha_s}}\sqrt{\frac{2 \, \eta}{\eta-s_{10}-y_0}} \, I_1\bigl(\sqrt{\alpha_s}\sqrt{2 \, \eta(\eta-s_{10}-y_0)}\bigr) \notag \\
     &\;\;\; - \frac{b^{\mathrm{NS}}}{\sqrt{\alpha_s}}\sqrt{\frac{2(\eta-s_{10}-y_0)}{\eta}} \, I_1\bigl(\sqrt{\alpha_s}\sqrt{2\eta(\eta-s_{10}-y_0)}\bigr) - b^{\mathrm{NS}}y_0\, I_0\bigl(\sqrt{\alpha_s}\sqrt{2\eta(\eta-s_{10}-y_0)}\bigr) \notag \\
     &\;\;\; + \;c^{\mathrm{NS}}\, I_0\bigl(\sqrt{\alpha_s}\sqrt{2\eta(\eta-s_{10}-y_0)}\bigr). \notag 
\end{align}
%

    \begin{figure}[t!]
    \begin{centering}
    \includegraphics[width=500pt]{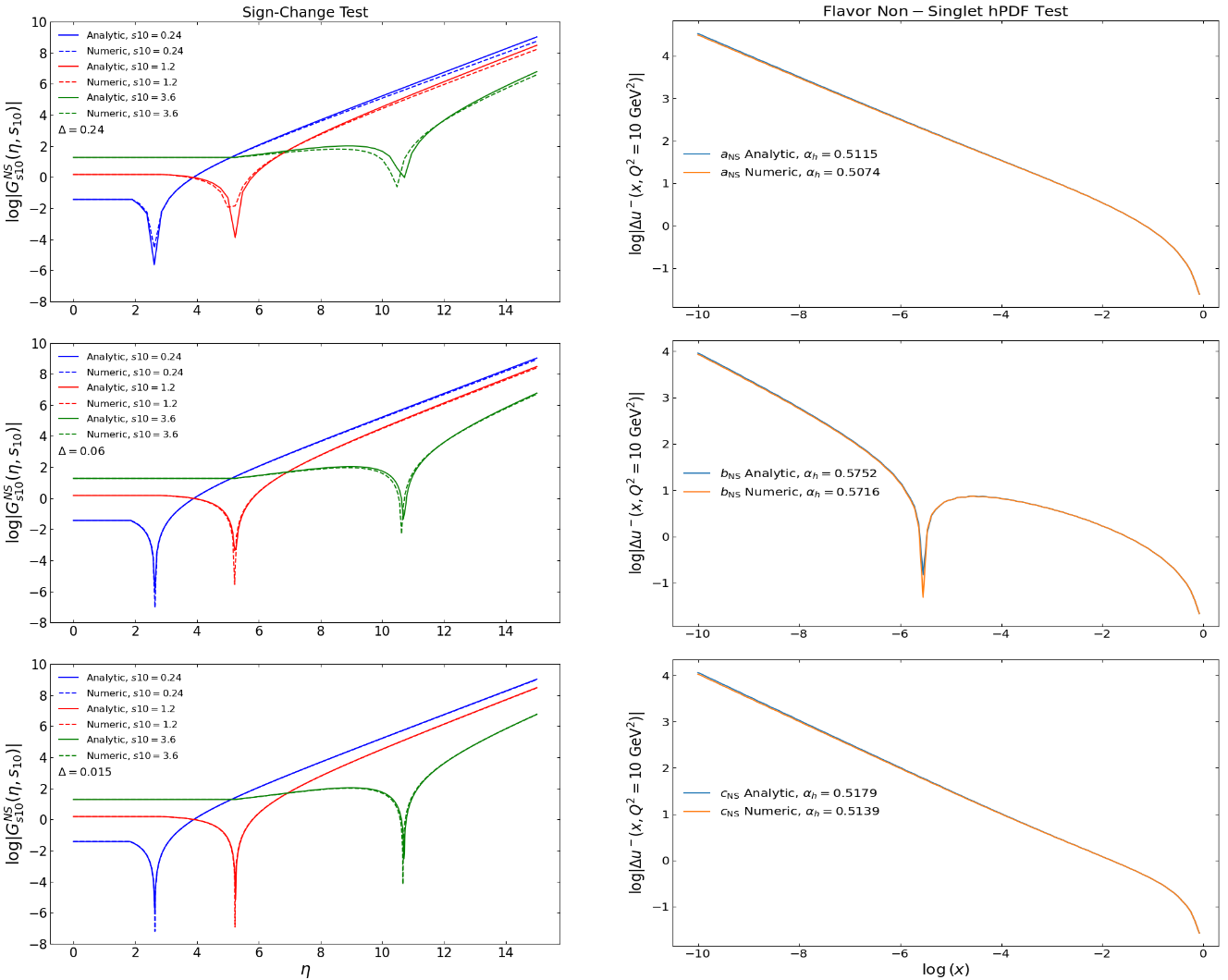}
    \caption{ (Left) A plot of (the logarithm of) the $s_{10}$ contribution to $G_u^{\mathrm{NS}}$ (parameterized by $b_u^{\mathrm{NS}}$) as a function of $\eta$. Each color represents a different fixed value of $s_{10}$. The location of the sign-change in the amplitude, indicated by the cusp, appears to vary with $s_{10}$. Smaller step sizes lead to convergence of the sign-change between the analytic and numeric solutions, and $\Delta\eta=\Delta s_{10} = \Delta < 0.06$ retains small-$x$ agreement. (Right) A plot of (the logarithm of) each $\Delta u^-$ basis functions (parameterized by $a_u^{\mathrm{NS}}$, $b_u^{\mathrm{NS}}$, and $c_u^{\mathrm{NS}}$) as a function of $\log{(x)}$. Each plot depicts the asymptotic agreement between the numeric and analytic solutions, as well as a measure of the intercept $\alpha_h$.
        \label{Signchange_test}
    }
    \end{centering}
    \end{figure}

The first place to start our comparisons would be the dipole amplitudes themselves. There are three properties of the flavor nonsinglet dipole amplitudes that we can use to cross-check the numerical solution:~the general shape of the amplitudes, a sign change in the $s_{10}$ contributions due to the positive starting point and negative growth, and the asymptotic behavior at small $x$. The last property is also useful for checking the implementation of our hPDF calculation, since the dipole amplitudes and hPDFs should have the same asymptotics.

We show in \fig{NS_Crosscheck} high-resolution (small step-size) numerical solutions of the polarized dipole amplitudes, as functions of $\eta$ for a fixed $s_{10}$, compared to their analytic counterparts. The general shape and growth of the flavor nonsinglet amplitudes (see the left panels in \fig{NS_Crosscheck}) shows a good agreement between the numerical and analytic solutions with a reasonably small step-size of $\Delta\eta=\Delta s_{10} = \Delta = 0.03$. One can see that the analytic solution grows in magnitude slightly faster than the numeric solution. The logarithm of the absolute value of the dipole amplitudes, plotted in the right panels of \fig{NS_Crosscheck}, reveals further quantitative agreement, where we see that the numerical intercept $\alpha_h$ converges to within 1.4\% of the analytic solution. The logarithmic scale also allows us to compare the two solutions' large-$x$ (low $\eta$) behaviors using the location of the sign change (the cusp) in the $b_{NS}$ contribution (the middle right panel of \fig{NS_Crosscheck}). The lower the fixed $s_{10}$ value, the lower the sign change.  We see in \fig{NS_Crosscheck} that when $s_{10}=\mathrm{const} = 0.3$ the sign changes coincide just above $\eta=2.5$, implying that our numerical solution is equally valid as $x\to x_0$. Furthermore, we can delay the sign-change by increasing $s_{10}$ for these plots, and that will allow us to to determine the necessary resolution for retaining agreement as $x$ becomes small. This test is given by the left-hand panel of \fig{Signchange_test}, which informs us that a resolution of $\Delta\eta=\Delta s_{10} = \Delta < 0.06$ will retain analytic agreement at the dipole amplitude level. We routinely use $\Delta \leq 0.025$ for our numerics and global analysis.

The polarized dipole amplitude-level agreement gives us confidence to compare how each solution impacts our observables $\Delta q^-$. We employ the plots on the right-hand panel of \fig{Signchange_test} to extract the intercept of the $\ln {|\Delta u^-|}$ basis functions and confirm that the  hPDFs asymptotics given by the analytic and numerical dipole amplitudes match within 1\%, and are consistent with the intercept that was computed at the dipole amplitude level. This completes the cross-check of our numerical solution for the flavor nonsinglet evolution equations.

\section{Convergence testing of numerical solutions}\label{Convergence}

The discretization defined in Appendix \ref{Disc_apdx} is very useful for solving complicated integral equations which are very difficult if not impossible to solve analytically. The numerical solution is rather straightforward to derive, but it has the same faults as any discrete function, namely the fact that the accuracy of a numerical solution is dependent on the resolution,  {\it i.e.}, the step size. In our case, we have two different variables to work with ($\eta$, $s_{10}$), which results in a two-dimensional grid ($G[i,\,j]$) for our numerical solution to compute. To simplify the discretization, we defined the step sizes for $\eta$ and $s_{10}$ to be the same, $\Delta\eta=\Delta s_{10} \equiv \Delta$. The requirement we impose on our numerical solution to confirm its validity is that as the step-size decreases, the computed values should converge to a single output.

\begin{figure}[h!]
    \begin{centering}
    \includegraphics[width=500pt]
    {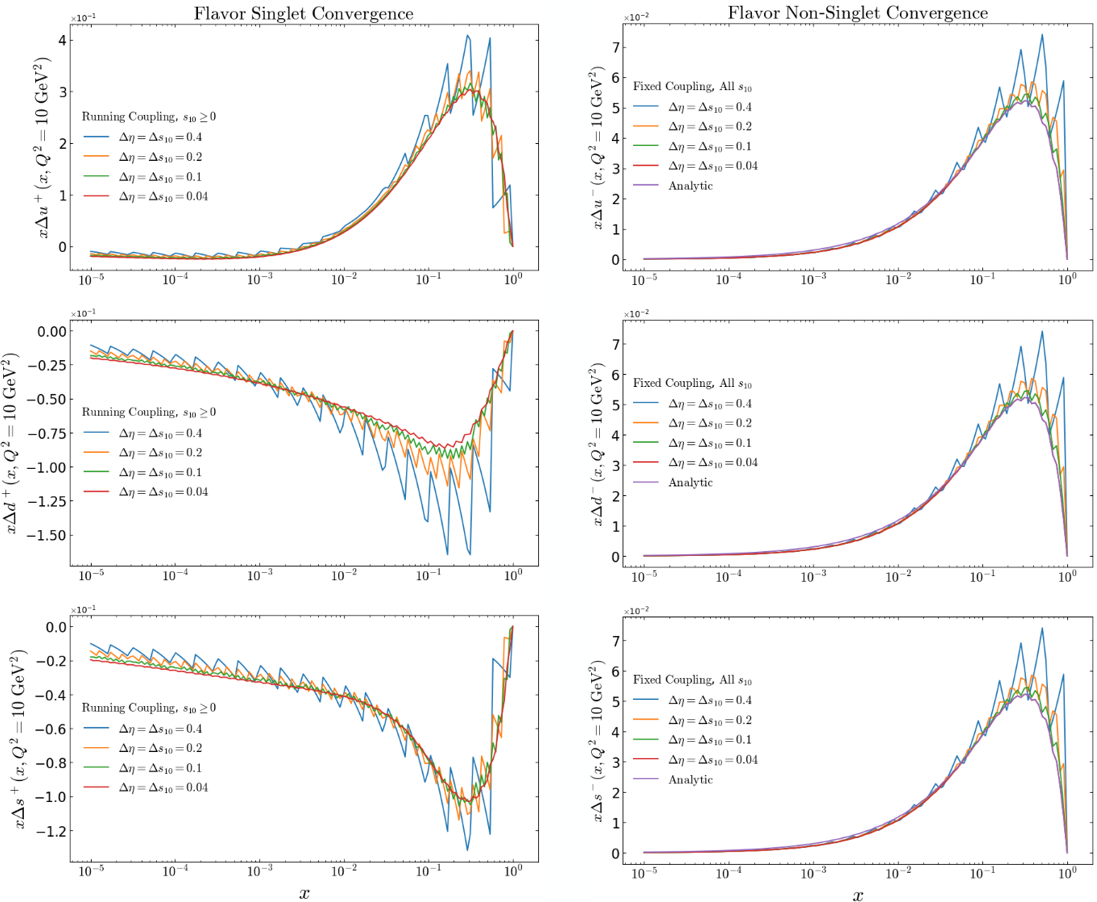}
    \caption{(Left) A numerical computation of $x\Delta q^+(x)$ for a test state of initial conditions. The graph shows the same numerical solution for various choices of step size, $\Delta = \Delta\eta = \Delta s_{10}$. As the step-size $\Delta$ decreases, our numerical solution converges to a single result. (Right) A numerical computation of $x\,\Delta q^-(x)$ that shows the convergence to a single output as $\Delta$ decreases.  For both $x\Delta q^+(x)$ and $x\Delta q^-(x)$, the single output is described by the analytic solution \eqref{e:NS_analytic}.
        \label{Convergences}
    }
    \end{centering}
    \end{figure}

We have tested each of our flavor singlet basis functions (\fig{xhpdfu_medx}) as well as the flavor nonsinglet basis functions (not shown). However, the results can be summarized by their subsequent implementation in calculating the hPDFs $\Delta q^+(x)$ and $\Delta q^-(x)$. The left-hand panel of \fig{Convergences} shows $x\Delta u^+(x)$ for a ``test state" of initial conditions. We define a test state simply as any replica that has been confirmed to fit data with $\tfrac{\chi^2}{N_{\mathrm{pts}}}\approx1$. This hPDF was plotted multiple times for varying step-sizes, and it is clear that as the step-size decreases the solutions converge to a single output. 

The same convergence test was conducted on $x\Delta q^-(x)$ and is displayed in the right panel of \fig{Convergences}. In this case there is also an analytic solution, as discussed in Appendix \ref{NS_Crosscheck_apdx}. We find not only a convergence of the numerical solution to a single output as $\Delta$ becomes smaller, but also that the converged output is exactly that of the analytic solution.  
We note here that \fig{Convergences} is a demonstration of the convergence. The results discussed in Sec.~\ref{sec:results} were computed using much higher resolutions, $\Delta \approx 0.02$. 

\end{document}